\documentclass[authoryear,preprint,3p,11pt]{elsarticle}

\usepackage{tikz}
\usetikzlibrary{arrows,intersections}
\usepackage{natbib}
\usepackage{epsfig}
\usepackage{float}
\usepackage{graphicx}
\usepackage{subcaption}
\usepackage{epstopdf}
\usepackage[colorlinks,linkcolor=blue,hyperindex,CJKbookmarks]{hyperref}
\usepackage{array}
\usepackage{latexsym,bm,amsmath,amssymb,amsfonts,amsthm,mathrsfs}
\usepackage{mathrsfs}
\usepackage{xcolor}
\usepackage[toc,page,title,titletoc,header]{appendix}
\usepackage{threeparttable}
\usepackage{longtable}
\usepackage{lscape}
\usepackage{multirow}
\usepackage{url}
\usepackage{comment}
\usepackage{soul}
\usepackage{tabulary}
\usepackage[displaymath,mathlines,edtable]{lineno}
\usepackage{lipsum}
\usepackage{cuted}
\usepackage[margins]{trackchanges}
\usepackage{booktabs}

\usepackage{etoolbox}
\newcommand*\linenomathpatch[1]{%
  \cspreto{#1}{\linenomath}%
  \cspreto{#1*}{\linenomath}%
  \csappto{end#1}{\endlinenomath}%
  \csappto{end#1*}{\endlinenomath}%
}
\linenomathpatch{equation}
\linenomathpatch{gather}
\linenomathpatch{multline}
\linenomathpatch{align}
\linenomathpatch{alignat}
\linenomathpatch{flalign}

\newcommand{\EQ}{\begin{eqnarray}}
\newcommand{\EN}{\end{eqnarray}}
\newcommand{\EQQ}{\begin{eqnarray*}}
\newcommand{\ENN}{\end{eqnarray*}}

\newcommand{\bremark}{\begin{remark} \begin{rm} }
\newcommand{\eremark}{ \end{rm} \rule{1mm}{2mm}\end{remark} }

\newcommand{\basm}{\begin{assumption} \begin{rm}}
\newcommand{\easm}{\end{rm}\end{assumption}}

\newcommand{\btheorem}{\begin{theorem} \begin{rm} }
\newcommand{\etheorem}{ \end{rm} \rule{1mm}{2mm}\end{theorem} }

\newcommand{\blemma}{\begin{lemma} \begin{rm} }
\newcommand{\elemma}{ \end{rm} \rule{1mm}{2mm}\end{lemma} }

\newcommand{\bcorollary}{\medskip\begin{corollary} \begin{rm} }
\newcommand{\ecorollary}{ \end{rm} \rule{1mm}{2mm}\end{corollary} }

\newcommand{\bdefinition}{\medskip\begin{definition}\begin{rm} }
\newcommand{\edefinition}{ \end{rm} \rule{1mm}{2mm}\end{definition} }

\newcommand{\bproposition}{\medskip\begin{proposition} \begin{rm} }
\newcommand{\eproposition}{ \end{rm} \rule{1mm}{2mm}\end{proposition} }

\newcommand{\bexample}{\begin{example} \begin{rm} }
\newcommand{\eexample}{ \end{rm} \rule{1mm}{2mm}\end{example} }

\newcommand{\bproof}{\begin{proof} \begin{rm} }
\newcommand{\eproof}{ \end{rm} \rule{1mm}{2mm}\end{proof} }

\newtheorem{theorem}{\bf Theorem}[section]
\newtheorem{lemma}{\bf Lemma}[section]
\newtheorem{definition}{\bf Definition}[section]
\newtheorem{remark}{\bf Remark}[section]
\newtheorem{corollary}{\bf Corollary}[section]
\newtheorem{proposition}{\bf Proposition}[section]
\newtheorem{example}{\bf Example}[section]
\newtheorem{assumption}{\bf Assumption}[section]

\newcommand{\MCFourPanels}[4]{%
    \begingroup
    \makebox[\linewidth][c]{%
        \subcaptionbox{$\alpha_{\mathrm{pr}}^q=0.2$}[0.485\linewidth]{%
            \includegraphics[width=\linewidth]{#1}%
        }%
        \hspace{0.02\linewidth}%
        \subcaptionbox{$\alpha_{\mathrm{pr}}^q=0.4$}[0.485\linewidth]{%
            \includegraphics[width=\linewidth]{#2}%
        }%
    }%
    \par\vspace{0.6em}%
    \makebox[\linewidth][c]{%
        \subcaptionbox{$\alpha_{\mathrm{pr}}^q=0.6$}[0.485\linewidth]{%
            \includegraphics[width=\linewidth]{#3}%
        }%
        \hspace{0.02\linewidth}%
        \subcaptionbox{$\alpha_{\mathrm{pr}}^q=0.8$}[0.485\linewidth]{%
            \includegraphics[width=\linewidth]{#4}%
        }%
    }%
    \endgroup
}

\newcommand{\MCClassPanels}[3]{%
    \begin{subfigure}[t]{0.325\linewidth}
        \centering
        \includegraphics[
            width=\linewidth,
            trim=30bp 0bp 40bp 0bp,
            clip
        ]{#1}
        \caption{Aggregate.}
    \end{subfigure}%
    \hfill
    \begin{subfigure}[t]{0.325\linewidth}
        \centering
        \includegraphics[
            width=\linewidth,
            trim=30bp 0bp 40bp 0bp,
            clip
        ]{#2}
        \caption{Priority class.}
    \end{subfigure}%
    \hfill
    \begin{subfigure}[t]{0.325\linewidth}
        \centering
        \includegraphics[
            width=\linewidth,
            trim=30bp 0bp 40bp 0bp,
            clip
        ]{#3}
        \caption{Standard class.}
    \end{subfigure}%
}

\usepackage[ruled,linesnumbered,lined,boxed,commentsnumbered]{algorithm2e}

\usepackage{autobreak}
\allowdisplaybreaks

\graphicspath{{fig/}}

\journal{\LaTeX}

\begin{document}
\begin{sloppypar}
\begin{frontmatter}

\title{Network Traffic Dynamics in Urban Low-Altitude Air Transport with Projected-Space Macroscopic Fundamental Diagrams}

\author[epfl]{Ran Chen}
\ead{ran.chen@epfl.ch}
\author[epfl]{Can Chen\corref{cor}}
\ead{can.chen@epfl.ch}
\cortext[cor]{Corresponding author}
\author[epfl]{Nikolas Geroliminis}
\ead{nikolas.geroliminis@epfl.ch}

\address[epfl]{Urban Transport Systems Laboratory, École Polytechnique Fédérale de Lausanne, Switzerland}

\begin{abstract}

The growth of Low-Altitude Air Transport (LAAT) systems calls for effective macroscopic modeling tools to manage dense drone traffic. The Macroscopic Fundamental Diagram (MFD) offers a promising framework for this purpose, but its conventional formulations do not directly account for the three-dimensional interactions and path deviations that characterize drone operations. To address this challenge, we introduce a projected-space framework that represents drone motion using projected speed and shortest origin--destination (OD) distances, eliminating the dependence on realized trip lengths that are unknown at trip generation. Under suitable assumptions, we prove the steady-state exactness of the projected-space accumulation-based model and establish deterministic finite-window error bounds for the trip-based estimator. The projected-space accumulation-based formulation is further extended to heterogeneous multi-class systems to accommodate different drone characteristics and operational priorities. Microscopic simulations validate the projected-space formulations under stationary and transient traffic conditions, including hysteresis. Moreover, simulations involving heterogeneous priority classes demonstrate that mixed-priority operations can improve network outflow under the investigated scenarios. These findings establish a theoretical foundation for projected-space macroscopic modeling of LAAT systems and support its application to real-time traffic estimation and future airspace management and control.

\end{abstract}


\end{frontmatter}


\section{Introduction}
\label{sec:intro}

Urban Air Mobility (UAM) has attracted increasing interest as an emerging component of urban transportation systems, supported by advances in unmanned aerial vehicles (UAVs), electric vertical take-off and landing aircraft, and autonomous flight technologies \citep{straubinger2020overview,sengupta2025uam}. Beyond monitoring, where UAVs equipped with high-resolution cameras have provided new opportunities for urban traffic monitoring and data collection \citep{barmpounakis2020new,paipuri2021empirical,kim2024monitoring,xiong2025multi}, the development of low-altitude air transport (LAAT) systems may enable new passenger and cargo services, particularly for time-sensitive applications such as medical supply delivery and urban logistics \citep{li2020overview,archetti2022optimization}. The prospect of large-scale LAAT operations, however, raises important challenges concerning airspace congestion, traffic efficiency, and the coordination of interacting aerial vehicles. Recent studies have further highlighted system-level challenges from complementary perspectives, including capacity limitations induced by dense multi-agent interactions and safety-oriented network planning for vertiports and contingency landing infrastructure \citep{zhao2026joint,sun2027sharp}. While trajectory-based approaches can describe individual aircraft movements and interactions \citep{safadi2023macroscopic,weng2026real}, they do not directly provide a compact representation of network-wide congestion dynamics, motivating the development of macroscopic traffic models for large-scale LAAT management.

The Macroscopic Fundamental Diagram (MFD), which relates network accumulation to aggregate traffic production or outflow, provides a compact and computationally efficient framework for modeling and managing large-scale transportation systems \citep{daganzo2007urban,geroliminis2008existence}. Originally developed for urban road networks, MFD-based models have been widely applied to traffic state estimation and perimeter control \citep{haddad2015robust,sirmatel2019nonlinear,li2021perimeter,chen2022data,ding2026state}, route guidance \citep{sirmatel2018economic,chen2024iterative,wang2026regional}, pricing \citep{yang2019heterogeneity,loder2022optimal}, and public transport operations \citep{ramezani2018dynamic,beojone2023dynamic,li2026two}. Recent studies have also investigated MFD estimation under sparse sensor coverage \citep{maiti2025scaling}. In conventional MFD-based traffic dynamics, the accumulation-based (AB) model represents network outflow using traffic production and a representative trip length. However, its accuracy may deteriorate under transient traffic conditions, where trip completions depend on the recent evolution of the traffic state \citep{mariotte2017macroscopic,leclercq2019macroscopic}. Trip-based (TB) and bathtub models address this limitation by explicitly representing trip progression and heterogeneous travel distances \citep{arnott2013bathtub,daganzo2015distance,lamotte2018morning}. Subsequent developments have incorporated remaining travel distance as an additional state \citep{sirmatel2021modeling}, introduced time delays into AB models \citep{haddad2020adaptive}, formulated agent-based bathtub dynamics in a relative space defined by the remaining distance to destinations \citep{martinez2024relative}, and developed MFD calibration and uncertainty quantification across congestion loading and recovery phases \citep{ma2026calibration}.

\begin{figure}[htbp]
    \centering
    \includegraphics[width=0.9\textwidth]{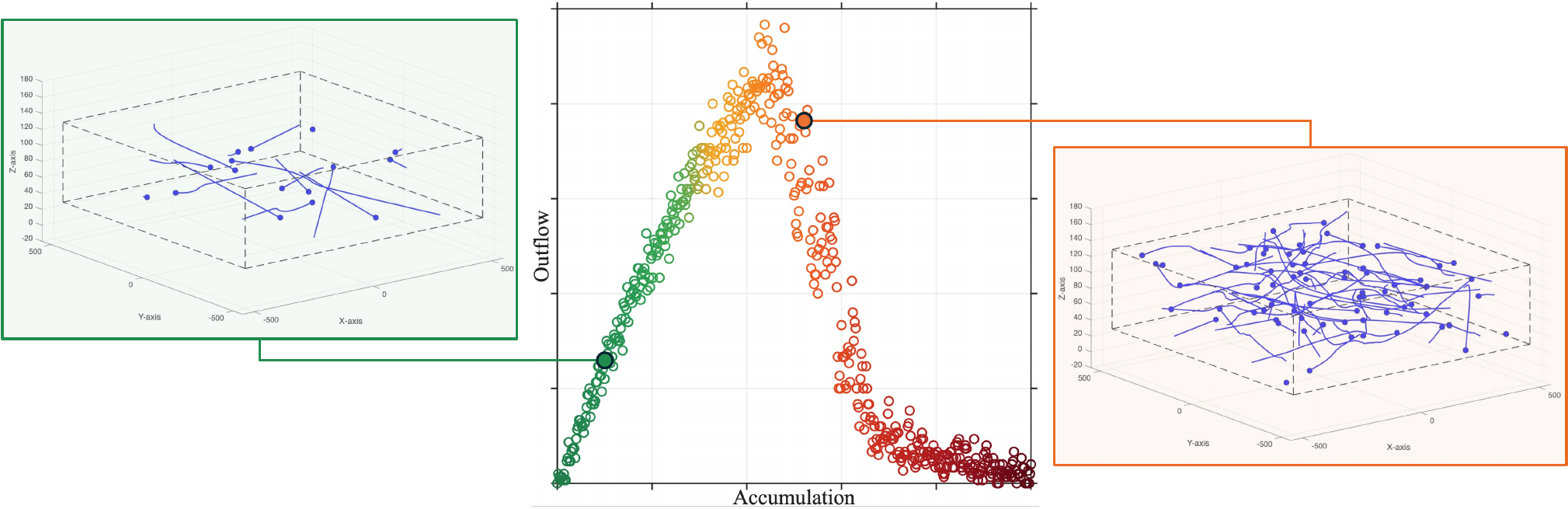}
    \caption{Illustration of microscopic drone trajectories and macroscopic traffic performance in LAAT. The central plot shows the outflow--accumulation MFD, while the surrounding panels illustrate drone trajectories under lower (left) and higher (right) traffic accumulations. }
    \label{fig:XaSimu}
\end{figure}

More recently, MFD-based modeling has been extended to low-altitude aerial transportation. \citet{haddad2021traffic} introduced an MFD-based modeling and feedback-control framework for LAAT, while \citet{safadi2023macroscopic} developed a microscopic--macroscopic simulation framework demonstrating the existence of LAAT MFDs under different operational configurations. \citet{cummings2024airspace} investigated macroscopic flow relationships in unstructured urban airspace, and \citet{knoop2025airplane} examined empirical aviation MFDs using effective travel distances to characterize route inefficiency. Further developments have addressed LAAT network equilibrium and traffic management \citep{weng2025urban}, route guidance with decentralized collision avoidance \citep{weng2026real}, and structured airspace for heterogeneous aerial traffic \citep{fu2026sky}. Nevertheless, LAAT differs fundamentally from conventional road traffic: multidirectional movement in three-dimensional airspace requires frequent collision-avoidance maneuvers, introducing substantial trajectory deviations and traffic-dependent travel distances. For illustration, \autoref{fig:XaSimu} presents an LAAT MFD obtained from a detailed microscopic simulator \citep{safadi2023macroscopic}, together with representative drone trajectories below and above the critical accumulation, respectively.
Drone trajectories exhibit deviations from their shortest OD paths in both regimes, with substantially more pronounced detours under high accumulation due to increasingly frequent collision-avoidance interactions, highlighting the challenge of relating microscopic drone motion to macroscopic traffic performance.

Beyond aggregate traffic efficiency, LAAT systems should accommodate different service requirements, motivating priority-based traffic management. Emergency missions, such as medical deliveries, search-and-rescue, and firefighting support, often require faster responses than routine commercial operations. For example, \citet{chen2026medical} investigated medical drone pre-positioning for cardiac-arrest response, while priority airspace access for public-safety drones has also been demonstrated \citep{nasa2026priority}. In ground transportation, multi-class MFDs have been developed to characterize heterogeneous traffic interactions \citep{geroliminis2014bimodal,huang2022bicycle}, with recent extensions to priority-aware perimeter control and tolling \citep{chen2026perimeter}. In aerial transportation, priority has been incorporated into conflict resolution \citep{jover2022priority,liu2026priority}, flight authorization \citep{carraminana2025authorization}, and strategic traffic flow optimization \citep{du2025mission}. However, these aerial studies primarily address operational-level priority mechanisms, while the resulting macroscopic dynamics of interacting priority classes remain relatively unexplored.

Despite these advances, two challenges remain for macroscopic LAAT modeling.
First, conventional MFD models rely on the mean trip length $L$, which is commonly assumed to remain constant in most literature despite its potential variability across traffic conditions \citep{batista2022exploring}.
This reliance presents two fundamental limitations in LAAT systems: \textbf{(i) trip lengths are only known at completion}, as they depend on subsequent trajectories and collision-avoidance maneuvers; and \textbf{(ii) trip lengths can vary substantially under congestion}, as increasingly frequent interactions lead to path elongation and greater variability across trips.
Consequently, the mean trip length cannot be directly observed at departure, while its dependence on accumulation and other traffic conditions is difficult to characterize, limiting the practical applicability of conventional MFD models.
Moreover, drones may continue moving under severe congestion without making effective progress toward their destinations, similar to pedestrians moving in very congested crowd conditions, complicating the relationship between space-mean speed and network outflow, which is an important physical principle of car-traffic MFD dynamics.

The second challenge is that priority-dependent collision avoidance introduces class-specific speeds, travel times, and trip lengths, even when drones share identical nominal characteristics and OD distributions.
This further complicates multi-class aggregation, as conventional formulations require class-specific mean trip lengths that are neither known at departure nor necessarily stable across operating conditions.
Although effective-distance and remaining-distance representations have been explored in previous studies \citep{sirmatel2021modeling,martinez2024relative,knoop2025airplane}, they do not explicitly account for instantaneous OD-directed progress under arbitrary three-dimensional collision-avoidance maneuvers.
A representation that separates effective progression from trip length is therefore desirable for both macroscopic outflow estimation and multi-class traffic modeling, providing a foundation for future priority-based control.

To address these challenges, this paper develops a projected-space macroscopic modeling framework for LAAT systems. For each drone, the instantaneous velocity is projected onto the fixed direction of its shortest OD path, defining a projected speed that measures effective progression along that direction, as if the drones were moving towards the destination without any interaction. The corresponding projected trip distance equals the shortest OD distance, which is known at departure and remains independent of subsequent trajectory deviations. This representation preserves the effects of collision avoidance through the projected-speed term while avoiding the need to predict the eventual realized travel distance. Furthermore, when different classes share the same expected shortest OD distance, their outflows can be aggregated using this common distance and their class-specific projected speeds, even if their realized trajectories and motion characteristics differ.

The main contributions of this paper are threefold. First, we establish the projected-space representation for multidirectional LAAT traffic and derive the corresponding AB MFD formulation. We identify the conditions under which the projected AB relation is exact in stationary traffic and investigate the limitations of its finite-window implementation. Numerical comparisons with conventional AB estimation demonstrate the advantages of using observable shortest OD distances and projected speeds, particularly under severe congestion. 
Second, we develop a projected TB formulation that explicitly tracks remaining trip progression in projected space and examine its ability to characterize transient traffic dynamics.
Unlike the conventional TB formulation, which requires knowledge of the trip length determined by the unknown future trajectory, the projected formulation initializes the remaining distance using the known shortest OD distance and updates it through estimated projected speeds, without requiring knowledge of future collision-avoidance maneuvers.
Together, the AB and TB formulations provide complementary descriptions of stationary and time-varying LAAT traffic states.
Third, we extend the projected-space framework to heterogeneous multi-class traffic and derive both an aggregate outflow relation and a relationship between generation and accumulation class fractions. An important feature of this extension is that class-specific motion characteristics can be incorporated through their projected speeds while retaining a common expected shortest OD distance, whereas conventional realized-distance formulations generally require class-dependent distance information. Using priority-based collision avoidance as a representative application, we numerically verify the multi-class relations and investigate how priority composition influences class-specific traffic characteristics and aggregate network capacity. These results establish a macroscopic modeling foundation for future priority-based LAAT traffic control.

Despite the current focus on LAAT, the projected-space representation may also be relevant to other mobility systems characterized by multidirectional movement and interaction-dependent trajectory deviations.
For instance, pedestrian crowds exhibit flexible and multidirectional movements, with trajectories that may change dynamically in response to surrounding pedestrians and obstacles \citep{alahi2014robust,hoogendoorn2018macroscopic}.
Non-lane-based bicycle traffic also involves flexible lateral movements and interaction-induced detours, making future trajectories difficult to predict \citep{brunner2024microscopic,wang2025detour}, while lane width, overtaking behavior, and cyclist heterogeneity can significantly influence macroscopic bicycle flow \citep{ni2024bicycle}.
Similar characteristics arise in autonomous mobile robot fleets operating in shared spaces, where collision avoidance and congestion can require alternative paths toward destinations \citep{fragapane2021planning,trinh2022navigation}.
Emerging lane-free vehicle traffic also allows greater flexibility in lateral movements and decentralized interactions, introducing additional deviations from nominal travel directions \citep{papageorgiou2021lane,beza2025lanefree}.
In these systems, distinguishing physical movement from effective destination-oriented progression may provide a useful basis for future macroscopic modeling and control.
The proposed projected-space MFD framework therefore has the potential to be applied to such mobility systems to explicitly account for interaction-induced trajectory deviations and their effects on network-level traffic dynamics.

The remainder of the paper is organized as follows. \autoref{sec:meth} introduces the projected-space representation and develops the corresponding macroscopic modeling framework. It first establishes the projected AB outflow relation and its statistical foundations, then formulates the projected TB model for transient traffic, and finally extends the framework to heterogeneous multi-class operations. \autoref{sec:result} presents numerical experiments based on microscopic LAAT simulations. The experiments first investigate the relationships between true and projected speeds and compare conventional and projected-space outflow estimation under progressive loading. Subsequent experiments evaluate the proposed formulations under transient traffic conditions, followed by a priority-based multi-class study that verifies the theoretical relations, characterizes class-specific traffic behavior, and examines the influence of priority composition on aggregate network capacity. Finally, \autoref{sec:concl} summarizes the main findings, discusses the limitations of the proposed framework, and outlines future research directions, including priority-based macroscopic traffic control.

\section{Methodology}
\label{sec:meth}

This section develops a projected-space macroscopic modeling framework for LAAT systems to address the difficulty of observing trip lengths when drone trajectories are continuously affected by traffic interactions and collision-avoidance maneuvers. The central idea is to represent drone motion through its instantaneous progress along the OD direction, thereby replacing the trajectory-dependent trip length with the shortest OD distance, which is known at departure. This representation enables macroscopic outflow estimation and trip-completion prediction without prior knowledge of realized trajectories, while accommodating heterogeneous drone classes with distinct motion characteristics and operational rules.

We first establish the underlying kinematic representation in \autoref{subsec:projected_space}. Based on this representation, \autoref{subsec:projected_AB} develops an accumulation-based formulation for stationary network outflow estimation, while \autoref{subsec:projected_TB} introduces a trip-based formulation for representing transient trip-completion dynamics. Finally, \autoref{subsec:multi_class} extends both formulations to heterogeneous multi-class systems and establishes the relationship between class compositions at generation and within the active population.

Throughout this section, the term \textit{formulation} distinguishes the proposed projected-space formulations from their conventional true-motion counterparts. Rather than modifying the underlying conservation principles, these variants reformulate the corresponding macroscopic models in terms of OD-directed progress and departure-observable distances. For ease of reference, \autoref{tab:notation} summarizes the main notation\footnote{In \autoref{tab:notation}, superscripts $\mathrm{conv}$ and $\mathrm{proj}$ distinguish the conventional and projected-space variants; subscripts $\mathrm{AB}$, $\mathrm{TB}$, and $\mathrm{MC}$ denote accumulation-based, trip-based, and multi-class formulations.} used throughout the methodology, while additional notation specific to the numerical experiments is introduced in the corresponding sections. Unless otherwise specified, bold symbols (e.g., $\mathbf{n}$) throughout the paper denote vector-valued variables.

\begingroup
\footnotesize
\renewcommand{\arraystretch}{1.12}
\setlength{\LTleft}{0pt}
\setlength{\LTright}{0pt}

\begin{longtable}{@{}p{0.32\linewidth}p{0.46\linewidth}p{0.13\linewidth}@{}}
\caption{Main notation used in the projected-space LAAT macroscopic models.}
\label{tab:notation}\\
\hline
Notation & Description & Unit \\
\hline
\endfirsthead

\caption[]{Main notation (continued).}\\
\hline
Notation & Description & Unit \\
\hline
\endhead

\hline
\endfoot

\multicolumn{3}{@{}l}{\textit{General quantities and projected-space kinematics}} \\
\hline
$i,\ t,\ \tau$ & Trip index and clock-time variables & --, s \\
$t_{0,i},\ T_i,\ c_i$ & Departure time, travel duration, and actual completion time & s \\
$\mathbf o_i,\mathbf d_i,\mathbf x_i(t)$ & Origin, destination, and instantaneous position vectors & m \\
$l_i$ & trip length & m \\
$\mathbf v_i(t),\ v_i(t)$ & Real velocity vector and its magnitude & m/s \\
$\mathbf s_i,\ s_i$ & OD displacement vector and shortest OD distance & m \\
$\theta_i(t),\ u_i(t)$ & Deviation angle and signed OD-projected speed & rad, m/s \\
$\mathcal A(t)$ & Set of active trips at time $t$ & -- \\
$W$ & Measurement-window duration & s \\
$f(\cdot),\ g(\cdot )$ & Generation and active snapshot distributions & -- \\[3pt]

\hline
\multicolumn{3}{@{}l}{\textit{Accumulation-based (AB) formulation}} \\
\hline
$n(t),\ q(t),\ G(t)$ & Accumulation, generation rate, and completion rate & a.c., a.c./s, a.c./s \\
$G(n)$ & Stationary accumulation--outflow relation & a.c./s \\
$V(n),\ U(n)$ & Space-mean speeds and space-mean projected speeds & m/s, m/s \\
$L,\ S$ & Mean trip lengths and mean shortest OD distances & m, m \\
$\eta(n)$ & Mean alignment efficiency & -- \\
$\mathcal D(t,W),\ \mathcal C(t,W)$ & Departure and completion sets over $(t-W,t]$ & --, -- \\
$\bar n(t,W)$ & Time-averaged accumulation & a.c. \\
$\bar V(t,W),\ \bar U(t,W)$ & Finite-window average speeds and projected speeds & m/s, m/s \\
$\bar L_D,\ \bar S_D$ & Departure length means and departure shortest-distance mean & m, m \\
$\hat G_{\mathrm{AB}}^{\mathrm{conv}},\ \hat G_{\mathrm{AB}}^{\mathrm{proj}}$ & Conventional and projected AB outflow estimates & a.c./s, a.c./s \\
[3pt]

\hline
\multicolumn{3}{@{}l}{\textit{Trip-based (TB) formulation and error bounds}} \\
\hline
$r_i(t),\ \hat r_i(t)$ & True and estimated remaining trip length & m, m \\
$p_i(t),\ \hat p_i(t)$ & True and estimated remaining projected progress & m, m \\
$\hat V(t),\ \hat U(t)$ & Space-mean speeds used in trip-state propagation & m/s, m/s \\
$\hat T_i^{\mathrm{conv}},\ \hat T_i^{\mathrm{proj}}$ & Predicted conventional and projected travel times & s, s \\
$\widehat{\mathcal C}_{\mathrm{TB}}^{\mathrm{conv}},\ \widehat{\mathcal C}_{\mathrm{TB}}^{\mathrm{proj}}$ & Estimated completion sets for the two TB variants & --, -- \\
$\hat G_{\mathrm{TB}}^{\mathrm{conv}},\ \hat G_{\mathrm{TB}}^{\mathrm{proj}}$ & Conventional and projected TB outflow estimates & a.c./s, a.c./s \\
$\varepsilon_V,\ \varepsilon_U$ & Maximum space-mean true/projected speed discrepancies & m/s, m/s \\
$V_{\max},\ V_{\mathrm{lb}},\ U_{\mathrm{lb}}$ & Maximum speed and speed lower bounds & m/s, m/s, m/s \\
$T_{\mathrm{ub}},\ \delta_r,\ \delta_p$ & Travel-time upper bound and TB completion-time error bounds & s, s, s \\
$G_{\max}$ & Upper completion-rate parameter in the count envelope & a.c./s \\
$B_{\max}$ & Discrete-completion allowance in the count envelope & a.c. \\[3pt]

\hline
\multicolumn{3}{@{}l}{\textit{Multi-class (MC) extension}} \\
\hline
$\mathcal M,\ m$ & Set of drone classes and class index & --, -- \\
$\mathcal A_m(t),\ n_m(t)$ & Set of active class-$m$ trips and their accumulation & --, a.c. \\
$\mathbf n,\ n_{\mathrm{tot}}$ & Class accumulation vector and total accumulation & a.c., a.c. \\
$U_m(\mathbf n),\ \hat U_m(t)$ & Class-specific stationary and assigned projected speeds & m/s, m/s \\
$S_m,\ G_m(\mathbf n)$ & Mean shortest OD distance and stationary class outflow & m, a.c./s \\
$q_m,\ q_{\mathrm{tot}}$ & Class-specific and total generation rates & a.c./s, a.c./s \\
$\alpha_m^q,\ \alpha_m^n$ & Generation and active-accumulation class fractions & --, -- \\
$\varepsilon_U^m,\ U_{\mathrm{lb}}^m$ & Class-specific projected-speed discrepancy and lower bound & m/s, m/s \\
$\delta_p^m,\ \delta_p^{\mathcal M}$ & Class-specific and worst-class completion-time error bounds & s, s \\
$\hat G_{\mathrm{TB},m}^{\mathrm{proj}},\ \hat G_{\mathrm{TB,MC}}^{\mathrm{proj}}$ & Class-specific and summed projected TB outflow estimates & a.c./s, a.c./s \\[3pt]
\end{longtable}
\endgroup

\subsection{Projected-space representation}
\label{subsec:projected_space}

In LAAT systems, drones may continuously adjust their trajectories to avoid collisions with other drones, resulting in substantial deviations from their shortest OD paths. Consequently, the actual distance traveled by a drone is not generally determined at departure, but depends on subsequent traffic conditions and interactions. This presents a practical challenge for conventional true-motion macroscopic models, in which trip length is an essential quantity for relating traffic accumulation, speed, and trip completion. Although the space-mean speed and position of an active drone can be observed, its eventual trip length remains unknown until completion, making its direct use in real-time macroscopic estimation difficult.

To formalize this distinction, let $\mathbf{x}_i(t)\in\mathbb{R}^3$ denote the position of drone $i$ at time $t$, and let $\mathbf{o}_i$ and $\mathbf{d}_i$ denote its origin and destination vectors, respectively. The instantaneous velocity vector $\mathbf{v}_i(t)$ [m/s] and speed $v_{i}(t)$ [m/s] are defined as
\begin{equation}
    \mathbf{v}_i(t)
    :=
    \dot{\mathbf{x}}_i(t),
    \qquad
    v_i(t)
    :=
    \|\mathbf{v}_i(t)\|.
\end{equation}
We assume that every trip considered in this study is completed within a finite travel time. Let drone $i$ enter the airspace at time $t_{0,i}$ and complete its trip after $T_i<\infty$. Its trip length is then defined as
\begin{equation}
\label{eq:real_length}
    l_i
    :=
    \int_{t_{0,i}}^{t_{0,i}+T_i}
    v_i(\tau)\,d\tau.
\end{equation}
Unlike a prescribed trip length in conventional road-network models, $l_i$ [m] is an endogenous outcome of the realized trajectory and may vary substantially with congestion and collision-avoidance behavior. Therefore, while the total distance traveled up to the current time can be measured, the remaining real distance and eventual total trip length generally cannot be determined without information about future motion. This limitation motivates an alternative representation in which the distance required for trip completion is specified independently of the subsequent trajectory.

\begin{figure}[t]
    \centering
    \includegraphics[width=0.48\linewidth]{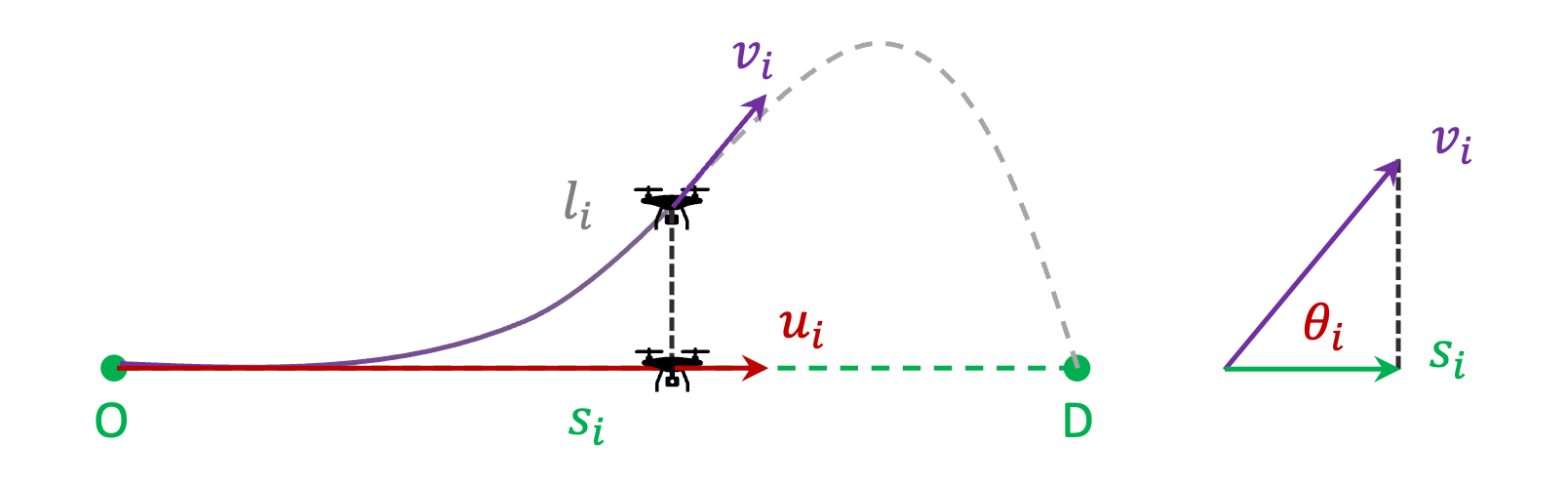}
    \caption{Kinematic illustration of projected speed relative to real velocity and the OD direction. The trip length $l_i$ depends on the complete trajectory, whereas the shortest OD distance $s_i$ is determined at departure.}
    \label{fig:Veff}
\end{figure}

We therefore construct the \textbf{projected-space representation} by defining the OD displacement vector $\mathbf{s}_i$ and the corresponding shortest OD distance $s_{i}$ in meters as
\begin{equation}
    \mathbf{s}_i:=
    \mathbf{d}_i-\mathbf{o}_i,
    \qquad
    s_i
    :=
    \|\mathbf{s}_i\|>0.
\end{equation}
Notably, $s_i$ can be immediately defined once OD is fixed. Let $\theta_i(t)$ [rad] denote the deviation angle between the instantaneous velocity vector $\mathbf{v}_i(t)$ and the fixed OD direction $\mathbf{s}_i$. The instantaneous projected speed $u_{i}(t)$ [m/s] is defined as the component of the velocity vector along this direction:
\begin{equation}
\label{eq:projected_speed}
    u_i(t):
    =
    \frac{\mathbf{s}_i^\top\mathbf{v}_i(t)}{s_i}
    =
    v_i(t)\cos\theta_i(t),
\end{equation}
where $\mathbf{z}^\top$ represents the transpose of the vector $\mathbf{z}$. \autoref{fig:Veff} illustrates the relationship between real and projected motion. The projected speed $u_i(t)$ represents the signed instantaneous rate of progress along the fixed OD direction. Specifically, $u_i(t)>0$ indicates positive OD-directed progress, whereas $u_i(t)<0$ indicates a backward component of motion along the OD direction. When $u_i(t)=0$, the drone makes no instantaneous OD-directed progress, either because it is stationary or because its velocity is perpendicular to the OD direction. Since the OD direction remains fixed throughout a trip, integrating Eq.~\eqref{eq:projected_speed} from departure to any time $t\in[t_{0,i},t_{0,i}+T_i]$ gives
\begin{equation}
\label{eq:progress_at_t}
\begin{aligned}
    \int_{t_{0,i}}^{t}u_i(\tau)\,d\tau
    &=
    \int_{t_{0,i}}^{t}
    \frac{\mathbf{s}_i^\top\mathbf{v}_i(\tau)}{s_i}\,d\tau\\
    &=
    \frac{\mathbf{s}_i^\top}{s_i}
    \left[
        \mathbf{x}_i(t)-\mathbf{x}_i(t_{0,i})
    \right]\\
    &=
    \frac{\mathbf{s}_i^\top}{s_i}
    \left[
        \mathbf{x}_i(t)-\mathbf{o}_i
    \right].
\end{aligned}
\end{equation}
Assuming that the drone departs from $\mathbf{o}_i$ and completes its trip at $\mathbf{d}_i$, the position at completion satisfies
\begin{equation}
\label{eq:sisquare}
\begin{aligned}
    \mathbf{s}_i^\top
    \left[
        \mathbf{x}_i(t_{0,i}+T_i)-\mathbf{o}_i
    \right]
    =
    \mathbf{s}_i^\top
    \left(\mathbf{d}_i-\mathbf{o}_i\right)=
    \mathbf{s}_i^\top\mathbf{s}_i
    =
    s_i^2.
\end{aligned}
\end{equation}
Combining Eqs.~\eqref{eq:progress_at_t} and \eqref{eq:sisquare}, we obtain the following kinematic identity:
\begin{equation}
\label{eq:projected_preservation}
    s_i
    =
    \int_{t_{0,i}}^{t_{0,i}+T_i}
    u_i(\tau)\,d\tau.
\end{equation}
Eq.~\eqref{eq:projected_preservation} reveals that regardless of trajectory deviations or variations in instantaneous speed, the cumulative progress of a completed trip equals its shortest OD distance. Thus, the same completed trip can be described through either the true-motion quantities $(l_i,v_i)$ or the projected-space quantities $(s_i,u_i)$, with both representations satisfying their corresponding distance-integral identities.

The distinction between these representations lies primarily in the observability and dependence of their distance variables. In true-motion space, the trip length $l_i$ depends on the complete trajectory and can only be determined retrospectively. In projected space, the shortest OD distance $s_i$ is fixed by the OD pair and is available at departure, while the effects of congestion, detours, and collision avoidance are reflected in the projected speed $u_i(t)$. Moreover, unlike the nonnegative speed $v_i(t)$, the projected speed is signed and may temporarily become negative during avoidance maneuvers. Nevertheless, as established by Eq.~\eqref{eq:progress_at_t}, every trip is assumed to reach its destination within a finite travel time $T_i$, so its cumulative projected progress at completion necessarily equals $s_i>0$. Therefore, temporary negative projected speeds do not invalidate Eq.~\eqref{eq:projected_preservation}, as the identity concerns the net progress over the entire completed trip.

The projected-space representation therefore separates the distance required for trip completion from the actual motion through which this progress is achieved. This distinction is particularly useful for macroscopic modeling: the distance component can be obtained directly from generated OD pairs, while the motion component can be estimated from the active population. As developed in the following subsections, this common representation supports both stationary AB outflow relations and transient TB formulations, and can be extended to multi-class systems in which different drone classes experience distinct collision-avoidance behaviors and trajectory deviations.

\subsection{Accumulation-based formulation}
\label{subsec:projected_AB}

The accumulation-based (AB) model describes network dynamics through the conservation of active trips. Let $n(t)$ [a.c.] denote the number of active drones in the airspace, $q(t)$ [a.c./s] the trip generation rate, and $G(t)$ [a.c./s] the trip completion rate at time $t$. The accumulation dynamics satisfy
\begin{equation}
    \label{eq:AB_conservation}
    \dot n(t)=q(t)-G(t).
\end{equation}
The AB formulation closes this conservation equation by expressing network outflow as a function of accumulation, $G(t)=G(n(t))$.
By describing network performance through aggregate quantities, it provides a computationally efficient, low-dimensional representation suitable for macroscopic monitoring, prediction, and control, although its accuracy is generally more reliable under stationary or slowly varying traffic conditions.
We first revisit the conventional formulation and its practical limitations before developing the corresponding projected-space formulation.

\subsubsection{Conventional accumulation-based model}
\label{subsubsec:real_motion_AB}

The conventional accumulation-based outflow relation under stationary conditions is expressed as
\begin{equation}
    \label{eq:ABMFD}
    G(n)
    =
    n\frac{V(n)}{L},
\end{equation}
where $G(n)$ [a.c./s] denotes the stationary outflow at accumulation $n$, and $V(n)$ [m/s] and $L$ [m] denote the space-mean speed and mean trip length, respectively.
Stationary conditions refer to an operating state in which accumulation and the underlying traffic statistics remain stable. When these quantities vary sufficiently slowly, the AB relation can also be applied as a quasi-stationary approximation. Following conventional MFD notation, the space-mean speed $V(n)$ is expressed as a function of $n$, while the mean trip length is simply denoted by $L$.

To establish the distributional interpretation of Eq.~\eqref{eq:ABMFD}, let $T$, $l$, and $v$ denote the random quantities associated with individual travel time, trip length, and instantaneous speed, respectively.
Following \cite{sirmatel2021modeling}, we distinguish the \textit{generation distribution} $f(\cdot)$, which characterizes newly generated trips, from the \textit{snapshot distribution} $g(\cdot)$, which characterizes the active drone population observed at an arbitrary instant.
These distributions generally differ because trips with longer travel times contribute more frequently to the snapshot population.
For a given stationary operating condition, both distributions refer to the same underlying traffic conditions. When conditioning on accumulation $n$ is required, we use $f(\cdot\mid n)$ and $g(\cdot\mid n)$, respectively.
Consistent with conventional MFD modeling, we explicitly retain the accumulation dependence of the snapshot speed distribution, while leaving the dependence of the generation trip-length distribution on the operating conditions implicit.
Specifically, let $f_L(l)$ denote the marginal distribution of trip length among generated trips, and $g_V(v\mid n)$ the marginal distribution of instantaneous speed among active drones at accumulation $n$.
The mean quantities are defined as
\begin{equation}
\label{eq:real_motion_representative_quantities}
    L
    :=
    \mathbb E_f[l]
    =
    \int l\,f_L(l)\,dl,
    \qquad
    V(n)
    :=
    \mathbb E_g[v\mid n]
    =
    \int v\,g_V(v\mid n)\,dv.
\end{equation}
The following proposition establishes the conventional AB relation under these distributional definitions.

\begin{proposition}[Steady-state accumulation-based outflow under distributional consistency]
\label{prop:steady_AB_heterogeneous}
Consider a stationary system with constant accumulation $n$, in which all generated trips eventually complete with finite mean travel time. Assume that the mean trip length and snapshot mean speed are also finite, and that each trip satisfies Eq.~\eqref{eq:real_length}. Then, the stationary outflow satisfies
\begin{equation}
    \label{eq:steady_AB_heterogeneous_general}
    G(n)
    =
    n
    \frac{\mathbb E_g[v\mid n]}
    {\mathbb E_f[l]}
    =
    n\frac{V(n)}{L}.
\end{equation}
\end{proposition}

The proof of \autoref{prop:steady_AB_heterogeneous} is provided in \autoref{app:proof_AB_heterogeneous}, together with an illustrative example distinguishing the generation and snapshot distributions. The relation follows from stationary distance conservation: the total distance traveled per unit time by active drones equals the trip completion rate multiplied by the mean trip length of generated trips. Importantly, it permits heterogeneous trip lengths and time-varying within-trip speeds, without requiring independence between trip length and speed. The mean quantities in Eq.~\eqref{eq:ABMFD} are therefore evaluated under different distributions: trip length under the generation distribution and instantaneous speed under the snapshot distribution.

In practice, traffic conditions often evolve over time, and the stationary assumptions underlying Eq.~\eqref{eq:ABMFD} may not hold; thus, estimation errors are expected when traffic conditions change substantially.
Nevertheless, the AB formulation remains a computationally tractable macroscopic approximation for large-scale traffic modeling. For empirical estimation, the mean quantities are approximated over a finite measurement window $(t-W,t]$ of duration $W$ [s], for $t\geq W$.
Let $\mathcal A(\tau)$ denote the set of active drones at time $\tau$. Following \cite{edie1963discussion}, the average network accumulation over this window is defined as
\begin{equation}
\label{eq:finite_accumulation}
    \bar n(t,W)
    =
    \frac{1}{W}
    \int_{t-W}^{t}
    |\mathcal A(\tau)|\,d\tau.
\end{equation}
When traffic conditions vary sufficiently slowly, the space-mean speed can be approximated by
\begin{equation}
\label{eq:finite_real_speed}
    V(\bar n(t,W))
    \approx
    \bar V(t,W)
    =
    \frac{
        \displaystyle\int_{t-W}^{t}
        \sum_{i\in\mathcal A(\tau)}
        v_i(\tau)\,d\tau
    }{
        \displaystyle\int_{t-W}^{t}
        |\mathcal A(\tau)|\,d\tau
    }.
\end{equation}
As $W\to0$, $\bar V(t,W)$ converges to the instantaneous mean speed of active drones.
This snapshot-based averaging is consistent with the theoretical definition $V(n)=\mathbb E_g[v\mid n]$, where the expectation is taken over the snapshot population.
In general, an appropriate window length $W$ should be sufficiently long to smooth short-term fluctuations, but sufficiently short to maintain approximately stationary traffic conditions within the measurement window \citep{leclercq2014macroscopic}.

The mean trip length, on the other hand, is defined under the generation distribution and is therefore naturally approximated using trips entering the airspace during the measurement window. Let
\begin{equation}
\label{eq:mathcalDtM}
    \mathcal D(t,W)
    =
    \left\{
        i:
        t-W<t_{0,i}\leq t
    \right\}
\end{equation}
denote the non-empty set of trips that depart during $(t-W,t]$. The corresponding departure-based sample mean of trip length is
\begin{equation}
\label{eq:Lhatdep}
    L
    \approx
    \bar L_D(t,W)
    =
    \frac{1}{|\mathcal D(t,W)|}
    \sum_{i\in\mathcal D(t,W)}l_i.
\end{equation}
Substituting these empirical averages into the conventional AB relation gives the estimated network outflow:
\begin{equation}
\label{eq:conventional_AB_estimator}
    G(\bar n(t,W))\approx\hat G_{\mathrm{AB}}^{\mathrm{conv}}(t,W)
    =
    \bar n(t,W)
    \frac{\bar V(t,W)}{\bar L_D(t,W)}.
\end{equation}
Notably, Eq.~\eqref{eq:conventional_AB_estimator} is consistent with \autoref{prop:steady_AB_heterogeneous} and with conventional MFD analysis in ground transportation, where trip lengths may be specified through predetermined routes, OD information, or calibrated trip-length distributions \citep{lamotte2018dynamic,batista2019estimation}. In LAAT systems, however, the trip length $l_i$ depends on subsequent trajectories and collision-avoidance maneuvers, and is therefore generally unknown at departure. Consequently, evaluating $\bar L_D(t,W)$ requires retrospective knowledge of the complete trajectories of departing trips, preventing its direct use for real-time estimation. This limitation motivates the projected-space formulation developed in the next subsection, which replaces trip length with the shortest OD distance, known at generation, and uses projected speed to preserve the corresponding trip-level progress identity.

\begin{remark}[Completion-based estimation of mean trip length]
\label{remark:depandcom}
A practical alternative to the departure-based sample mean $\bar L_D(t,W)$, which is not directly available in real time, is to use the lengths of recently completed trips. Let
\begin{equation}
\label{eq:mathcalCtM}
    \mathcal C(t,W)
    =
    \left\{
        i:
        t-W<t_{0,i}+T_i\leq t
    \right\}
\end{equation}
denote the set of trips completed during $(t-W,t]$. When this set is non-empty, the corresponding completion-based sample mean is
\begin{equation}
\label{eq:Lhatcom}
    \bar L_C(t,W)
    =
    \frac{1}{|\mathcal C(t,W)|}
    \sum_{i\in\mathcal C(t,W)}l_i.
\end{equation}
Under stationary and stable operating conditions, provided that all generated trips eventually complete, the generation and completion distributions of trip length coincide. Hence, both $\bar L_D(t,W)$ and $\bar L_C(t,W)$ converge to $\mathbb E_f[l]$ over sufficiently long observation periods. However, if the complete set $\mathcal C(t,W)$ is observable, the outflow over the same window can already be measured directly as $G(t,W)=|\mathcal C(t,W)|/W$, without requiring an AB model, implying that using completion-based trip lengths to reconstruct the outflow over that window offers limited additional value. Therefore, completion-based observations remain useful for retrospective model calibration and validation, but do not resolve the need for an outflow model that can predict traffic dynamics using information available before trip completion.
In contrast, the projected-space formulation introduced next uses shortest OD distances available at departure, providing a more practical basis for real-time estimation and model-based control.
\end{remark}

\subsubsection{Projected-space accumulation-based formulation}
\label{subsubsec:proj_AB}

The projected-space AB formulation addresses the observability limitation of trip lengths by replacing real trip length with OD-directed progress. Let $s$, $u$ and $\theta$ denote the random quantities associated with the individual shortest OD distance, projected speed and directional deviation angle introduced in \autoref{subsec:projected_space}. We define the mean shortest OD distance $S$ as the expectation of $s$ under the generation distribution:
\begin{equation}
\label{eq:S=Efs}
    S
    :=
    \mathbb E_f[s].
\end{equation}
We assume that this expectation remains constant across stationary operating states, without requiring the underlying OD distribution to be invariant.
Unlike the mean trip length $L$, $S$ is unaffected by subsequent collision-avoidance maneuvers. Then, the space-mean projected speed is defined under the snapshot distribution as
\begin{equation}
    U(n)
    :=
    \mathbb E_g[u\mid n].
\end{equation}
We further introduce the mean alignment efficiency as the ratio between the two space-mean speeds:
\begin{equation}
\label{eq:eta}
    \eta(n)
    :=
    \frac{U(n)}{V(n)}.
\end{equation}
This quantity characterizes the speed-weighted mean directional alignment of drone motion with the OD direction.
Individual drones may temporarily exhibit negative directional alignment, while the operating states considered in this study are characterized by positive macroscopic space-mean projected speeds, thus $\eta(n)\in(0,1]$. 
If $v$ and $\cos\theta$ are conditionally independent under the snapshot distribution at accumulation $n$, the mean alignment efficiency coincides with the mean of the cosine of the deviation angle:
\begin{equation}
\label{eq:eta_independence}
\begin{aligned}
    \eta(n)
   = \frac{U(n)}{V(n)} 
    = \frac{\mathbb E_g[v\cos\theta\mid n]}{\mathbb E_g[v\mid n]} 
    = \frac{\mathbb E_g[v\mid n]\mathbb E_g[\cos\theta\mid n]}{\mathbb E_g[v\mid n]} 
    = \mathbb E_g[\cos\theta\mid n].
\end{aligned}
\end{equation}
The projected-space AB relation is summarized in the following proposition.
\begin{proposition}[Steady-state accumulation-based outflow in projected space]
\label{prop:steady_projected_AB}
Consider a stationary system with constant accumulation $n$, in which all generated trips eventually complete with finite mean travel time. Assume that the mean shortest OD distance and snapshot expected projected speed are finite, with $S>0$, and that each trip satisfies Eq.~\eqref{eq:projected_preservation}.
Then, the stationary outflow satisfies
\begin{equation}
\label{eq:steady_projected_AB}
    G(n)
    =
    n
    \frac{\mathbb E_g[u\mid n]}
    {\mathbb E_f[s]}
    =
    n
    \frac{U(n)}{S},
\end{equation}
where the last equality follows from the definitions of $U(n)$ and $S$.
\end{proposition}

The proof of \autoref{prop:steady_projected_AB} is provided in \autoref{app:proof_projectedoutflow}. Eq.~\eqref{eq:steady_projected_AB} establishes the stationary projected-space AB relation through conservation of OD-directed progress. Identical to the conventional AB formulation, the space-mean projected speed is defined over the snapshot population, whereas the mean shortest OD distance is defined over generated trips.

For finite-window estimation, the mean shortest OD distance can be obtained directly from the departure set $\mathcal D(t,W)$ defined in Eq.~\eqref{eq:mathcalDtM}:
\begin{equation}
\label{eq:Shatdep}
    S
    \approx
    \bar S_D(t,W)
    =
    \frac{1}{|\mathcal D(t,W)|}
    \sum_{i\in\mathcal D(t,W)}s_i.
\end{equation}
Similarly, the space-mean projected speed can be approximated by the space-mean projected speed over the measurement window:
\begin{equation}
\label{eq:finite_projected_speed}
    U(\bar n(t,W))
    \approx
    \bar U(t,W)
    =
    \frac{
        \displaystyle\int_{t-W}^{t}
        \sum_{i\in\mathcal A(\tau)}
        u_i(\tau)\,d\tau
    }{
        \displaystyle\int_{t-W}^{t}
        |\mathcal A(\tau)|\,d\tau
    }.
\end{equation}
Using the average accumulation $\bar n(t,W)$ defined in Eq.~\eqref{eq:finite_accumulation}, the resulting projected-space AB outflow estimator is
\begin{equation}
\label{eq:projected_AB_estimator}
    \hat G_{\mathrm{AB}}^{\mathrm{proj}}(t,W)
    =
    \bar n(t,W)
    \frac{\bar U(t,W)}{\bar S_D(t,W)}.
\end{equation}

Therefore, both the conventional and projected-space AB estimators use the same average accumulation and measurement window, with their distance expectations approximated from the same departure set.
Their key practical distinctions lie in the \textbf{observability} and \textbf{traffic-state dependence} of the distance components $L$ and $S$.
The conventional estimator requires the trip lengths of departing trips, which are generally unknown until completion and may vary substantially with accumulation and other operating conditions, particularly under congestion.
In contrast, the projected-space estimator uses shortest OD distances that are immediately available at departure and are unaffected by subsequent collision-avoidance interactions, although their mean may change with the generated OD distribution.
Consequently, the proposed projected-space estimator can be evaluated using observed drone motion and generated OD information, without waiting for trip completion or requiring calibration of a potentially highly state-dependent mean trip length.

Despite their different distance representations, both AB formulations share the advantage of reducing complex network traffic dynamics to a computationally efficient, low-dimensional macroscopic model. By relying on aggregate quantities rather than individual trip states, they offer a simple and practical framework for large-scale traffic estimation and control. However, their limitations become more pronounced under rapidly changing traffic conditions, as the stationary AB relations cannot explicitly capture the progression of individual trips. This motivates the trip-based formulations introduced next, which retain individual trip states to better represent transient traffic dynamics.


\subsection{Trip-based formulation}
\label{subsec:projected_TB}

As a complement to the accumulation-based formulation, the trip-based (TB) model represents transient network dynamics by explicitly tracking the remaining distance of individual trips. This allows the TB formulation to retain trip-level heterogeneity and account for the history of trip progression, making it particularly relevant to non-stationary traffic conditions, despite at the cost of increased data and computational requirements for initializing and updating individual trip states \citep{mariotte2017macroscopic,batista2019estimation}. In this subsection, we first revisit the conventional TB estimator and establish a finite-window bound on the outflow-estimation error arising from the representative-speed approximation.
We then develop the corresponding projected-space formulation, which enables the initialization of individual trip states using known shortest OD distances instead of trip lengths that are unavailable at departure, and establish its corresponding finite-window error bound.

\subsubsection{Conventional trip-based outflow estimator}

Let $r_i(t)$ [m] denote the remaining realized distance of $i$ at time $t$ in the conventional true-motion representation. Given its trip length $l_i$ and instantaneous space-mean speed $v_i(t)$, the exact remaining-distance state is
\begin{equation}
\label{eq:r_i(t)}
    r_i(t)
    =
    l_i
    -
    \int_{t_{0,i}}^{t}
    v_i(\tau)\,d\tau.
\end{equation}
A trip is completed when its remaining distance reaches zero. In conventional ground-transportation applications, $l_i$ is typically assumed to be known or specified at departure, for example through a selected route or an OD-based trip-length assignment. Since individual speed trajectories are generally not explicitly tracked in macroscopic models, conventional TB formulations commonly propagate trips using an accumulation-dependent network space-mean speed, while retaining heterogeneous trip lengths \citep{fosgerau2015congestion,leclercq2017dynamic}. Let $\hat V(t)=V(n(t))$ [m/s] denote this space-mean speed estimate used for trip progression at time $t$. The estimated remaining distance of trip $i$ is then
\begin{equation}
\label{eq:rhat_i(t)}
    \hat r_i(t)
    =
    l_i
    -
    \int_{t_{0,i}}^{t}
    \hat V(\tau)\,d\tau.
\end{equation}
The estimated travel time $\hat T_i^\mathrm{conv}$ [s] under the conventional TB formulation is defined by the first passage of the cumulative distance through $l_i$:
\begin{equation}
\label{eq:That_r}
    \hat T_i^\mathrm{conv}
    =
    \inf
    \left\{
        T\geq 0:
        \int_{t_{0,i}}^{t_{0,i}+T}
        \hat V(\tau)\,d\tau
        \geq l_i
    \right\},
\end{equation}
provided that the set is non-empty. The inequality also accommodates discrete-time implementation, in which the accumulated distance may exceed $l_i$ during the final update. For the measurement window $(t-W,t]$, the estimated completion set is defined as
\begin{equation}
\label{eq:TB_conv_completion_set}
    \widehat{\mathcal C}_{\mathrm{TB}}^\mathrm{conv}(t,W)
    =
    \left\{
        i:
        t-W<t_{0,i}+\hat T_i^\mathrm{conv}\leq t
    \right\}.
\end{equation}
The conventional TB outflow estimate is therefore
\begin{equation}
\label{eq:TB_conv_estimator}
    \hat G_{\mathrm{TB}}^\mathrm{conv}(t,W)
    =
    \frac{
        \left|
            \widehat{\mathcal C}_{\mathrm{TB}}^\mathrm{conv}(t,W)
        \right|
    }{W}.
\end{equation}
Unlike the stationary AB estimator, which evaluates outflow through aggregate distance production, the TB estimator explicitly determines the estimated completion time of each trip. Its accuracy consequently depends on the discrepancy between the space-mean speed and the instantaneous speeds experienced by individual drones. The following theorem quantifies how this discrepancy affects finite-window outflow estimation.

We next establish a finite-window outflow-error bound for the conventional TB estimator under an admissible operating state sufficiently away from gridlock. In the following theorem, $\tau$ denotes an arbitrary clock time during system operation, whereas $t$ denotes the ending time of the measurement window $(t-W,t]$.

\begin{theorem}[Finite-window outflow-error bound for the conventional TB estimator]
\label{thm:TB_outflow_error_bound}
Consider a traffic system operating in a state where all trips complete within a uniformly bounded travel time and the space-mean speed remains strictly positive. Specifically, suppose that there exist finite constants $T_{\mathrm{ub}}>0$, $V_{\mathrm{lb}}>0$, and $V_{\max}\geq V_{\mathrm{lb}}$ such that
\begin{equation}
\label{eq:boundedtraveltime_r}
    T_i\leq T_{\mathrm{ub}},
    \qquad \forall i,
\end{equation}
and
\begin{equation}
\label{eq:representative_speed_lower_bound_r}
    0\leq v_i(\tau)\leq V_{\max},
    \qquad
    V_{\mathrm{lb}}\leq\hat V(\tau)\leq V_{\max},
    \qquad
    \forall \tau,\ i\in\mathcal A(\tau).
\end{equation}
The maximum discrepancy between the individual space-mean speeds and the space-mean speed is therefore well-defined and bounded:
\begin{equation}
\label{eq:representative_speed_discrepancy_r}
    \varepsilon_V
    =
    \sup_{\tau:\mathcal A(\tau)\neq\varnothing}
    \max_{i\in\mathcal A(\tau)}
    \left|
        \hat V(\tau)-v_i(\tau)
    \right|
    \leq V_{\max}.
\end{equation}
Furthermore, assume that the number of actual trip completions within any time window satisfies
\begin{equation}
\label{eq:TB_completion_envelope}
    |\mathcal C(\xi,h)|
    \leq
    G_{\max}h+B_{\max},
    \qquad
    \forall \xi> h>0,
\end{equation}
where $\mathcal C(\xi,h)$ is the set of trips completed during $(\xi-h,\xi]$, and $G_{\max}>0$ and $B_{\max}\geq1$ are finite constants representing an upper completion-rate envelope and an allowance for discrete completion events, respectively. Then, for every measurement window $(t-W,t]$ of duration $W>0$, the conventional TB outflow-estimation error satisfies
\begin{equation}
\label{eq:TB_outflow_error_bound}
    \left|
        \hat G_{\mathrm{TB}}^\mathrm{conv}(t,W)
        -
        G(t,W)
    \right|
    \leq
    \frac{2(\delta_rG_{\max}+B_{\max})}{W},
    \qquad
    \delta_r
    =
    \frac{\varepsilon_VT_{\mathrm{ub}}}{V_{\mathrm{lb}}},
\end{equation}
where $G(t,W)=|\mathcal C(t,W)|/W$ denotes the observed outflow over the measurement window.
\end{theorem}

The proof is provided in \autoref{app:proof_TB_outflow_error_bound}. The assumptions characterize an operating state sufficiently away from gridlock, with bounded travel times, physically limited drone speeds, and a strictly positive space-mean speed. Under these conditions, the discrepancy between individual and space-mean speeds induces a completion-time estimation error bounded by $\delta_r$. Consequently, differences between estimated and actual completion counts can arise only from trips whose completion times lie within $\delta_r$ of the measurement-window boundaries. The completion-count envelope in Eq.~\eqref{eq:TB_completion_envelope}, which accounts for possible discrete completion events, then yields the finite-window outflow-error bound in Eq.~\eqref{eq:TB_outflow_error_bound}.

\autoref{thm:TB_outflow_error_bound} reveals that the theoretical error bound decreases with the measurement-window length $W$, suggesting a trade-off between temporal resolution and robustness to completion-time estimation errors. However, this does not guarantee a corresponding decrease in the actual or expected estimation error, as the completion-count discrepancies may themselves vary with $W$ and the prevailing traffic conditions. Similar considerations arise in practical traffic monitoring, where measurement noise has been documented in loop-detector flow data \citep{I210}, highlighting the importance of selecting an appropriate aggregation window. Despite this bounded-error property, the conventional TB estimator requires the trip length $l_i$ to initialize each trip state. In LAAT systems, $l_i$ depends on future collision-avoidance maneuvers and is therefore generally unavailable at departure. To overcome this limitation, we introduce a projected-space TB formulation that initializes each trip using its shortest OD distance $s_i$, which is known at generation.

\subsubsection{Projected-space trip-based formulation}

In the projected-space TB formulation, we replace the remaining distance used in the conventional formulation with the \textbf{remaining OD-directed progress}. Specifically, let $p_i(t)$ [m] denote the remaining progress of trip $i$ at time $t$, defined as the shortest OD distance minus its cumulative progress:
\begin{equation}
\label{eq:p_i(t)}
    p_i(t)
    =
    s_i
    -
    \int_{t_{0,i}}^{t}
    u_i(\tau)\,d\tau.
\end{equation}
Unlike the conventional remaining realized distance $r_i(t)$, the remaining progress $p_i(t)$ may temporarily increase when the drone moves backward along its OD direction due to collision avoidance. Moreover, although the projected-distance identity in Eq.~\eqref{eq:projected_preservation} guarantees that $p_i(t_{0,i}+T_i)=0$ at physical arrival, reaching zero progress at an earlier time does not necessarily imply that the drone has reached its destination. Let $\hat U(t)=U(n(t))$ [m/s] denote the space-mean projected speed estimate assigned to active trips by the TB model. Analogously to the conventional TB estimator, the estimated remaining progress is
\begin{equation}
\label{eq:phat_i(t)}
    \hat p_i(t)
    =
    s_i
    -
    \int_{t_{0,i}}^{t}
    \hat U(\tau)\,d\tau.
\end{equation}
The estimated travel time $\hat T_i^{\mathrm{proj}}$ [s] is defined by the first passage of cumulative progress through $s_i$:
\begin{equation}
\label{eq:That_p}
    \hat T_i^{\mathrm{proj}}
    =
    \inf
    \left\{
        T\geq0:
        \int_{t_{0,i}}^{t_{0,i}+T}
        \hat U(\tau)\,d\tau
        \geq s_i
    \right\},
\end{equation}
provided that the set is non-empty. For a measurement window $(t-W,t]$, the estimated completion set in the projected space is defined as
\begin{equation}
\label{eq:ChatTB}
    \widehat{\mathcal C}_{\mathrm{TB}}^{\mathrm{proj}}(t,W)
    =
    \left\{
        i:
        t-W<t_{0,i}+\hat T_i^{\mathrm{proj}}\leq t
    \right\}.
\end{equation}
The corresponding projected-space TB outflow estimate is
\begin{equation}
\label{eq:GhatTB}
    \hat G_{\mathrm{TB}}^{\mathrm{proj}}(t,W)
    =
    \frac{
        \left|
            \widehat{\mathcal C}_{\mathrm{TB}}^{\mathrm{proj}}(t,W)
        \right|
    }{W}.
\end{equation}

We next establish a finite-window outflow-error bound for the projected-space TB estimator under operating conditions analogous to those considered in \autoref{thm:TB_outflow_error_bound}. In particular, we restrict the theoretical analysis to a state sufficiently away from gridlock, where travel times are bounded and the space-mean projected speed remains strictly positive. As in the conventional formulation, $\tau$ denotes an arbitrary operational time, while $t$ denotes the ending time of the measurement window.

\begin{theorem}[Finite-window outflow-error bound for the projected-space TB estimator]
\label{thm:TB_outflow_error_bound_proj}
Consider a traffic system operating in a state with uniformly bounded travel times and strictly positive space-mean projected speed. Specifically, suppose that there exist finite constants $T_{\mathrm{ub}}>0$, $U_{\mathrm{lb}}>0$, and $V_{\max}\geq U_{\mathrm{lb}}$ such that
\begin{equation}
\label{eq:boundedtraveltime_p}
    T_i\leq T_{\mathrm{ub}},
    \qquad \forall i,
\end{equation}
and
\begin{equation}
\label{eq:projected_speed_bounds_p}
    0\leq v_i(\tau)\leq V_{\max},
    \qquad
    U_{\mathrm{lb}}\leq\hat U(\tau)\leq V_{\max},
    \qquad
    \forall \tau,\ i\in\mathcal A(\tau).
\end{equation}
As $u_i(\tau)$ can be negative, the maximum discrepancy between individual and space-mean projected speeds is therefore well-defined and bounded:
\begin{equation}
\label{eq:projected_speed_discrepancy_p}
    \varepsilon_U
    =
    \sup_{\tau:\mathcal A(\tau)\neq\varnothing}
    \max_{i\in\mathcal A(\tau)}
    \left|
        \hat U(\tau)-u_i(\tau)
    \right|
    \leq 2V_{\max}.
\end{equation}
Furthermore, assume that the actual completion counts satisfy the completion-count envelope in Eq.~\eqref{eq:TB_completion_envelope}, namely,
\begin{equation}
    N_C(\xi,h)
    \leq
    G_{\max}h+B_{\max},
    \qquad
    \forall \xi> h>0.
\end{equation}
Then, for every measurement window $(t-W,t]$ of duration $W>0$, the projected-space TB outflow-estimation error satisfies
\begin{equation}
\label{eq:projected_TB_outflow_error_bound}
    \left|
        \hat G_{\mathrm{TB}}^{\mathrm{proj}}(t,W)
        -
        G(t,W)
    \right|
    \leq
    \frac{
        2(\delta_pG_{\max}+B_{\max})
    }{W},
    \qquad
    \delta_p
    =
    \frac{\varepsilon_U T_{\mathrm{ub}}}
    {U_{\mathrm{lb}}},
\end{equation}
where $G(t,W)=N_C(t,W)/W$ denotes the observed outflow over the measurement window.
\end{theorem}

The proof is provided in \autoref{app:proof_TB_outflow_error_bound_proj}. Analogously to \autoref{thm:TB_outflow_error_bound}, the projected-space TB estimator admits a finite-window outflow-error bound, with the real-speed discrepancy $\varepsilon_V$ replaced by the projected-speed discrepancy $\varepsilon_U$. The latter accounts for both individual speed variations and directional deviations arising from collision-avoidance maneuvers. Since the space-mean projected speed may be lower than the space-mean speed, the projected-space error bound can be more conservative, particularly under congested conditions with reduced OD-directed progress. Nevertheless, the two bounds do not imply a general ordering of the actual estimation errors.

In summary, the key advantage of projected-space TB formulation is that \textbf{individual trip states can be initialized using the departure-observable shortest OD distance $s_i$, rather than the unknown trip length $l_i$}. Together with the projected-space AB formulation, this provides a consistent framework for modeling stationary outflow and transient trip-completion dynamics without prior knowledge of realized trajectories.

\subsection{Extension to multi-class systems}
\label{subsec:multi_class}

The projected-space framework can be extended to heterogeneous traffic systems comprising multiple drone classes. These classes may differ in their physical characteristics, speed distributions, or operational rules, including speed limits, access permissions, collision-avoidance policies, and priority levels. Such differences affect the motion of individual drones and may lead to distinct class-specific macroscopic traffic dynamics. In this subsection, we establish the projected-space formulations for a general multi-class system and derive the relationship between the class compositions at generation and within the active population.

\subsubsection{Multi-class accumulation-based formulation}
\label{subsubsec:mehod_mc_ab}
Let $\mathcal M=\{1,\ldots,M\}$ denote the set of drone classes. Define the stacked accumulation vector as
\begin{equation}
    \mathbf n
    = \mathrm{col}\left(n_m\right)_{m\in\mathcal M},
\end{equation}
where $n_m=|\mathcal A_m|$, and $\mathcal A_m$ denotes the set of active class-$m$ drones. The total accumulation is
\begin{equation}
\label{eq:multi_class_total_accumulation}
    n_\mathrm{tot}
    =
    \mathbf 1^\top\mathbf n
    =
    \sum_{m\in\mathcal M}n_m.
\end{equation}
Each drone follows the projected-space representation introduced previously, while its instantaneous projected speed may depend on its class and the prevailing traffic conditions. Under steady-state operation, let $g(\cdot\mid\mathbf n,m)$ denote the snapshot distribution conditional on class $m$ and the accumulation vector $\mathbf n$. The space-mean projected speed of class $m$ is defined as
\begin{equation}
\label{eq:class_projected_speed}
    U_m(\mathbf n)
    =
    \mathbb E_g[u\mid\mathbf n,m].
\end{equation}
The dependence on the full accumulation vector is important because the motion of class-$m$ drones may be affected by interactions with drones from other classes. Consequently, two traffic states with the same total accumulation but different class compositions, namely different class fractions, may exhibit different space-mean projected speeds.

For each class $m$, define the mean shortest OD distance under the generation distribution as
\begin{equation}
\label{eq:class_shortest_distance_multiclass}
    S_m
    =
    \mathbb E_f[s\mid m],
\end{equation}
where the class-specific generation distribution is assumed invariant across stationary operating states. Applying the projected-space AB relation to each class gives
\begin{equation}
\label{eq:Gkbfn}
    G_m(\mathbf n)
    =
    n_m\frac{U_m(\mathbf n)}{S_m},\quad\text{and}\quad \dot{n}_m(t) = q_m(t)-G_m(\mathbf{n}(t)).
\end{equation}
Here, $G_m(\mathbf n)$ and $q_m$ [a.c./s] denote the stationary class-specific outflow and inflow, respectively. We assume that class membership remains fixed throughout each trip and that the conditions of \autoref{prop:steady_projected_AB} hold separately for each class. The following proposition establishes the aggregate outflow relation.

\begin{proposition}[Aggregate projected-space AB relation under multiple classes]
\label{prop:multi_class_projected_AB}
Consider a stationary multi-class system with classes $m\in\mathcal M$, in which the projected-space AB relation in Eq.~\eqref{eq:Gkbfn} holds separately for each class. Suppose that all classes share the same expected shortest OD distance at trip generation,
\begin{equation}
\label{eq:common_shortest_distance_multiclass}
    S_m
    =
    \mathbb E_f[s\mid m]
    =
    S,
    \qquad \forall m\in\mathcal M.
\end{equation}
Then, the total outflow satisfies
\begin{equation}
\label{eq:multi_class_projected_AB_aggregate}
    G_\mathrm{tot}(\mathbf n)=\sum_{m\in \mathcal M}G_m(\mathbf{n})
    =
    \frac{n_\mathrm{tot}U(\mathbf n)}{S},
\end{equation}
where $n_\mathrm{tot}=\sum_{m\in\mathcal M}n_m$ is the total accumulation, and $U(\mathbf n)$ is the space-mean projected speed over the entire active population, regardless of class.
\end{proposition}

The proof is provided in \autoref{app:proof_multi_class_AB}. The proposition establishes that, under a common mean shortest OD distance $S$, the aggregate multi-class AB relation preserves the algebraic structure of its single-class counterpart. However, the space-mean projected speed $U(\mathbf n)$ generally depends on the full class accumulation vector rather than solely on the total accumulation $n_{\mathrm{tot}}$, since class composition and inter-class interactions can affect individual motion. Consequently, the multi-class MFD is characterized by $G_{\mathrm{tot}}(\mathbf n)$ rather than a scalar function of total accumulation alone.

In addition, in multi-class traffic systems, the fractions of different classes among generated trips generally differ from those observed in the active population. This distinction arises because classes with different movement characteristics may have different travel times and, consequently, different representations in the snapshot population. For each class $m\in\mathcal M$, suppose $q_m\approx G_m(\mathbf n)$ under steady-state flow conservation. Define the generation and accumulation class fractions, respectively, as
\begin{equation}
\label{eq:multi_class_fractions}
    \alpha_m^q
    =
    \frac{q_m}{q_{\mathrm{tot}}},
    \qquad
    \alpha_m^n
    =
    \frac{n_m}{n_\mathrm{tot}},
\end{equation}
where
\[    q_{\mathrm{tot}}
    =
    \sum_{m\in\mathcal M}q_m,
    \qquad
    n_\mathrm{tot}=\sum_{m\in\mathcal M}n_m,
\]
satisfying
\begin{equation}
    \dot{n}_\mathrm{tot}(t)=q_\mathrm{tot}(t)-G_\mathrm{tot}(\mathbf{n}(t)).
\end{equation}
The relationship between these two class compositions is established in the following proposition.

\begin{proposition}[Stationary relationship between generation and accumulation class fractions]
\label{prop:multi_class_fraction_relation}
Consider a stationary multi-class system satisfying the class-specific projected-space AB relation in Eq.~\eqref{eq:Gkbfn}. Assume that class membership remains unchanged during each trip, all generated trips eventually complete, and each class has finite mean travel time and positive space-mean projected speed $U_m(\mathbf n)$. Then, for every class with positive stationary generation rate,
\begin{equation}
\label{eq:multi_class_fraction_forward}
    \alpha_m^n
    =
    \frac{
        \displaystyle
        \alpha_m^q\frac{S_m}{U_m(\mathbf n)}
    }{
        \displaystyle
        \sum_{l\in\mathcal M}
        \alpha_l^q\frac{S_l}{U_l(\mathbf n)}
    }
    \qquad m\in\mathcal M.
\end{equation}
In particular, if $S_m=S$ for all classes, the relationships reduce to
\begin{equation}
\label{eq:multi_class_fraction_common_S}
\begin{aligned}
    \alpha_m^n
    &=
    \frac{
        \alpha_m^q/U_m(\mathbf n)
    }{
        \displaystyle\sum_{l\in\mathcal M}
        \alpha_l^q/U_l(\mathbf n)
    }.
\end{aligned}
\end{equation}
\end{proposition}

The proof is provided in \autoref{app:proof_multi_class_fractions}. The relationship follows from the class-specific expected travel time $\mathbb E_f[T\mid\mathbf n,m]=S_m/U_m(\mathbf n)$, and indicates that \textbf{classes with longer mean travel times are overrepresented in the active population relative to their generation fractions}. Under a common mean shortest OD distance $S$, the relationship depends solely on class-specific space-mean projected speeds. In contrast, the conventional formulation involves the class-specific mean trip length $L_m$, which may differ across classes even when their shortest OD-distance expectations are identical. Therefore, even if $L_m$ were known, the simplified relationship in Eq.~\eqref{eq:multi_class_fraction_common_S} could not generally be recovered by replacing $U_m$ with the space-mean speeds $V_m$. This distinction is particularly relevant to the priority-based setting later introduced in \autoref{subsec:result_multiclass}, where different collision-avoidance behaviors may produce different expected trip lengths.

\subsubsection{Multi-class trip-based formulation}

The projected-space TB formulation can be extended to multi-class systems in two ways. First, the heterogeneous traffic system can be treated as a mixed population, with all trips propagated using the aggregate space-mean projected speed $\hat U(\tau)$, regardless of their class labels. Since \autoref{thm:TB_outflow_error_bound_proj} applies at the individual-trip level, the same error bound remains valid provided that its assumptions hold over the entire active population. Thus, class-specific macroscopic quantities are not required for the aggregate projected-space TB estimator.

Alternatively, when class-specific space-mean projected speeds are available, trips can be propagated according to their respective classes. Let $m(i)\in\mathcal M$ denote the class of trip $i$, and let $\hat U_m(\tau)$ denote the space-mean projected speed assigned to class $m$ at time $\tau$. The class-specific estimated remaining progress is
\begin{equation}
\label{eq:multi_class_projected_remaining_progress}
    \hat p_{i,\mathrm{MC}}(t)
    =
    s_i
    -
    \int_{t_{0,i}}^{t}
    \hat U_{m(i)}(\tau)\,d\tau.
\end{equation}
The corresponding estimated travel time $\hat T_{i,\mathrm{MC}}^{\mathrm{proj}}$ is defined by the first-passage condition in Eq.~\eqref{eq:That_p}, with $\hat U(\tau)$ replaced by $\hat U_{m(i)}(\tau)$. For a measurement window $(t-W,t]$, the class-$m$ estimated completion set is
\begin{equation}
\label{eq:multi_class_ChatTB}
    \widehat{\mathcal C}_{\mathrm{TB},m}^{\mathrm{proj}}(t,W)
    =
    \left\{
        i:
        m(i)=m,\,
        t-W<t_{0,i}+\hat T_{i,\mathrm{MC}}^{\mathrm{proj}}\leq t
    \right\},
    \qquad m\in\mathcal M.
\end{equation}
The corresponding class-specific outflow estimate is
\begin{equation}
\label{eq:multi_class_GhatTB}
    \hat G_{\mathrm{TB},m}^{\mathrm{proj}}(t,W)
    =
    \frac{
        \left|
            \widehat{\mathcal C}_{\mathrm{TB},m}^{\mathrm{proj}}(t,W)
        \right|
    }{W},
    \qquad m\in\mathcal M.
\end{equation}
Summing over all classes gives the total outflow estimate of the class-specific formulation:
\begin{equation}
\label{eq:multi_class_projected_TB_total_outflow}
    \hat G_{\mathrm{TB,MC}}^{\mathrm{proj}}(t,W)
    =
    \sum_{m\in\mathcal M}
    \hat G_{\mathrm{TB},m}^{\mathrm{proj}}(t,W).
\end{equation}
The finite-window error analysis can be extended accordingly. Suppose that the assumptions of \autoref{thm:TB_outflow_error_bound_proj} hold, with each class additionally satisfying a positive projected-speed bound
\begin{equation}
    0<U_{\mathrm{lb}}^m
    \leq
    \hat U_m(\tau)
    \leq
    V_{\max},
    \qquad m\in\mathcal M.
\end{equation}
Define the class-specific projected-speed discrepancy as
\begin{equation}
\label{eq:multi_class_projected_speed_discrepancy}
    \varepsilon_U^m
    =
    \sup_{\tau:\mathcal A_m(\tau)\neq\varnothing}
    \max_{i\in\mathcal A_m(\tau)}
    \left|
        \hat U_m(\tau)-u_i(\tau)
    \right|.
\end{equation}
Analogously to the proof of \autoref{thm:TB_outflow_error_bound_proj}, the completion-time estimation error of each class-$m$ trip is bounded by
\begin{equation}
\label{eq:multi_class_delta_p_k}
    \left|
        c_i-\hat c_{i,\mathrm{MC}}^{\mathrm{proj}}
    \right|
    \leq
    \delta_p^m
    =
    \frac{
        \varepsilon_U^m T_{\mathrm{ub}}
    }{
        U_{\mathrm{lb}}^m
    },
    \qquad m(i)=m,
\end{equation}
where $\hat c_{i,\mathrm{MC}}^{\mathrm{proj}}=t_{0,i}+\hat T_{i,\mathrm{MC}}^{\mathrm{proj}}$. Defining the maximum class-specific completion-time error bound as
\begin{equation}
\label{eq:multi_class_delta_p}
    \delta_p^{\mathcal M}
    =
    \max_{m\in\mathcal M}\delta_p^m,
\end{equation}
the completion-count argument in \autoref{app:proof_TB_outflow_error_bound_proj} applies directly to the combined class-specific estimator. Under the same completion-count envelope in Eq.~\eqref{eq:TB_completion_envelope}, we obtain
\begin{equation}
\label{eq:multi_class_projected_TB_error_bound}
    \left|
        \hat G_{\mathrm{TB,MC}}^{\mathrm{proj}}(t,W)
        -
        G(t,W)
    \right|
    \leq
    \frac{
        2(\delta_p^{\mathcal M}G_{\max}+B_{\max})
    }{W}.
\end{equation}
Compared with the pooled formulation, class-specific space-mean speeds can reduce the completion-time error bound when they better capture the motion of drones within each class. In particular, if $\delta_p^{\mathcal M}\leq\delta_p$, the class-specific formulation yields an equal or tighter outflow-error bound, although this improvement is not guaranteed in general.

Finally, unlike the aggregate multi-class AB relation in \autoref{prop:multi_class_projected_AB}, which assumes a common expected shortest OD distance $S$, the projected-space TB formulation does not require identical class-specific expectations $S_m$. Since each trip is initialized using its own shortest OD distance $s_i$, the TB formulation accommodates heterogeneous OD-distance distributions without requiring trip lengths at departure.

\section{Results and Discussion}
\label{sec:result}

This section evaluates the proposed projected-space macroscopic modeling framework through numerical experiments. We first describe the microscopic simulation environment, followed by separate evaluations of the projected-space accumulation-based (AB) and trip-based (TB) formulations. The AB formulation is evaluated across quasi-stationary traffic states generated by stepwise increases in the accumulation ceiling, whereas the TB formulation is examined under time-varying traffic conditions to assess its ability to capture transient trip progression. Finally, the multi-class extension is investigated through priority-based traffic scenarios.

\subsection{Simulation setup}
\label{subsec:config}

The numerical experiments are conducted using the open-source \textit{XaMobility} framework developed by \citet{safadi2023macroscopic}, which simulates three-dimensional drone traffic with decentralized collision avoidance based on a force-field (FF) mechanism \citep{quan2021practical}. Each drone follows a nominal straight-line route connecting its current position and destination, while its realized trajectory is continuously adjusted in response to surrounding traffic to maintain safe separation. The continuous-time dynamics are given by
\begin{equation}
\label{eq:dronedynamics}
\begin{aligned}
    \dot{\mathbf{x}}_i(t)
    &=
    \mathbf{v}_i(t),\\
    \dot{\mathbf{v}}_i(t)
    &=
    -\mathbf{K}
    \left(
        \mathbf{v}_i(t)-\mathbf{v}_{i,c}(t)
    \right),
\end{aligned}
\end{equation}
where $\mathbf{x}_i(t),\mathbf{v}_i(t)\in\mathbb R^3$ denote the position and velocity vectors of drone $i$, respectively, $\mathbf K$ is a pre-defined control gain matrix, and $\mathbf{v}_{i,c}(t)$ is the commanded velocity determined by the FF controller as:
\begin{equation}
\label{eq:v_command_ff}
    \mathbf{v}_{i,c}(t)
    =
    \mathrm{sat}\left(
        \mathrm{sat}\left(
            \phi_{i,\mathrm{att}}
            (\mathbf{x}_i(t),\mathbf{v}_i(t)),
            V_{\max}
        \right)
        +
        \sum_{j\in\mathcal N_i(t)}
        \phi_{i,\mathrm{rep}}
        (\mathbf{x}_i(t),\mathbf{x}_j(t)),
        V_{\max}
    \right).
\end{equation}
The operator $\mathrm{sat}(\cdot,V_{\max})$ limits the commanded speed to $V_{\max}$, while $\phi_{i,\mathrm{att}}$ denotes the attractive component directing drone $i$ toward its destination, and $\phi_{i,\mathrm{rep}}$ represents the repulsive effect of neighboring drones.
Specifically, the neighboring drone set $\mathcal N_i(t)$ is defined as
\begin{equation}
    \mathcal N_i(t)
    =
    \left\{
    j\in\mathcal A(t)\setminus\{i\}:
    \|\mathbf x_i(t)-\mathbf x_j(t)\|_2
    \leq R_{\mathrm{detect}}
    \right\},
\end{equation}
where the detection radius $R_{\mathrm{detect}}$ is set to $67.5$ [m], and the magnitude of the repulsive force increases as the Euclidean distance $\|\mathbf x_i(t)-\mathbf x_j(t)\|_2$ decreases. The dynamics are discretized using the forward Euler method with a time step of $T_s=0.5$ [s].

The simulated airspace has dimensions $1000\times1000\times100$ [m$^3$]. Origins and destinations are randomly generated through Monte Carlo sampling within the airspace, with OD pairs separated by less than 90 [m] discarded. This randomized OD setting aims to maintain a spatially distributed travel demand without introducing additional concentrations associated with predefined vertiport locations, allowing us to focus on the effects of drone interactions and congestion. The resulting mean shortest OD distance is approximately $S=530$ [m]. The maximum commanded speed is set to $V_{\max}=20$ [m/s]. Accounting for acceleration and deceleration, the space-mean speed in uncongested conditions is approximately 16 [m/s], corresponding to typical free-flow travel times on the order of 40 [s]. Unless otherwise specified, the baseline measurement-window duration is $W=1$ [min], slightly larger than the typical free-flow travel time. With the microscopic simulation environment established, the following subsections evaluate the proposed accumulation-based and trip-based formulations under different traffic loading conditions, followed by their extension to priority-based multi-class traffic.

\subsection{Evaluation of the projected-space accumulation-based formulation}
\label{subsec:result_AB}

We first evaluate the proposed projected-space AB formulation against its conventional counterpart.
Since both theoretical outflow relations are established under stationary conditions, we employ a simple departure management strategy based on stepwise increases in the accumulation ceiling to obtain a broad range of quasi-stationary traffic states, covering both free-flow and congested regimes.
This procedure prevents uncontrolled accumulation growth, particularly in the congested regime, allowing us to assess the finite-window outflow estimation accuracy of both formulations under approximately stationary conditions.

\subsubsection{Loading profile through stepwise accumulation-ceiling increases}
\label{subsubsec:traffic_loading}

As mentioned previously, we design and implement a simple departure management through an accumulation ceiling $n_{\mathrm{lim}}$. A requested departure is admitted only when the current accumulation satisfies $n(t)<n_{\mathrm{lim}}$; otherwise, it is postponed until capacity becomes available, i.e., an active drone completes its trip. Let $q_{\mathrm{req}}(t)$ and $q_{\mathrm{act}}(t)$ denote the requested and actual admitted inflows, respectively. A sufficiently high requested departure rate is imposed as $q_{\mathrm{req}}(t)=3$ [a.c./s], so that admissions are primarily regulated by the accumulation ceiling. The accumulation ceiling $n_{\mathrm{lim}}$ is progressively increased from 1 to 301 [a.c.], each of which is maintained for 20 [min], to sample traffic conditions ranging from free-flow to severe congestion. The first measurement following each ceiling adjustment is excluded to reduce transitional effects.
The experiment is repeated over 10 independent simulation runs, each with randomly generated OD pairs.
The measurements retained after the initial transition period are pooled across runs to compute the mean and standard deviation at each accumulation level.
The observed network outflow is then calculated from the number of completed trips within each measurement window,
\begin{equation}
\label{eq:observed_outflow_results}
    G_0(t,W)
    =
    \frac{|\mathcal C(t,W)|}{W}.
\end{equation}

\begin{figure}[htbp]
    \centering
    \begin{subfigure}[t]{0.49\linewidth}
        \centering
        \includegraphics[width=\linewidth]{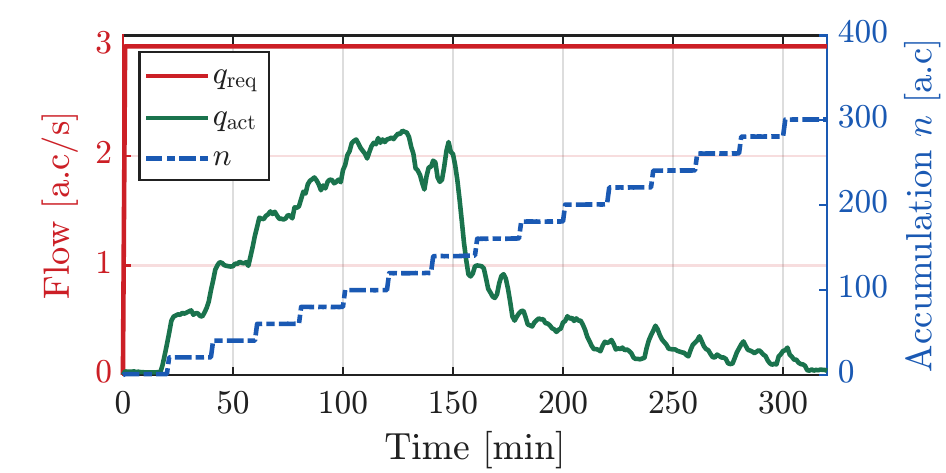}
        \caption{Actual inflow and accumulation.}
        \label{fig:loading_time}
    \end{subfigure}
    \hfill
    \begin{subfigure}[t]{0.49\linewidth}
        \centering
        \includegraphics[width=\linewidth]{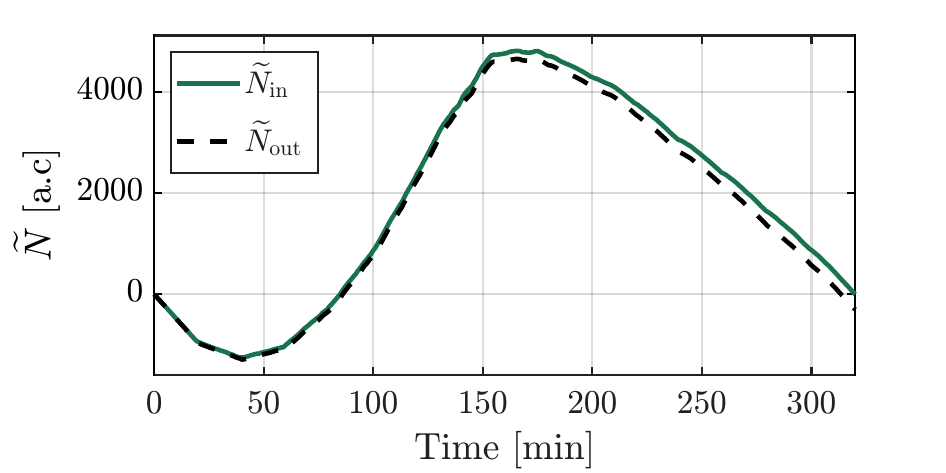}
        \caption{Oblique cumulative counts.}
        \label{fig:loading_oblique}
    \end{subfigure}
    \caption{Traffic evolution under stepwise accumulation-ceiling increases. (a) Requested inflow $q_\mathrm{req}$, actual inflow $q_{\mathrm{act}}$ and network accumulation $n$. (b) Oblique cumulative inflow $\widetilde N_{\mathrm{in}}$ and outflow $\widetilde N_{\mathrm{out}}$ following Eq.~\eqref{eq:loading_cumulative_counts} with $\bar{q}_\mathrm{act}=0.82$ [a.c/s].}
    \label{fig:progressive_loading_profile}
\end{figure}

\autoref{fig:progressive_loading_profile} illustrates the traffic evolution under stepwise accumulation-ceiling increases. In \autoref{fig:loading_time}, the actual inflow $q_{\mathrm{act}}$ initially increases as the accumulation ceiling is raised, but begins to decline after approximately 150 [min], despite further increases in the ceiling. Meanwhile, the network accumulation $n$ increases stepwise and remains nearly constant within each stage. To further examine the balance between inflow and outflow, \autoref{fig:loading_oblique} presents the oblique cumulative-count curves \citep{cassidy2002empirical}, defined as
\begin{equation}
\label{eq:loading_oblique}
\begin{aligned}
    \widetilde N_{\mathrm{in}}(t)
    &=N_{\mathrm{in}}(t)-\bar q_{\mathrm{act}}\cdot(t-t_0),\\
    \widetilde N_{\mathrm{out}}(t)
    &=N_{\mathrm{out}}(t)-\bar q_{\mathrm{act}}\cdot(t-t_0),
\end{aligned}
\end{equation}
where the cumulative admitted departures $N_{\mathrm{in}}$, completed trips $N_{\mathrm{out}}$, and mean admitted inflow $\bar q_{\mathrm{act}}$ are given by
\begin{equation}
\label{eq:loading_cumulative_counts}
\begin{aligned}
    N_{\mathrm{in}}(t)
    &=\int_{t_0}^{t}q_{\mathrm{act}}(\tau)\, d\tau,\\
    N_{\mathrm{out}}(t)
    &=\int_{t_0}^{t}G_0(\tau)\, d\tau,\\
    \bar q_{\mathrm{act}}
    &=\frac{1}{t_\mathrm{end}-t_0}\int_{t_0}^{T}q_{\mathrm{act}}(\tau)\, d\tau,
\end{aligned}
\end{equation}
with $t_0$ and $t_\mathrm{end}$ denoting the beginning and end of the experiment. Subtracting the same reference flow highlights variations in the cumulative counts while preserving their difference, $\widetilde N_{\mathrm{in}}(t)-\widetilde N_{\mathrm{out}}(t)=n(t)-n(t_0)$. The two curves remain nearly parallel within each accumulation stage, indicating that the admitted inflow approximately balances the observed outflow $G_0$. Their separation increases when the accumulation ceiling is raised, as additional departures are admitted to reach the new level. Together, these results demonstrate that the experiment produces a sequence of quasi-stationary operating conditions suitable for evaluating the AB formulations. Hence, under these quasi-stationary conditions, flow conservation and Little's law yield
\begin{equation}
\label{eq:loading_inflow}
    q_{\mathrm{act}}(n_{\mathrm{lim}})
    \approx
    G(n_{\mathrm{lim}})
    =
    \frac{
        n_{\mathrm{lim}}
    }{
        \mathbb E_f[T\mid n_{\mathrm{lim}}]
    }.
\end{equation}
Consequently, increasing the accumulation ceiling does not necessarily increase the admitted inflow, as higher accumulation can lead to longer travel times and reduced network throughput. This procedure allows both the increasing and decreasing branches of the MFD to be sampled under approximately stationary conditions.

The conventional and projected-space AB estimates are evaluated using Eq.~\eqref{eq:conventional_AB_estimator} and Eq.~\eqref{eq:projected_AB_estimator}, respectively.
For a consistent comparison, both estimators use the same departure set $\mathcal D(t,W)$ to compute their respective mean distance measures.
The conventional estimator is evaluated retrospectively using the trip lengths $l_i$, which become available only upon trip completion, whereas the projected-space estimator uses the shortest OD distances $s_i$, which are observable at departure.
The space-mean speeds $V$ and $U$ are computed over  $W=1$ [min].

\subsubsection{Physical interpretation and outflow MFD verification}
\label{subsubsec:speedMFD_singleclass}

We first examine the physical effects of congestion on drone motion and spatial interactions through speed, trip length, and nearest-neighbor distance (NND) distributions, where NND is defined as the Euclidean distance from each drone to its closest neighboring drone at a given time, followed by the evaluation of conventional and projected-space AB formulations in estimating network outflow.
At each accumulation ceiling, measurements are collected after the initial transition period to characterize the corresponding traffic conditions, with macroscopic quantities averaged and nearest-neighbor distances examined through their empirical distributions.

\begin{figure}[htbp]
    \centering
    \begin{subfigure}[t]{0.49\linewidth}
        \centering
        \includegraphics[width=\linewidth]{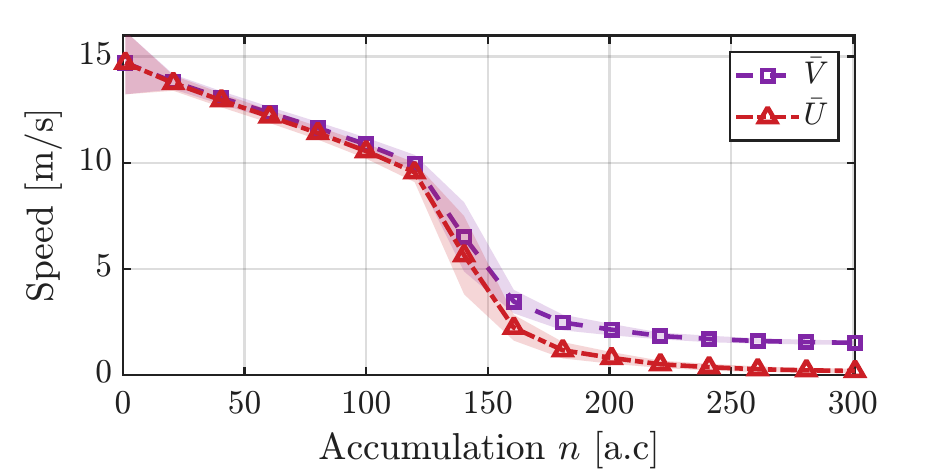}
        \caption{Space-mean speeds.}
        \label{fig:loading_speed_panel}
    \end{subfigure}
    \hfill
    \begin{subfigure}[t]{0.49\linewidth}
        \centering
        \includegraphics[width=\linewidth]{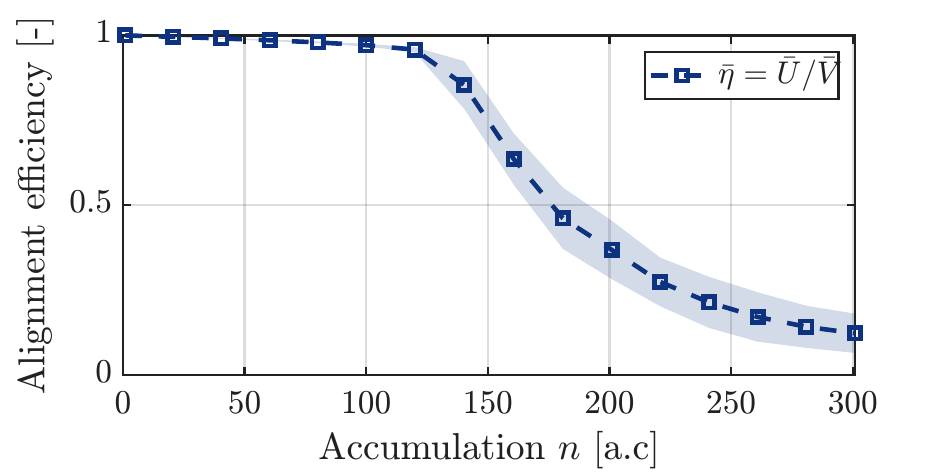}
        \caption{Alignment efficiency.}
        \label{fig:loading_alignment_panel}
    \end{subfigure}
    \caption{Space-mean speed, space-mean projected speed, and alignment efficiency under different accumulations.}
    \label{fig:loading_speed_alignment}
\end{figure}

\autoref{fig:loading_speed_alignment} illustrates the effects of network accumulation on the space-mean speeds and mean alignment efficiency $\bar{\eta}:=\bar U/\bar V$, following the definition in Eq.~\eqref{eq:eta}. Specifically, in \autoref{fig:loading_speed_panel}, the space-mean speed $\bar V$ and projected speed $\bar U$ behave nearly identically at low accumulations ($n\lesssim120$). Both decrease gradually and approximately linearly with increasing accumulation, as collision-avoidance interactions become more frequent but remain relatively mild. Within this regime, drones generally maintain trajectories close to their nominal OD directions, resulting in only a small difference between the two speeds. Beyond this range, both speeds drop sharply as collision-avoidance interactions intensify. The projected speed, however, decreases more rapidly than the space-mean speed, reflecting the growing proportion of drone motion that does not contribute to effective OD-directed progress. Under severe congestion, $\bar V$ approaches a nonzero value, suggesting that drones continue moving in search of available space, whereas $\bar U$ approaches zero as their net progress toward destinations becomes negligible. Nevertheless, a near-zero mean projected speed does not strictly imply zero trip completions, since positive progress by some drones may be offset by others moving away from their destinations during avoidance maneuvers. \autoref{fig:loading_alignment_panel} further illustrates this divergence through the mean alignment efficiency $\bar\eta$.
The alignment efficiency remains close to 1 at low accumulations, but decreases sharply beyond $n=120$, approaching zero under severe congestion. 
This reduction reveals the declining effectiveness of physical drone movement in contributing to OD-directed progression, highlighting the importance of projected speed in characterizing network traffic production.

\begin{figure}[htbp]
    \centering
    \begin{subfigure}[t]{0.49\linewidth}
        \centering
        \includegraphics[width=\linewidth]{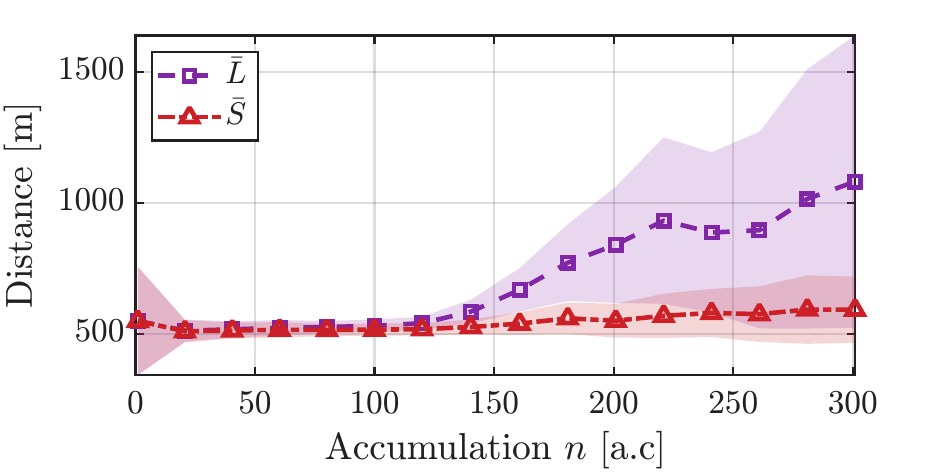}
        \caption{Mean trip length and shortest OD distance.}
        \label{fig:loading_distance_panel}
    \end{subfigure}
    \hfill
    \begin{subfigure}[t]{0.49\linewidth}
        \centering
        \includegraphics[width=\linewidth]{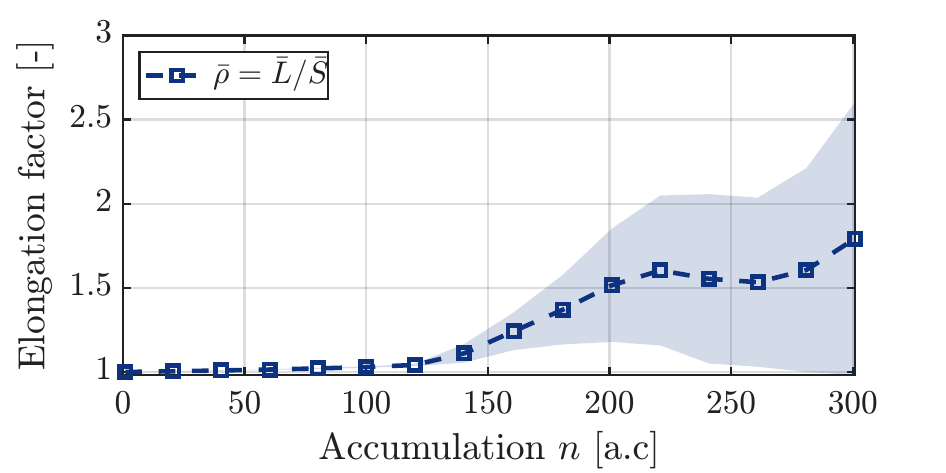}
        \caption{Path elongation factor.}
        \label{fig:loading_elongation_panel}
    \end{subfigure}
    \caption{Mean trip distances and path elongation factor under different accumulations.}
    \label{fig:loading_distance_elongation}
\end{figure}

\autoref{fig:loading_distance_elongation} presents the effects of congestion on drone detours by comparing the empirical mean trip length ${\bar L}$, mean shortest OD distance ${\bar S}$, and their ratio, defined as the mean path elongation factor $\bar\rho:= {\bar L}/{\bar S}$. \autoref{fig:loading_distance_panel} first compares the mean trip length $\bar L$ and mean shortest OD distance $\bar S$ across different accumulations.
At low accumulations, the two quantities remain close, as drones generally follow near-direct trajectories.
As congestion intensifies, $\bar L$ increases substantially due to collision-avoidance detours, approximately doubling at high accumulations, while $\bar S$ remains relatively stable.
The slight increase in $\bar S$ at high accumulations results from adjustments to the generated OD distribution introduced to maintain quasi-stationary operating conditions, as described in \autoref{subsec:config}, and should not be interpreted as an intrinsic dependence of $\bar{S}$ on accumulation.
Notably, the standard deviation of $\bar L$ across independent simulation runs also increases dramatically under severe congestion, whereas that of $\bar S$ remains much smaller.
Furthermore, a slight decrease in $\bar L$ is observed around $n=251$ [a.c.], despite the overall increasing trend.

\autoref{fig:loading_elongation_panel} further characterizes these deviations through $\bar {\rho}$.
The factor remains close to 1 at low accumulations but increases markedly beyond approximately $n=121$, reflecting the growing discrepancy between trip length and shortest OD distance.
Its standard deviation across independent simulation runs also increases substantially under severe congestion, primarily reflecting the greater variability of $\bar L$.
These observations indicate that trip length is highly sensitive to traffic interactions and operating conditions, suggesting that accumulation alone may not adequately characterize its variability or support a reliably identifiable relationship $L(n)$.
In contrast, shortest OD distances are determined at departure and remain unaffected by subsequent collision-avoidance maneuvers.
Although the variability of trip length does not invalidate the conventional AB relation under stationary conditions, it complicates the estimation and prediction of the mean trip length required by that formulation.

\begin{figure}[htbp]
    \centering
    \begin{subfigure}[t]{0.24\linewidth}
        \centering
        \includegraphics[width=\linewidth]{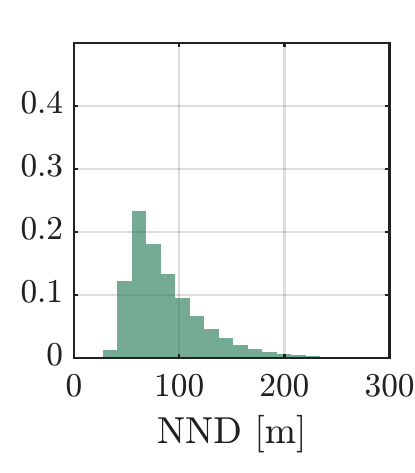}
        \caption{$n_{\mathrm{lim}}=41$ ($89.1\pm40.0$).}
        \label{fig:loading_nnd41}
    \end{subfigure}
    \hfill
    \begin{subfigure}[t]{0.24\linewidth}
        \centering
        \includegraphics[width=\linewidth]{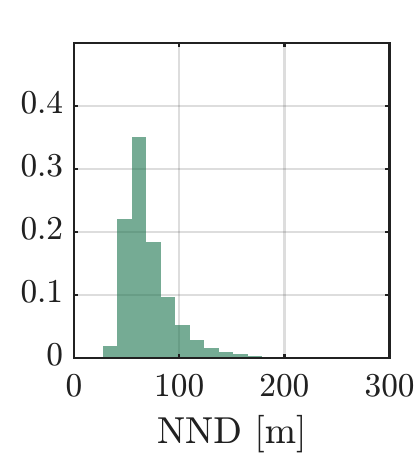}
        \caption{$n_{\mathrm{lim}}=81$ ($71.4\pm25.8$).}
        \label{fig:loading_nnd81}
    \end{subfigure}
    \hfill
    \begin{subfigure}[t]{0.24\linewidth}
        \centering
        \includegraphics[width=\linewidth]{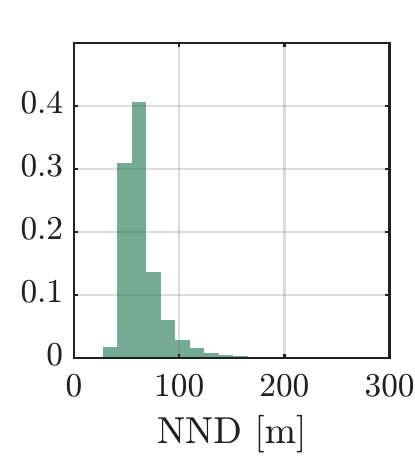}
        \caption{$n_{\mathrm{lim}}=121$ ($64.7\pm20.1$).}
        \label{fig:loading_nnd121}
    \end{subfigure}
    \hfill
    \begin{subfigure}[t]{0.24\linewidth}
        \centering
        \includegraphics[width=\linewidth]{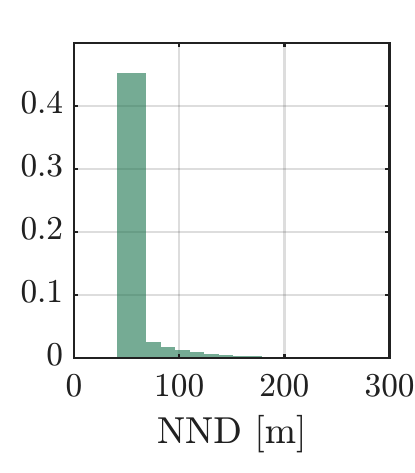}
        \caption{$n_{\mathrm{lim}}=161$ ($60.7\pm23.2$).}
        \label{fig:loading_nnd161}
    \end{subfigure}
    \caption{Empirical NND distributions under four accumulation ceilings $n_\mathrm{lim}\in\{41, 81, 121, 161\}$. Values in parentheses indicate the mean $\pm$ standard deviation of NND in meters.}
    \label{fig:loading_nnd_distributions}
\end{figure}

To characterize the local spatial interactions underlying these macroscopic observations, we further examine the NND, defined as the minimum Euclidean distance between each active drone and any other active drone at a given time:
\begin{equation}
\label{eq:NND}
    \mathrm{NND}_i(t)
    :=
    \min_{j\in\mathcal A(t),\,j\neq i}
    \left\|\mathbf x_i(t)-\mathbf x_j(t)\right\|_2.
\end{equation}
\autoref{fig:loading_nnd_distributions} presents the empirical NND distributions under $n_{\mathrm{lim}}\in\{41,81,121,161\}$, covering uncongested, critical, and congested operating conditions.
For each accumulation ceiling, only trips that both depart and complete within the corresponding 20-minute measurement period are considered, and the histograms aggregate observations from all 10 independent simulation runs. The corresponding means and standard deviations are summarized below the histograms. At $n=41$, the NND distribution is relatively broad and right-skewed, reflecting the sparse spatial arrangement of drones under uncongested conditions, with a mean of $89.1$ [m] and a standard deviation of $40.0$ [m].
As the accumulation ceiling increases to $81$, the distribution shifts toward smaller distances and becomes narrower, with the mean and standard deviation decreasing to $71.4$ and $25.8$ [m], respectively, although the airspace remains relatively uncongested.
This trend continues at higher accumulations, as the increasingly crowded airspace leads to smaller and generally more concentrated NND distributions.
At $n=121$, the mean NND reaches $64.7$ [m], close to the detection radius $R_{\mathrm{detect}}=67.5$ [m], and decreases further to $60.7$ [m] at $n=161$.

Importantly, $n=121$ marks a notable transition in local spacing, where the mean NND becomes comparable to the detection radius, indicating that drones increasingly fall within the detection ranges of others.
This observation is consistent with the substantial changes in speed, alignment efficiency, and path elongation observed in \autoref{fig:loading_speed_alignment} and \autoref{fig:loading_distance_elongation}.
As the NND decreases, the repulsive effects of nearby drones become stronger, while the increasing number of neighbors within the detection radius introduces additional repulsive components.
These competing repulsive effects can partially counteract the attractive component toward the destination, reducing the magnitude of the resulting commanded velocity and contributing to the decline in space-mean speed $\bar V$.
Meanwhile, the combined repulsive effects also cause greater deviations from the OD direction, leading to a more substantial reduction in space-mean projected speed $\bar U$ and, consequently, alignment efficiency $\bar\eta$.
These effects become increasingly evident as accumulation rises beyond $n=121$, where close-range interactions and collision-avoidance maneuvers become significantly more frequent.
The resulting directional deviations also require drones to travel longer paths toward their destinations, providing a physical explanation for the substantial increase in path elongation observed in \autoref{fig:loading_elongation_panel}. This transition also occurs near the critical accumulation identified from the observed outflow MFD, as discussed next.

\begin{figure}[htbp]
    \centering
    \begin{subfigure}[t]{0.24\linewidth}
        \centering
        \includegraphics[width=\linewidth]{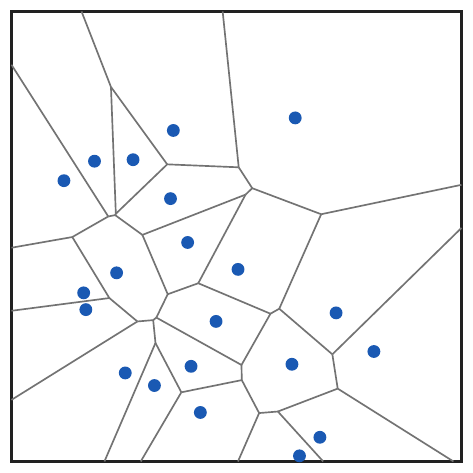}
        \caption{$n_{\mathrm{lim}}=41$.}
        \label{fig:loading_voronoi41}
    \end{subfigure}
    \hfill
    \begin{subfigure}[t]{0.24\linewidth}
        \centering
        \includegraphics[width=\linewidth]{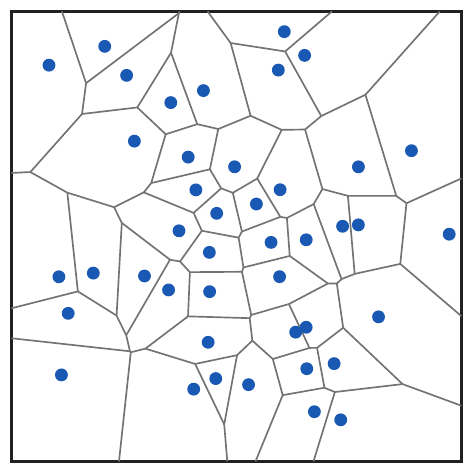}
        \caption{$n_{\mathrm{lim}}=81$.}
        \label{fig:loading_voronoi81}
    \end{subfigure}
    \hfill
    \begin{subfigure}[t]{0.24\linewidth}
        \centering
        \includegraphics[width=\linewidth]{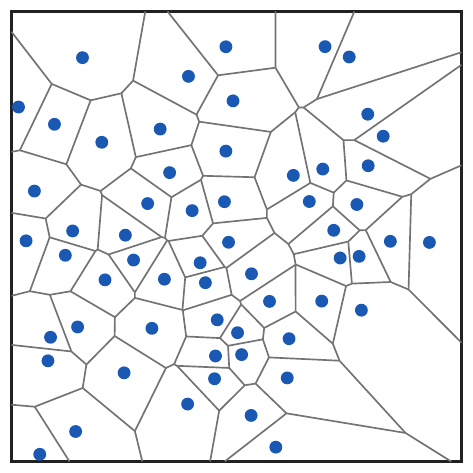}
        \caption{$n_{\mathrm{lim}}=121$.}
        \label{fig:loading_voronoi121}
    \end{subfigure}
    \hfill
    \begin{subfigure}[t]{0.24\linewidth}
        \centering
        \includegraphics[width=\linewidth]{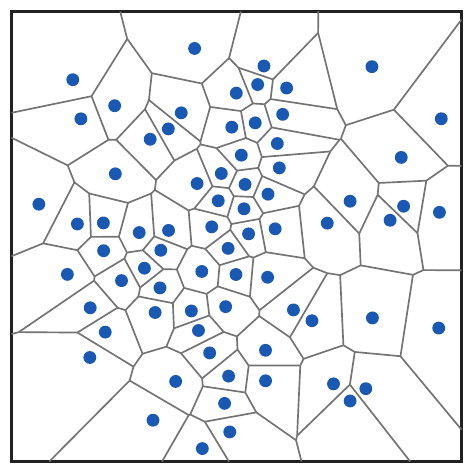}
        \caption{$n_{\mathrm{lim}}=161$.}
        \label{fig:loading_voronoi161}
    \end{subfigure}
    \caption{Top-view spatial snapshots under four accumulation ceilings. Blue dots indicate the horizontal positions of drones within the middle $50\%$ of the airspace altitude range, while black lines represent the corresponding two-dimensional Voronoi tessellation only for illustration.}
    \label{fig:loading_voronoi_snapshots}
\end{figure}

To further illustrate the spatial configurations underlying the NND distributions, \autoref{fig:loading_voronoi_snapshots} presents representative top-view snapshots of drones within the middle 50\% of the airspace altitude range, together with the corresponding two-dimensional Voronoi tessellations.
At low accumulation ceilings, drones are relatively sparsely distributed, resulting in larger Voronoi cells with considerable differences in size.
As the accumulation ceiling increases, the available space per drone decreases, leading to increasingly smaller cells, particularly in the central region of the airspace.
At $n_{\mathrm{lim}}=161$, most drones in the densely occupied central region are associated with relatively small Voronoi cells, while a few larger cells remain near the periphery.

\begin{figure}[htbp]
    \centering
    \begin{subfigure}[t]{0.49\linewidth}
        \centering
        \includegraphics[width=\linewidth]{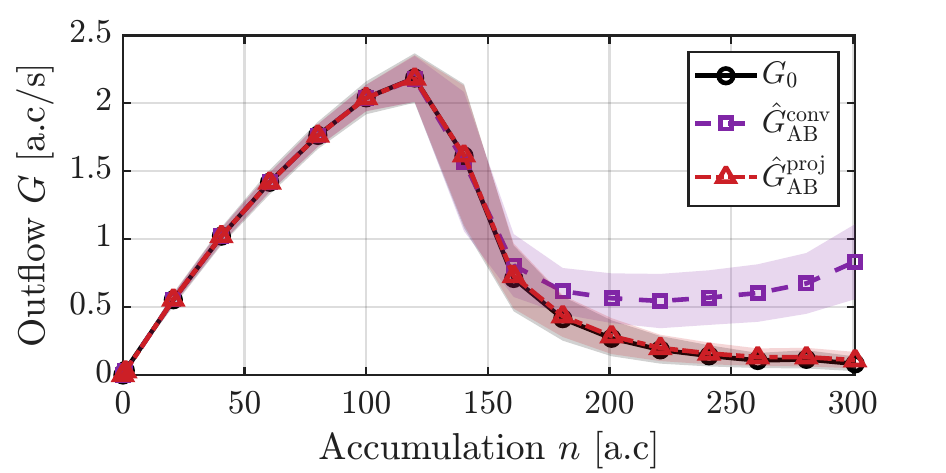}
        \caption{Departure-set estimation.}
        \label{fig:loading_mfd_departure}
    \end{subfigure}%
    \hfill
    \begin{subfigure}[t]{0.49\linewidth}
        \centering
        \includegraphics[width=\linewidth]{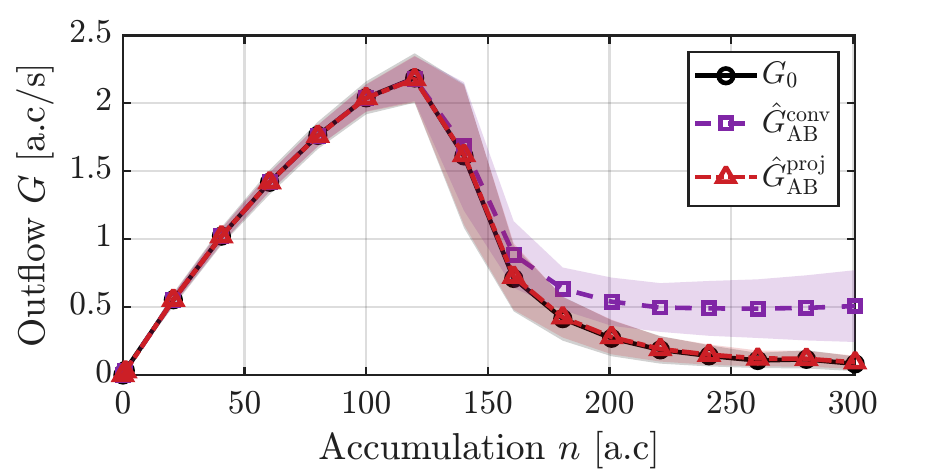}
        \caption{Completion-set estimation.}
        \label{fig:loading_mfd_completion}
    \end{subfigure}
    \caption{Observed and estimated outflow MFDs under stepwise accumulation-ceiling experiments. (a) Conventional and projected-space AB estimates using departure-based mean trip distances. (b) Corresponding estimates using completion-based mean trip distances. The observed outflow $G_0$ is shown in black, the conventional AB estimate in purple, and the projected-space AB estimate in red. Shaded bands indicate one standard deviation across independent simulation runs.}
    \label{fig:loading_outflow_comparison}
\end{figure}

Lastly, \autoref{fig:loading_outflow_comparison} compares the observed outflow $G_0$ with the conventional and projected-space AB estimates, $\hat G_{\mathrm{AB}}^{\mathrm{conv}}$ and $\hat G_{\mathrm{AB}}^{\mathrm{proj}}$, respectively, using two approaches to estimating mean trip distances.
\autoref{fig:loading_mfd_departure} uses the departure set, as adopted in the theoretical formulations, while \autoref{fig:loading_mfd_completion} uses the completion set discussed in \autoref{remark:depandcom}.

As shown in \autoref{fig:loading_outflow_comparison}, the observed outflow initially increases with accumulation along the free-flow branch, reaching its maximum at the critical accumulation
\(
n_{\mathrm{crit}}=\arg\max_n G_0(n)\approx121.
\)
Interestingly, the critical accumulation closely corresponds to the transition identified in the preceding NND analysis near $n_{\mathrm{lim}}=121$, where the mean NND approaches the detection radius $R_{\mathrm{detect}}$.
Under the force-field-based collision-avoidance mechanism adopted in the present model, this coincidence suggests a possible link between the onset of network congestion and the growing influence of individual collision avoidance.
Whether this relationship extends to other collision-avoidance mechanisms remains to be investigated in future research.

Turning to the outflow estimation results in \autoref{fig:loading_mfd_departure}, both AB estimators closely reproduce the observed outflow along the free-flow branch, but their performance diverges considerably beyond the critical accumulation.
The conventional AB estimator increasingly overestimates the observed outflow, remaining substantially above $G_0$ and even rising again under severe congestion.
In comparison, the proposed projected-space estimator closely follows the observed decreasing branch, including the near-gridlock regime, with only relatively small discrepancies at high accumulations.
Importantly, both AB formulations are theoretically consistent under stationary conditions, as established in \autoref{prop:steady_AB_heterogeneous} and \autoref{prop:steady_projected_AB}.
Their different numerical performance should therefore be interpreted in the context of finite measurement windows and deviations from stationarity.
Moreover, the departure-based conventional estimator requires the trip lengths of newly generated trips, which depend on subsequent collision-avoidance interactions and are therefore unavailable at departure.
Consequently, the conventional results in \autoref{fig:loading_mfd_departure} represent a retrospective benchmark rather than a directly implementable real-time estimator.
In contrast, the shortest OD distances used by the projected-space estimator are immediately available at departure, and its numerical estimates remain close to the observed outflow even under severe congestion.

Following \autoref{remark:depandcom}, we also examine the alternative of estimating the mean trip distance from recently completed trips, as shown in \autoref{fig:loading_mfd_completion}.
The projected-space estimator maintains a performance similar to its departure-based counterpart, closely reproducing the observed outflow across both free-flow and congested regimes.
For the conventional estimator, completion-based sampling moderately reduces the substantial overestimation observed with departure-based sampling at high accumulations, but significant discrepancies remain, particularly under severe congestion.
The considerable differences between the two conventional estimates further illustrate the sensitivity of trip-length-based estimation to the choice of sampling population.
In contrast, the projected-space estimates remain comparatively consistent across the two sampling approaches.

Nevertheless, completion-based sampling does not fully resolve the practical limitations of conventional AB estimation. Under changing traffic conditions, the trip lengths of recently completed trips may reflect earlier operating states and therefore introduce a temporal mismatch with the current traffic conditions.
Furthermore, if the complete set of trips completed within a measurement window is already observable, the corresponding outflow can be measured directly from the completion count, without requiring an AB estimator.
Completion-based trip lengths remain useful for retrospective calibration and validation, but provide limited additional value for independently estimating the outflow over the same observed window.
The departure-based projected-space formulation instead uses OD distances known at generation, offering a more practical basis for real-time outflow estimation and future model-based control.

Overall, the projected-space AB formulation provides substantially more accurate finite-window outflow estimates under the investigated congested conditions while avoiding the dependence on trip lengths that are unavailable at departure.
Its consistency across departure- and completion-based sampling further supports the robustness of the projected-space representation in these experiments.
Nevertheless, the stationary AB relations do not guarantee the same accuracy during rapidly evolving traffic conditions. This motivates the transient experiments and the trip-based estimators considered next.

\subsection{Evaluation of the projected-space trip-based formulation}
\label{subsec:result_TB}

We next evaluate the conventional and projected-space TB formulations under time-varying traffic conditions. Unlike the AB formulations, which estimate outflow using aggregate speed and trip-distance statistics, the TB formulations explicitly track individual trip progression to estimate completion times, thereby accounting for the evolution of traffic conditions experienced by active trips. As in the AB comparison, the conventional TB estimator is evaluated retrospectively using realized trip lengths, whereas the projected-space TB estimator uses shortest OD distances available at departure. The AB estimators are also included as benchmarks to highlight the advantages of explicitly accounting for trip progression under transient conditions, while the primary objective of this section remains to evaluate the accuracy and applicability of the TB formulations.

\subsubsection{Loading profile with abrupt demand variations}
\label{subsubsec:TB_traffic_loading}

To examine the limitations of AB estimation and evaluate the TB formulations under transient conditions, we consider an additional experiment with rapid loading and unloading, without the intermediate stabilization used in \autoref{subsubsec:traffic_loading}. In this setting, the requested departure rate $q_{\mathrm{req}}(t)$ increases from 0 to 2.5 [a.c./s] before decreasing gradually back to 0. The departure admission mechanism remains unchanged, but the accumulation ceiling is fixed at $n_{\mathrm{lim}}=150$ [a.c.], restricting departures whenever the ceiling is reached.

\begin{figure}[htbp]
    \centering
    \begin{subfigure}[t]{0.49\linewidth}
        \centering
        \includegraphics[width=\linewidth]{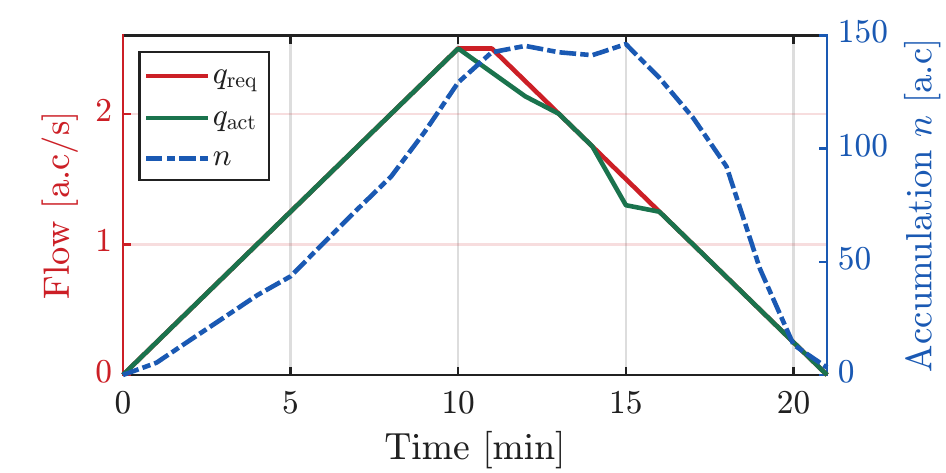}
        \caption{Actual inflow and accumulation.}
        \label{fig:TB_loading_time}
    \end{subfigure}
    \hfill
    \begin{subfigure}[t]{0.49\linewidth}
        \centering
        \includegraphics[width=\linewidth]{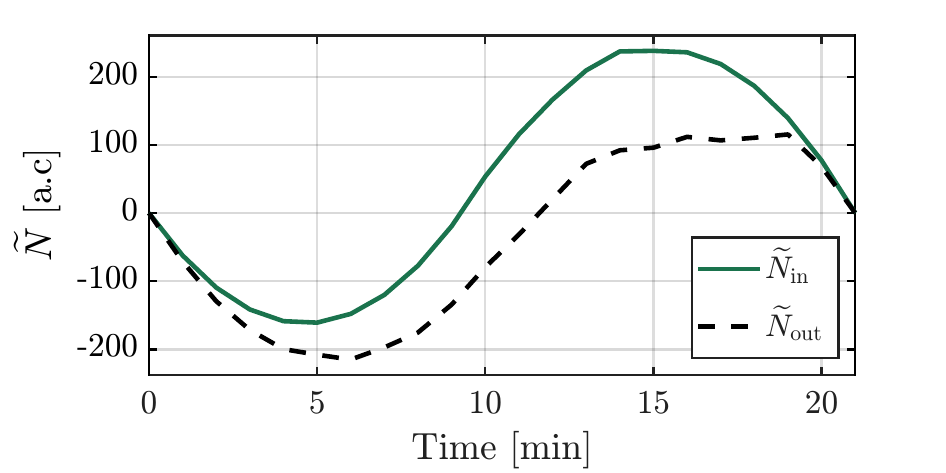}
        \caption{Oblique cumulative counts.}
        \label{fig:TB_loading_oblique}
    \end{subfigure}
    \caption{Transient traffic evolution under abrupt demand variations. (a) Requested inflow $q_\mathrm{req}$[m] actual inflow $q_{\mathrm{act}}$ and network accumulation $n$. (b) Oblique cumulative inflow $\widetilde N_{\mathrm{in}}$ and outflow $\widetilde N_{\mathrm{out}}$, with $\bar{q}_\mathrm{act}=1.29$ [a.c/s]. Results from one illustrative simulation realization.}
    \label{fig:TB_loading_profile}
\end{figure}

As shown in \autoref{fig:TB_loading_time}, the actual inflow $q_{\mathrm{act}}$ initially follows the requested inflow, while accumulation increases continuously. Following the demand reduction, accumulation briefly continues to rise, reaching approximately 150 [a.c.], before declining as trip completions exceed admitted departures. The corresponding oblique cumulative-count curves in \autoref{fig:TB_loading_oblique}, defined in Eq.~\eqref{eq:loading_oblique}, illustrate this evolution through their increasing separation during loading and decreasing separation during unloading. These trajectories are obtained from one illustrative realization among multiple independent runs, which may exhibit different traffic evolution due to variations in OD distributions and collision-avoidance interactions.

\begin{figure}[htbp]
    \centering
    \begin{subfigure}[t]{0.49\linewidth}
        \centering
        \includegraphics[width=\linewidth]{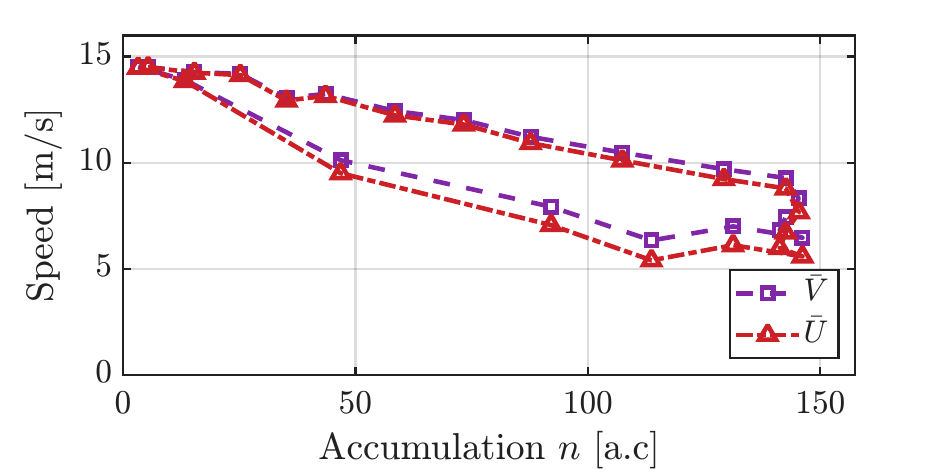}
        \caption{True and projected speed MFDs.}
        \label{fig:TB_hysteresis_speed}
    \end{subfigure}
    \hfill
    \begin{subfigure}[t]{0.49\linewidth}
        \centering
        \includegraphics[width=\linewidth]{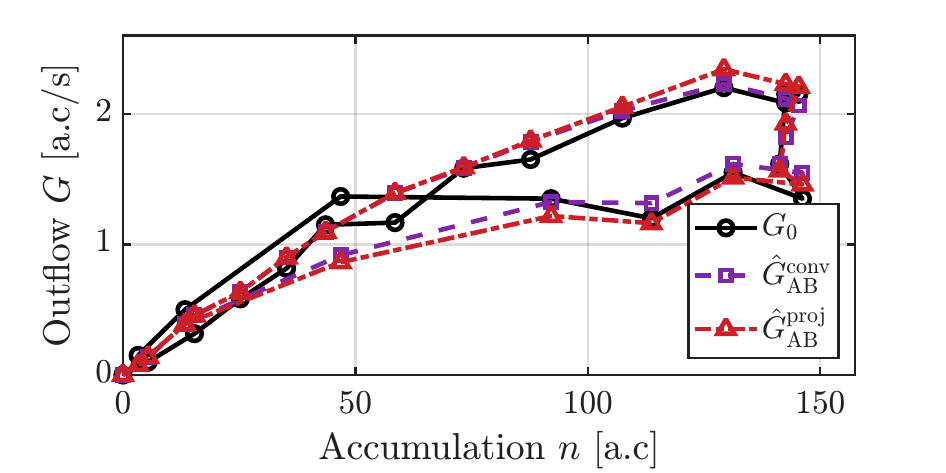}
        \caption{Observed and AB-estimated outflow MFDs.}
        \label{fig:TB_hysteresis_outflow}
    \end{subfigure}
    \caption{Illustrative hysteresis under transient loading. (a) space-mean speed $\bar V$ and projected speed $\bar U$. (b) Observed outflow $G_0$ and conventional and projected-space AB estimates. Results from one selected simulation realization.}
    \label{fig:TB_hysteresis}
\end{figure}

Since the AB formulations are theoretically established under stationary conditions, their accuracy may deteriorate under such transient traffic states. To illustrate this limitation, \autoref{fig:TB_hysteresis} presents the speed--accumulation and outflow--accumulation relationships from a selected realization. As shown in \autoref{fig:TB_hysteresis_speed}, both the true and projected speeds exhibit pronounced hysteresis (clockwise loop), with lower speeds during unloading at comparable accumulations before recovering toward their free-flow values. Similarly, \autoref{fig:TB_hysteresis_outflow} shows a distinct hysteresis loop in the observed outflow $G_0$. Although both AB estimators capture similar loop-like patterns, their estimated outflows deviate noticeably from the observations, particularly along the unloading branch. These results qualitatively illustrate the dependence of macroscopic traffic relationships on loading history and the limitations of AB estimation under transient conditions. A systematic quantitative comparison across independent runs is presented in the following subsection.

\subsubsection{Outflow estimation under transient dynamics}
\label{subsubsec:TB_transient_results}

We next compare the conventional and projected-space AB and TB estimators under the transient loading scenario introduced in \autoref{subsubsec:TB_traffic_loading}, where the experiment is repeated over 20 independent simulation runs using the same requested-demand profile. The AB estimates are obtained from Eq.~\eqref{eq:conventional_AB_estimator} and Eq.~\eqref{eq:projected_AB_estimator}, while the TB estimates are calculated using Eq.~\eqref{eq:TB_conv_estimator} and Eq.~\eqref{eq:GhatTB}, respectively, with $V(t)$ and $U(t)$ updated every $W=1$ [min]. 

\begin{figure}[htbp]
    \centering
    \begin{subfigure}[t]{0.49\linewidth}
        \centering
        \includegraphics[width=\linewidth]{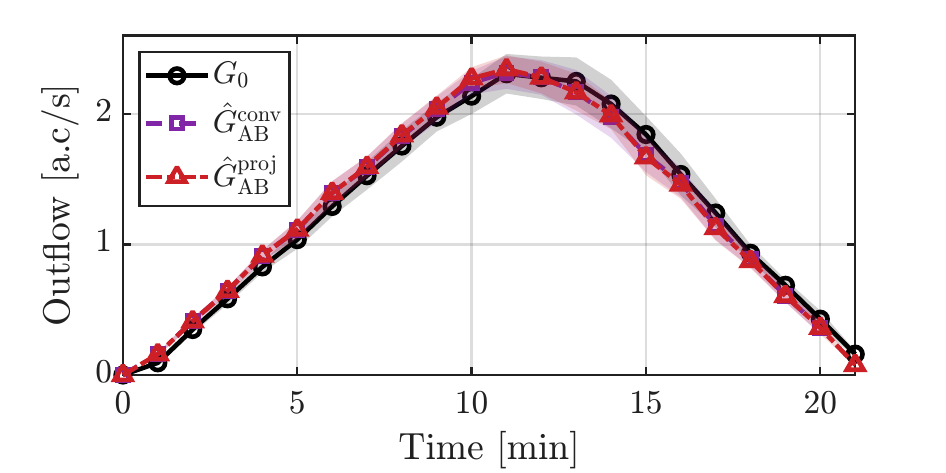}
        \caption{Accumulation-based estimates.}
        \label{fig:TB_AB_comparison}
    \end{subfigure}
    \hfill
    \begin{subfigure}[t]{0.49\linewidth}
        \centering
        \includegraphics[width=\linewidth]{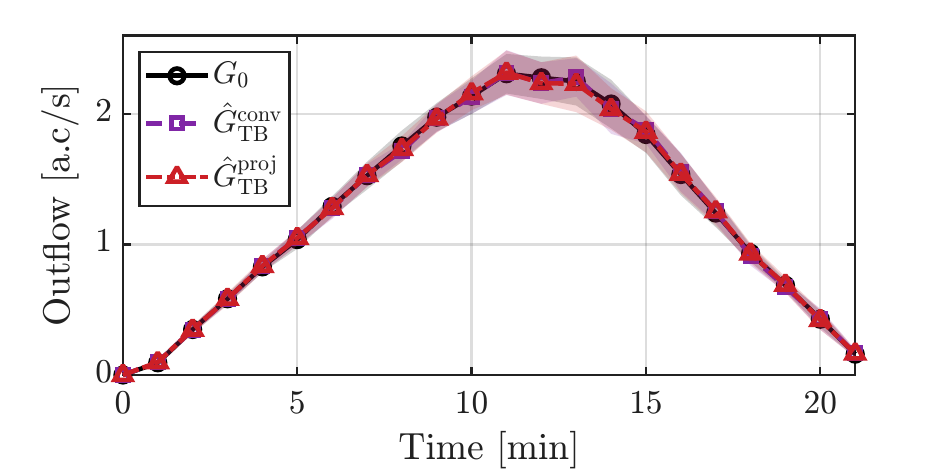}
        \caption{Trip-based estimates.}
        \label{fig:TB_TB_comparison}
    \end{subfigure}
    \caption{Observed and estimated outflows under transient loading. (a) Conventional and projected-space AB estimates. (b) Conventional and projected-space TB estimates. Curves represent the means across 20 independent simulation runs, while shaded bands indicate one standard deviation.}
    \label{fig:TB_outflow_comparison}
\end{figure}

\autoref{fig:TB_outflow_comparison} compares the observed outflow $G_0$ with the AB and TB estimates. As shown in \autoref{fig:TB_AB_comparison}, both AB estimators continuously overestimate the observed outflow during the loading phase and underestimate it during unloading. This systematic deviation reflects the limitations of instantaneous AB outflow estimation under rapidly evolving traffic conditions, where the current accumulation and space-mean speed may not adequately capture the progression of active trips. In contrast, \autoref{fig:TB_TB_comparison} shows that both TB estimators closely reproduce the observed outflow throughout the loading and unloading phases, including the onset of congestion and subsequent recovery. Their estimated trajectories are also nearly identical, indicating comparable performance under the investigated transient conditions.

\begin{figure}[htbp]
    \centering
    \begin{subfigure}[t]{0.49\linewidth}
        \centering
        \includegraphics[width=\linewidth]{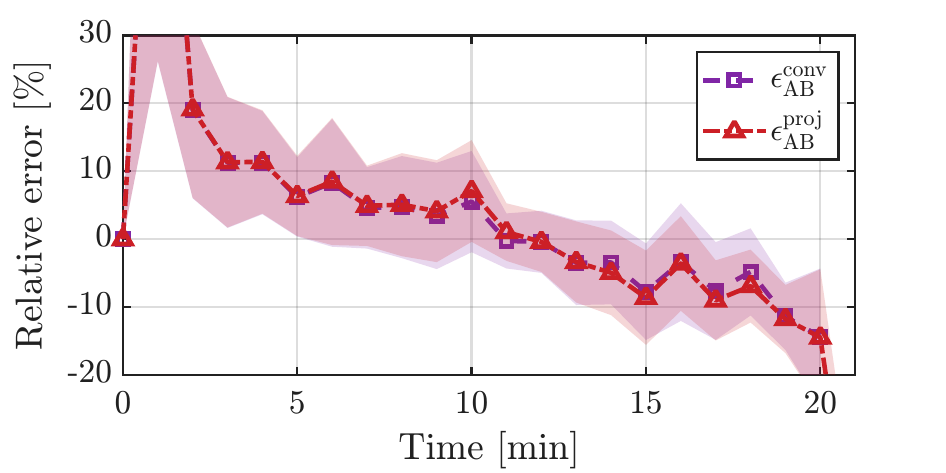}
        \caption{Accumulation-based estimation errors.}
        \label{fig:TB_AB_error}
    \end{subfigure}
    \hfill
    \begin{subfigure}[t]{0.49\linewidth}
        \centering
        \includegraphics[width=\linewidth]{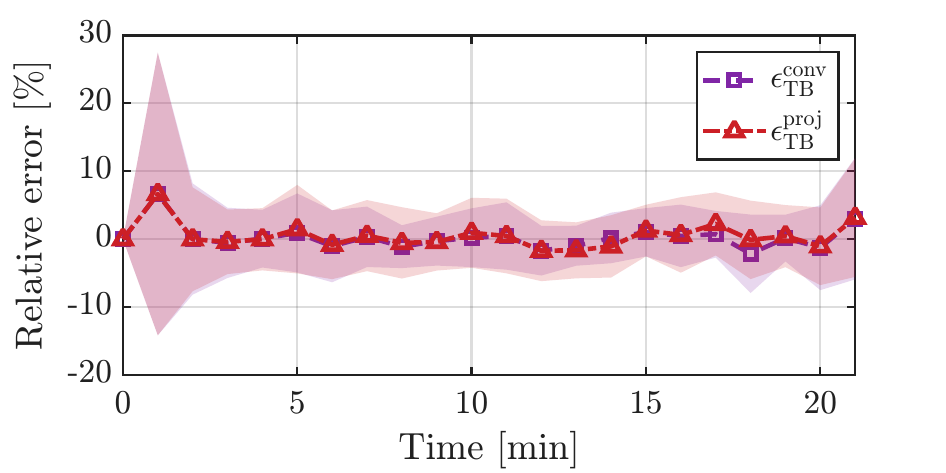}
        \caption{Trip-based estimation errors.}
        \label{fig:TB_TB_error}
    \end{subfigure}
    \caption{Relative outflow estimation errors under transient loading. (a) Conventional and projected-space AB estimators. (b) Conventional and projected-space TB estimators. Curves represent the means across 20 independent simulation runs, while shaded bands indicate one standard deviation. The vertical axes are limited to the range $[-30\%,30\%]$ for clearer visual comparison.}
    \label{fig:TB_error_comparison}
\end{figure}

To further quantify the estimation accuracy, we define the relative outflow estimation error as
\begin{equation}
\label{eq:relative_outflow_error}
    \epsilon(t,W)
    =
    \frac{\hat G(t,W)-G_0(t,W)}{G_0(t,W)}
    \times 100\%,
    \qquad G_0(t,W)>0,
\end{equation}
where $\hat G(t,W)$ denotes the corresponding AB or TB outflow estimate. Positive and negative values indicate overestimation and underestimation, respectively. \autoref{fig:TB_error_comparison} presents the mean and standard deviation of the relative errors across independent simulation runs.

As shown in \autoref{fig:TB_AB_error}, the AB estimators exhibit relatively large positive errors during the initial loading phase at $t$ from 0 to 12 [min], followed by negative errors during unloading. This pattern is consistent with the systematic deviations observed in \autoref{fig:TB_AB_comparison}. However, the sign of the estimation error is not determined solely by whether accumulation is increasing or decreasing; it also depends on the evolving traffic state and active-trip distribution. The observed relationship between loading direction and estimation bias therefore characterizes the present experiment rather than a general property of AB formulations. In comparison, \autoref{fig:TB_TB_error} shows that the TB formulations achieve substantially smaller errors under the investigated transient conditions.

These results highlight the different roles of the AB and TB formulations. While the AB models provide compact and theoretically consistent outflow relations under stationary conditions, their instantaneous estimates can exhibit substantial biases when traffic conditions change rapidly. The TB formulations instead account for the progression of individual trips, allowing them to better capture the effects of traffic evolution on trip completions. Importantly, although the conventional and projected-space TB estimators exhibit similarly high accuracy, the conventional estimator relies on realized trip lengths $l_i$, which are only known after trip completion and are therefore supplied retrospectively in this experiment. In contrast, the projected-space TB estimator uses the shortest OD distances $s_i$, which are available at departure. The projected-space formulation thus achieves comparable transient estimation accuracy without requiring prior knowledge of the realized trajectories, making it more suitable for real-time implementation.

\subsubsection{Sensitivity to the measurement window}
\label{subsubsec:TB_window_sensitivity}

As established in \autoref{thm:TB_outflow_error_bound} and \autoref{thm:TB_outflow_error_bound_proj}, the finite-window error bounds decrease proportionally to $1/W$ when the other terms are held fixed, although this does not guarantee a corresponding decrease in the actual estimation error. To examine this relationship numerically, we evaluate both TB estimators under the transient loading scenario in \autoref{subsubsec:TB_traffic_loading} but using measurement windows of $W\in\{0.25,0.5,1,2,4\}$ [min]. The absolute relative error $|\epsilon|$, with $\epsilon$ defined in Eq.~\eqref{eq:relative_outflow_error}, is calculated for each simulation run and averaged across multiple independent runs at each measurement time. The mean absolute relative error over the observation period is also reported for each window duration.

\begin{figure}[htbp]
    \centering
    \begin{subfigure}[t]{0.49\linewidth}
        \centering
        \includegraphics[width=\linewidth]{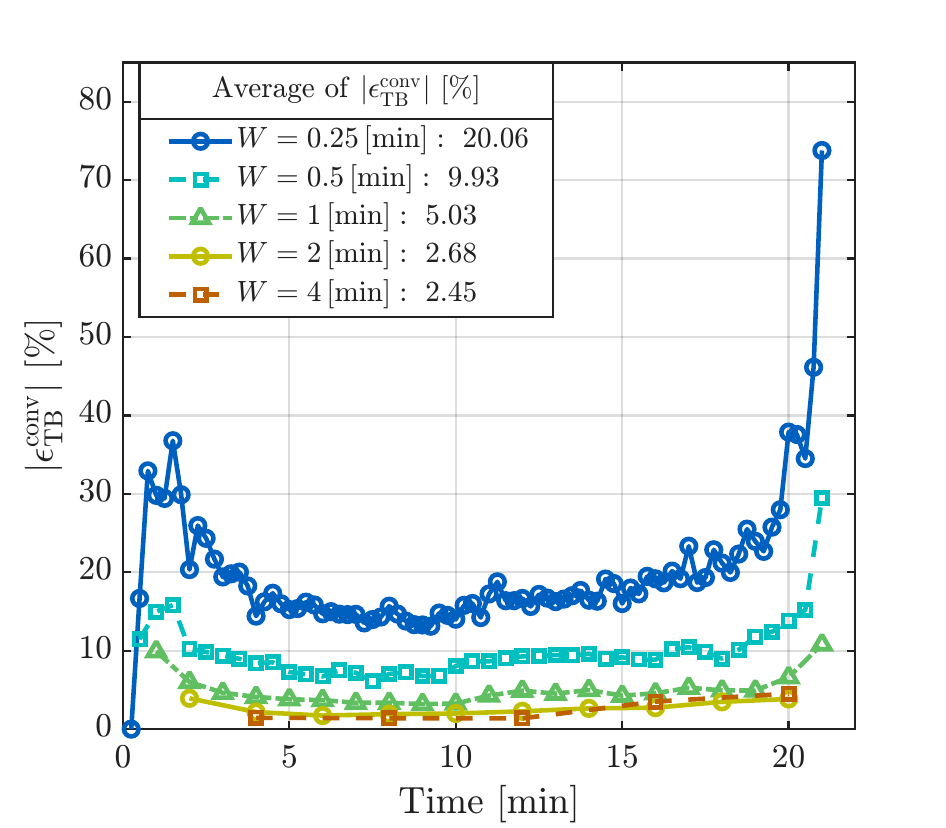}
        \caption{Conventional TB estimator.}
        \label{fig:TB_window_conv}
    \end{subfigure}
    \hfill
    \begin{subfigure}[t]{0.49\linewidth}
        \centering
        \includegraphics[width=\linewidth]{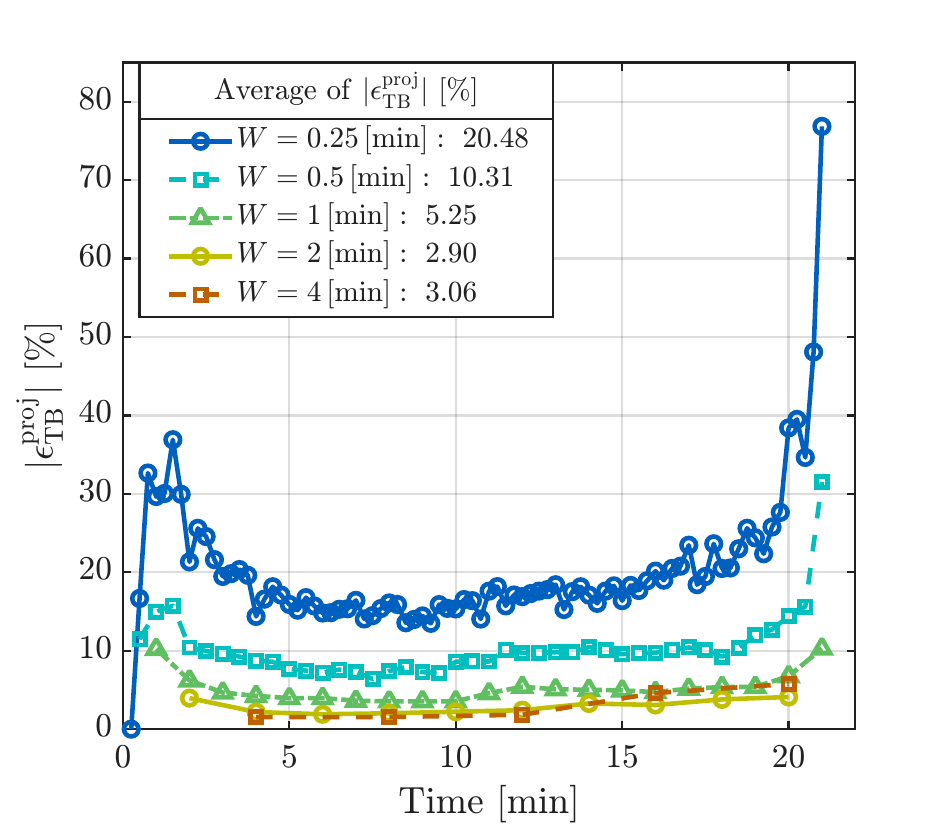}
        \caption{Projected-space TB estimator.}
        \label{fig:TB_window_proj}
    \end{subfigure}
    \caption{Sensitivity of absolute relative outflow estimation errors to the measurement-window duration $W$ under transient loading with inflow updated every minute. Curves represent the mean absolute relative errors across multiple independent simulation runs, while the legends report their temporal averages over the observation period.}
    \label{fig:TB_window_sensitivity}
\end{figure}

\autoref{fig:TB_window_sensitivity} presents the resulting error trajectories for the conventional and projected-space TB estimators. Both exhibit similar behavior, with substantially decreasing errors as $W$ increases from 0.25 to 2 [min]. Specifically, the mean absolute relative errors decrease from 20.06\% to 2.68\% for the conventional estimator and from 20.48\% to 2.90\% for the projected-space estimator. Over this range, doubling the measurement window approximately halves the mean absolute relative error, consistent with the inverse-window dependence suggested by the theoretical bounds. The errors are generally larger near the beginning and end of the experiment, where the observed outflow is relatively low and individual completion-time discrepancies have a greater influence on the relative error.

Further increasing $W$ from 2 to 4 [min] produces little additional improvement. The mean absolute relative error decreases slightly from 2.68\% to 2.45\% for the conventional estimator, while increasing from 2.90\% to 3.06\% for the projected-space estimator. This saturation can be understood from the mechanism underlying TB estimation errors. Discrepancies between estimated and actual trip completion times affect the estimated outflow primarily through trips whose completion times lie near the measurement-window boundaries. For shorter windows, these discrepancies can account for a substantial fraction of the completed trips, leading to errors that decrease approximately inversely with $W$. For longer windows, however, the actual error depends on the distribution and cancellation of completion-time misclassifications at the two boundaries, which need not decrease monotonically with increasing $W$. Moreover, the relative error also depends on the observed outflow in each window. Consequently, the theoretical $1/W$ dependence of the upper bounds does not imply a corresponding monotonic reduction in empirical estimation errors.

A further observation is that the conventional TB estimator consistently achieves slightly lower errors than its projected-space counterpart. This difference can be interpreted through the theoretical error bounds in Eqs.~\eqref{eq:TB_outflow_error_bound} and \eqref{eq:projected_TB_outflow_error_bound}, where the completion-time error parameters are given by $\delta_r=\varepsilon_V T_{\mathrm{ub}}/V_{\mathrm{lb}}$ and $\delta_p=\varepsilon_U T_{\mathrm{ub}}/U_{\mathrm{lb}}$, respectively. In dense traffic, drones frequently adjust their flight directions to avoid conflicts while maintaining relatively stable speed magnitudes. Since projected speed depends on both speed magnitude and alignment with the origin--destination direction, these directional changes can introduce greater variability in projected speed, potentially leading to $\varepsilon_U>\varepsilon_V$. Furthermore, the space-mean projected speed generally has a lower positive bound than the true speed, i.e., $U_{\mathrm{lb}}<V_{\mathrm{lb}}$. Together, these conditions yield $\delta_r<\delta_p$, providing a tighter theoretical error bound for the conventional estimator, consistent with the observed numerical results. Nevertheless, the conventional TB estimator requires realized trip lengths, which are generally unavailable at trip generation in LAAT systems. Its results therefore serve as a benchmark, whereas the projected-space TB estimator can operate using the known shortest origin--destination distances.

Overall, these results highlight the practical trade-off in selecting the measurement window. Although shorter windows provide higher temporal resolution, they can produce more fluctuating outflow estimates, particularly when few trips are completed within each interval. This observation supports the data-resolution considerations discussed earlier in connection with loop-detector measurements \citep{I210}, reinforcing that finer temporal resolution does not necessarily yield more accurate estimates. Under the investigated scenario, $W=2$ [min] provides a reasonable balance between temporal resolution and estimation accuracy, while longer windows offer limited additional benefits.

\subsection{Priority-based multi-class traffic}
\label{subsec:result_multiclass}

The multi-class projected-space framework developed in \autoref{subsec:multi_class} accommodates heterogeneous drone classes with distinct motion characteristics and operational rules. As a representative application, we consider a priority-based two-class system, consisting of priority ($\mathrm{pr}$) and standard ($\mathrm{st}$) drones, to investigate the effects of differentiated collision avoidance on macroscopic traffic behavior. While the theoretical framework allows an arbitrary number of classes, the numerical experiments focus on two classes to isolate the effects of priority assignment.

\begin{figure}[htbp]
    \centering
    \includegraphics[width=0.9\linewidth]{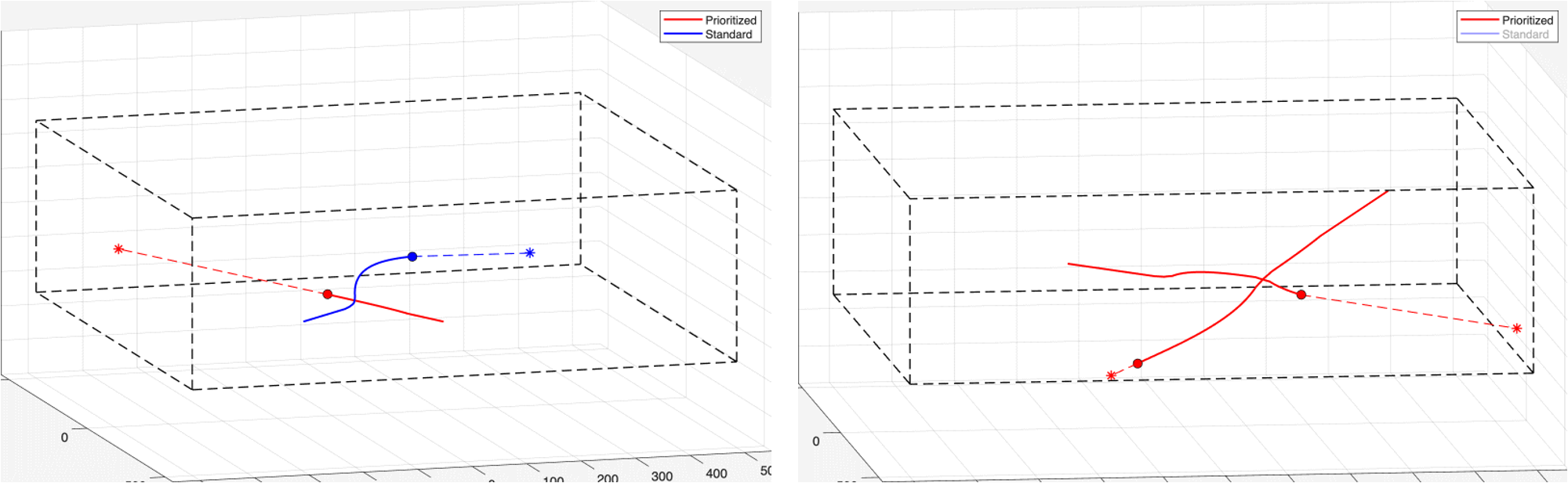}
    \caption{Illustration of priority-based collision avoidance in LAAT. Solid curves indicate traveled trajectories, solid dots current positions, eight-spoked asterisks destinations, and dashed lines direct paths to destinations without further avoidance. Left: Standard drones (blue) yield to priority drones (red). Right: Same-class drones mutually avoid collisions.}
    \label{fig:priority_demo}
\end{figure}

Priority is implemented through asymmetric collision-avoidance behavior. During interactions between different classes, priority drones maintain their intended trajectories whenever feasible, while standard drones undertake a greater share of the avoidance maneuvers. However, priority does not eliminate collision-avoidance requirements, as drones belonging to the same class must still adjust their trajectories when necessary. \autoref{fig:priority_demo} illustrates these interaction patterns. Such differentiated behavior may lead to class-dependent trajectory deviations, projected speeds, and travel times, thereby affecting both class-specific and aggregate traffic characteristics.

The experiments follow the same accumulation-ceiling setting described in \autoref{subsec:config}. Both classes share identical nominal flight characteristics and the same OD distribution, differing only in their collision-avoidance priorities. Consequently, their expected shortest OD distances are equal, i.e., $S_{\mathrm{pr}}=S_{\mathrm{st}}=S$, satisfying the common-distance assumption required for the aggregate projected-space formulation.

The numerical investigation is organized into three parts. First, the aggregate multi-class MFD and class-fraction relations are verified against simulation observations under different priority compositions. The subsequent analyses focus on the effects of priority composition, first examining the class-specific speed and outflow characteristics, and then evaluating the resulting changes in aggregate MFD characteristics and network capacity. In the following multi-class analyses, we focus exclusively on the projected-space formulation, whose advantages over the conventional formulation have been demonstrated in the preceding sections.

\subsubsection{Verification of the aggregate projected-space multi-class formulation}
\label{subsubsec:mc_verification}

We first evaluate the applicability of the projected-space formulation to heterogeneous priority-based traffic by numerically verifying two theoretical results: the aggregate multi-class MFD relation in \autoref{prop:multi_class_projected_AB} and the relationship between generation and accumulation class fractions in \autoref{prop:multi_class_fraction_relation}. Four representative generation priority fractions, $\alpha_{\mathrm{pr}}^q\in\{0.2,0.4,0.6,0.8\}$, are considered to assess the consistency of these relations across different priority compositions. The objective here is to verify the theoretical relations under different traffic scenarios, rather than to evaluate the performance effects of varying the priority fraction, which are investigated in the following subsections.

Since both classes share the same expected shortest OD distance $S$, \autoref{prop:multi_class_projected_AB} gives the stationary aggregate outflow relation
\begin{equation}
    G_{\mathrm{tot}}(\mathbf n)
    =
    \frac{n_{\mathrm{tot}}U(\mathbf n)}{S},
\end{equation}
where $n_{\mathrm{tot}}=n_{\mathrm{pr}}+n_{\mathrm{st}}$ and $U(\mathbf n)$ denotes the space-mean projected speed of the entire active population. The corresponding finite-window outflow estimate is
\begin{equation}
\label{eq:multiclass_AB_estimate_results_template}
    \hat G_{\mathrm{AB}}^{\mathrm{proj}}(t,W)
    =
    \frac{\bar n(t,W)\bar U(t,W)}{\bar S_D(t,W)}.
\end{equation}

\begin{figure}[htbp]
    \centering
    \MCFourPanels%
        {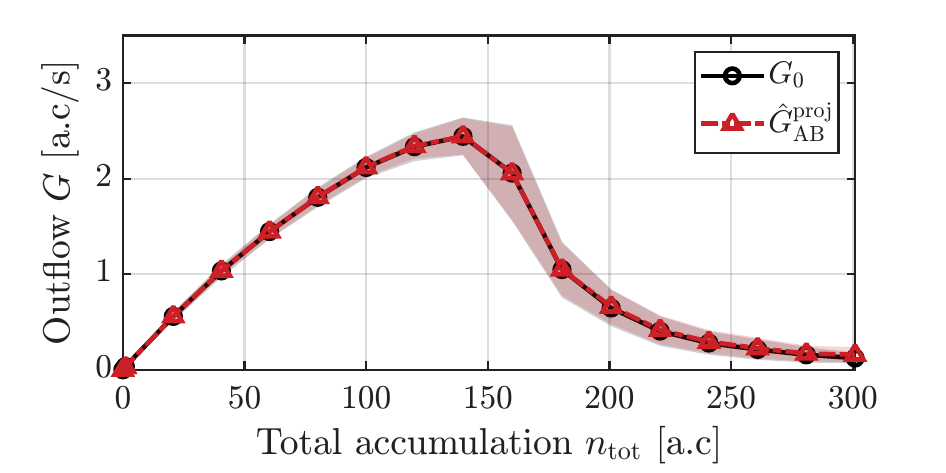}%
        {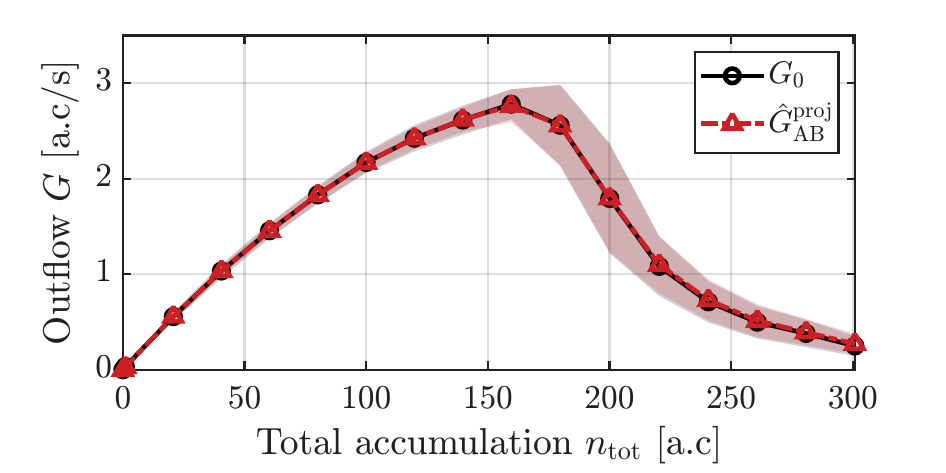}%
        {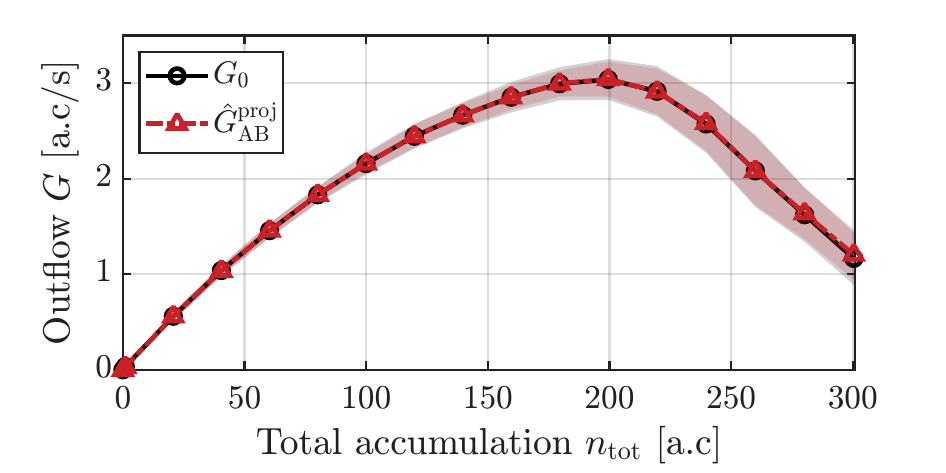}%
        {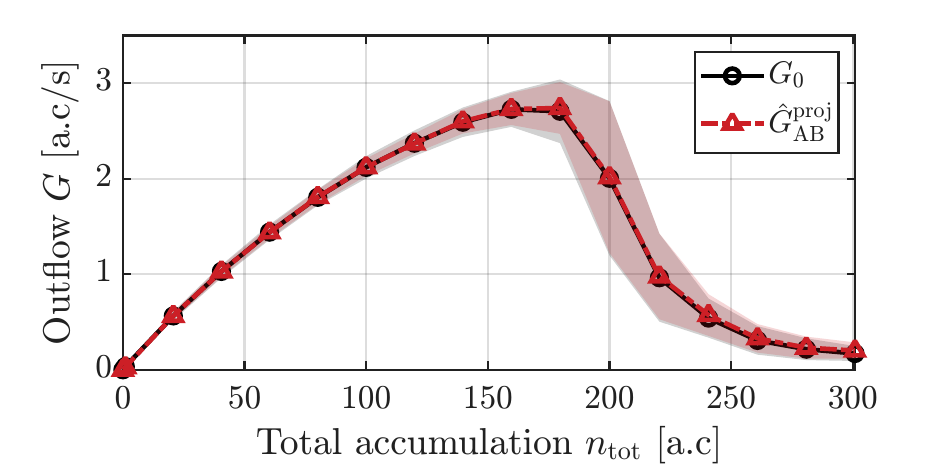}
    \caption{Verification of the aggregate projected-space MFD under generation priority fractions $\alpha_{\mathrm{pr}}^q=0.2$, $0.4$, $0.6$, and $0.8$. Solid curves represent the mean observed and estimated aggregate outflows over multiple independent simulation runs, while shaded bands indicate one standard deviation.}
    \label{fig:multiclass_AB_comparison}
\end{figure}

\autoref{fig:multiclass_AB_comparison} compares the observed total outflow $G_0$ with the projected-space estimate $\hat G_{\mathrm{AB}}^{\mathrm{proj}}$ for the four generation priority fractions. For both quantities, the mean and standard deviation across multiple independent simulation runs are presented, with the latter represented by shaded bands.
In all four cases, the estimated mean outflow closely follows the observations over both the increasing and decreasing branches of the MFD. Notably, the estimated standard deviations also agree closely with the observed ones, although this agreement is less apparent visually because the shaded bands nearly overlap, with that of $G_0$ largely obscured by that of $\hat G_{\mathrm{AB}}^{\mathrm{proj}}$. Some discrepancies become more noticeable at high accumulations, where the fixed measurement window becomes relatively short compared with the increasing average travel time and may therefore be insufficient to fully characterize the traffic state. Nevertheless, the absolute estimation errors remain small even under severe congestion. Overall, these results provide numerical support for the applicability of the aggregate projected-space relation to heterogeneous priority-based traffic, provided that the classes share the same expected shortest OD distance.

The aggregate outflow relation is complemented by the class-fraction relation, which characterizes the composition of the active drone population. While the generation priority fraction $\alpha_{\mathrm{pr}}^q$ is prescribed and remains constant within each experiment, the accumulation priority fraction $\alpha_{\mathrm{pr}}^n$ depends on the relative travel times of the two classes and may vary with congestion. Under the common-distance assumption, \autoref{prop:multi_class_fraction_relation} establishes the stationary relationship
\begin{equation}
\label{eq:alpha_priority_relation_template}
    \alpha_{\mathrm{pr}}^n
    =\frac{n_{\mathrm{pr}}}{n_{\mathrm{tot}}}
    =
    \frac{
        \alpha_{\mathrm{pr}}^q U_{\mathrm{st}}(\mathbf n)
    }{
        \alpha_{\mathrm{pr}}^q U_{\mathrm{st}}(\mathbf n)
        +(1-\alpha_{\mathrm{pr}}^q)U_{\mathrm{pr}}(\mathbf n)
    }.
\end{equation}
For numerical verification, the observed accumulation priority fraction and its estimate based on the finite-window space-mean projected speeds are respectively defined as
\begin{equation}
\label{eq:alpha_priority_estimate_template}
\begin{aligned}
    \alpha_{\mathrm{pr,0}}^{n}(t,W)
    &=
    \frac{\bar n_{\mathrm{pr}}(t,W)}
         {\bar n_{\mathrm{tot}}(t,W)},\\
    \hat\alpha_{\mathrm{pr}}^n(t,W)
    &=
    \frac{
        \alpha_{\mathrm{pr}}^q\bar U_{\mathrm{st}}(t,W)
    }{
        \alpha_{\mathrm{pr}}^q\bar U_{\mathrm{st}}(t,W)
        +(1-\alpha_{\mathrm{pr}}^q)\bar U_{\mathrm{pr}}(t,W)
    }.
\end{aligned}
\end{equation}

\begin{figure}[htbp]
    \centering
    \MCFourPanels%
        {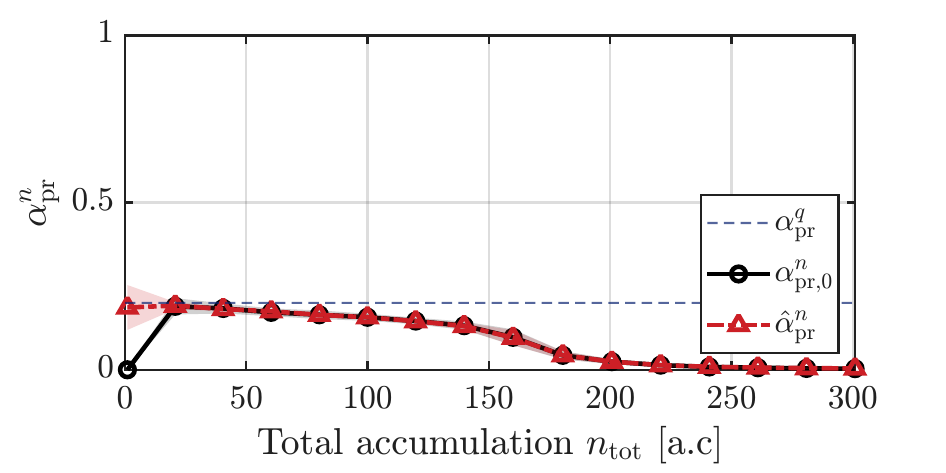}%
        {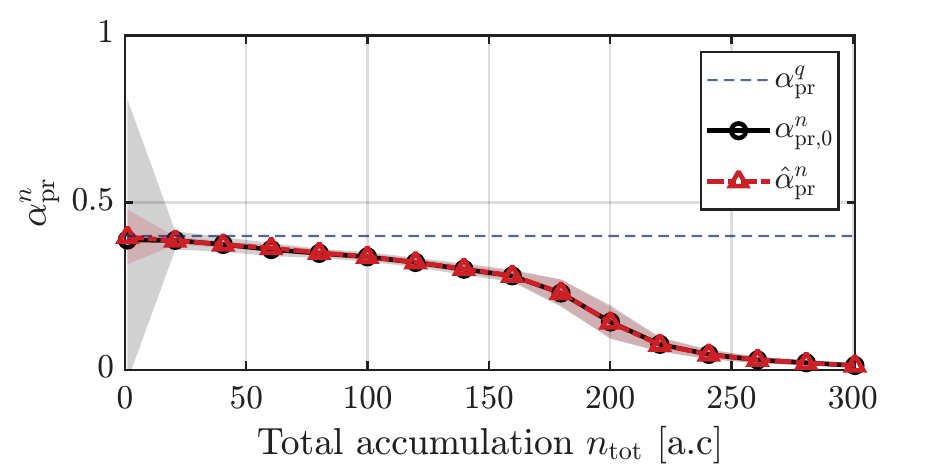}%
        {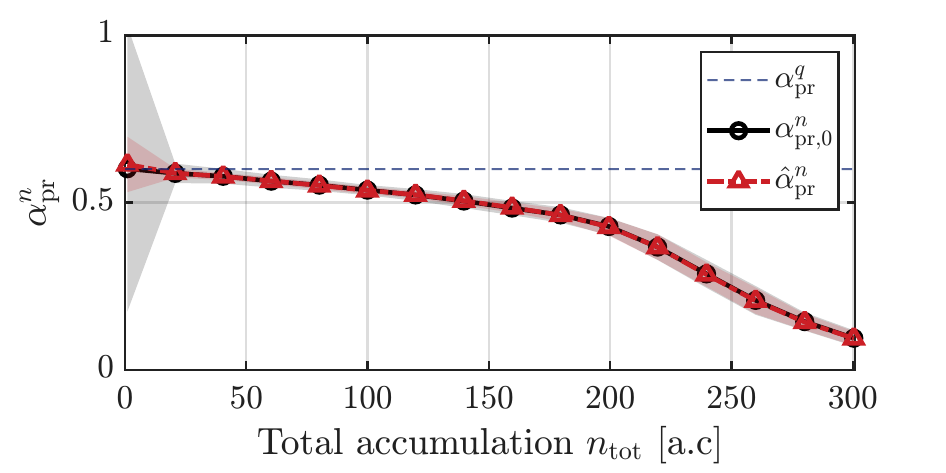}%
        {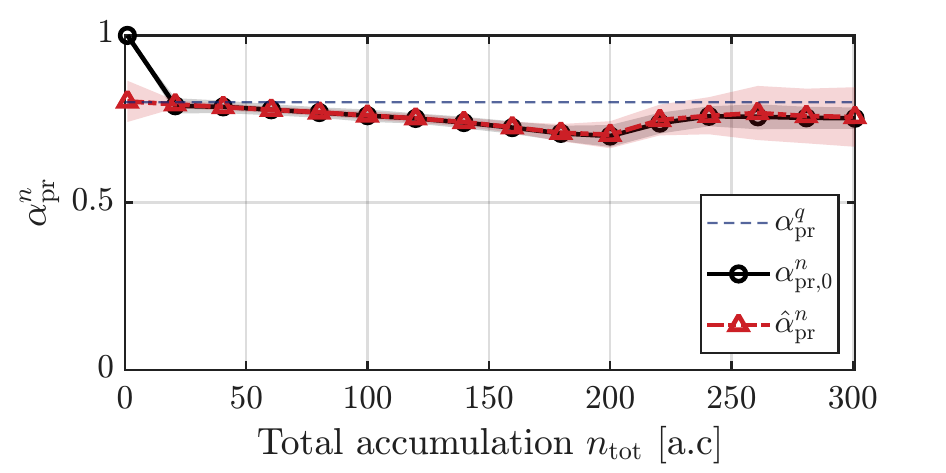}
    \caption{Verification of the relationship between generation and accumulation priority fractions for $\alpha_{\mathrm{pr}}^q=0.2$, $0.4$, $0.6$, and $0.8$. The observed accumulation priority fraction $\alpha_{\mathrm{pr,0}}^{n}$ is compared with its projected-speed-based estimate $\hat\alpha_{\mathrm{pr}}^n$ from Eq.~\eqref{eq:alpha_priority_estimate_template}. Shaded bands indicate one standard deviation across multiple independent simulation runs.}
    \label{fig:multiclass_fraction_verification}
\end{figure}

\autoref{fig:multiclass_fraction_verification} compares the observed accumulation priority fraction $\alpha_{\mathrm{pr,0}}^{n}$ with the estimate $\hat\alpha_{\mathrm{pr}}^n$ for the four generation priority fractions. In all cases, the observed fraction is initially close to the prescribed generation fraction $\alpha_{\mathrm{pr}}^q$, but decreases markedly as total accumulation increases. This reduction becomes particularly pronounced in the congested regime, where standard drones experience substantially lower OD-directed progress than priority drones. Consequently, standard drones remain in the airspace for longer durations and account for an increasingly large share of the active population, despite the unchanged generation composition.

The estimate $\hat\alpha_{\mathrm{pr}}^n$ closely reproduces these trends across all four cases, including the sharp reductions under severe congestion. Relatively larger discrepancies occur at very low accumulations, since the limited number of active drones makes the observed class fractions more susceptible to sampling variability. Nevertheless, the overall agreement provides numerical support for \autoref{prop:multi_class_fraction_relation}, demonstrating that the difference between generation and accumulation compositions can be captured through the class-specific space-mean projected speeds.

Together, the aggregate outflow and class-fraction comparisons provide numerical support for both multi-class propositions across different priority compositions. Despite the heterogeneous motion characteristics of priority and standard drones, the aggregate projected-space relation accurately estimates their combined outflow, while the class-fraction relation captures the resulting changes in active traffic composition. Having verified these relations, we next examine how varying the generation priority fraction affects the class-specific speed and outflow characteristics, before investigating its implications for aggregate network performance.


\subsubsection{Effects of priority composition on class-specific speeds and MFDs}
\label{subsubsec:mc_class_specific}

Following the verification of the multi-class projected-space formulation, we examine how priority composition affects the traffic characteristics of individual classes. We consider the same four generation priority fractions, $\alpha_{\mathrm{pr}}^q\in\{0.2,0.4,0.6,0.8\}$, with particular attention to the distinction between moderate priority compositions and the high-priority case in which priority drones themselves experience substantial congestion. In the following figures, aggregate quantities are plotted against the total accumulation $n_{\mathrm{tot}}$, whereas class-specific quantities are plotted against their respective accumulations $n_{\mathrm{pr}}$ and $n_{\mathrm{st}}$. This distinction is important because class-specific traffic characteristics depend on the full accumulation vector $\mathbf n$, rather than solely on the accumulation of the corresponding class.

\begin{figure}[htbp]
    \centering
    \MCClassPanels%
        {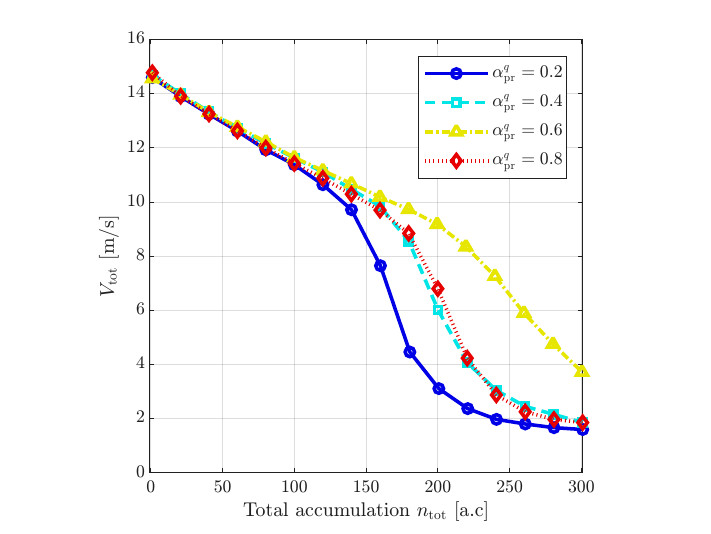}%
        {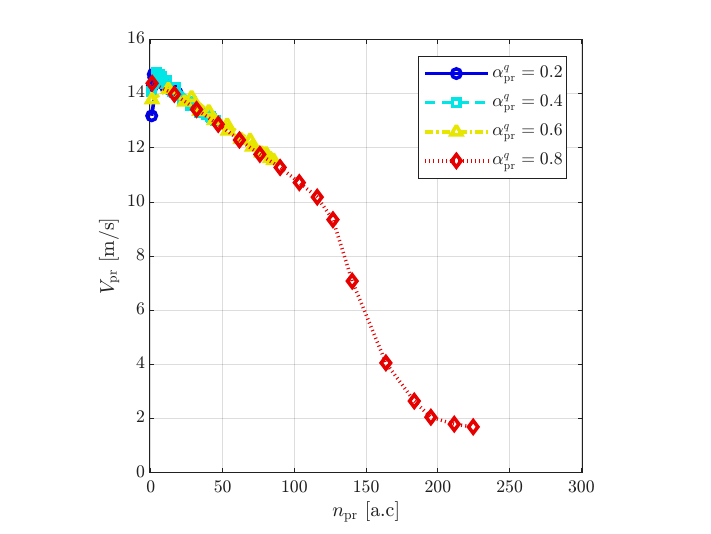}%
        {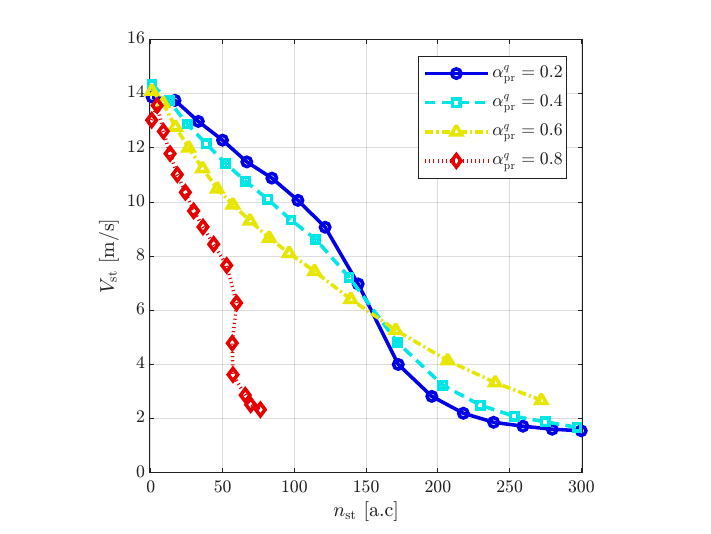}
    \caption{Space-mean speed MFDs under generation priority fractions $\alpha_{\mathrm{pr}}^q\in\{0.2,0.4,0.6,0.8\}$. (a) Aggregate speed $V_{\mathrm{tot}}$ versus total accumulation $n_{\mathrm{tot}}$. (b) Priority-class speed $V_{\mathrm{pr}}$ versus priority-class accumulation $n_{\mathrm{pr}}$. (c) Standard-class speed $V_{\mathrm{st}}$ versus standard-class accumulation $n_{\mathrm{st}}$.}
    \label{fig:multiclass_true_speeds}
\end{figure}

\autoref{fig:multiclass_true_speeds} compares the space-mean speed MFDs under the four priority compositions. At the aggregate level, increasing $\alpha_{\mathrm{pr}}^q$ from $0.2$ to $0.6$ generally delays the sharp speed reduction associated with congestion, allowing higher space-mean speeds to be maintained at larger total accumulations. However, this improvement is not monotonic, as the aggregate true-speed MFD deteriorates when the priority fraction increases further to $0.8$.

The two classes exhibit markedly different responses. The priority-class true-speed curves nearly overlap over their common accumulation ranges, indicating that priority composition has a relatively limited influence on their motion characteristics at comparable $n_{\mathrm{pr}}$. This is consistent with the collision-avoidance mechanism, under which priority drones are primarily required to avoid other priority drones and can maintain their intended motion during mixed-class encounters whenever feasible. For moderate priority fractions $\alpha_{\mathrm{pr}}^q\in\{0.2,0.4,0.6\}$, the priority-class accumulation remains relatively low, while the case $\alpha_{\mathrm{pr}}^q=0.8$ covers a much wider range and exhibits a pronounced speed reduction at higher $n_{\mathrm{pr}}$. Thus, the near-overlapping curves suggest a common underlying speed--accumulation pattern over the sampled ranges, rather than complete independence from the surrounding traffic state.

In contrast, the standard-class true-speed MFDs exhibit a more complex dependence on the generation priority fraction. For moderate priority fractions $\alpha_{\mathrm{pr}}^q\in\{0.2,0.4,0.6\}$, larger fractions generally yield lower $V_{\mathrm{st}}$ at a given $n_{\mathrm{st}}\lesssim150$, reflecting the greater avoidance burden on standard drones. Interestingly, this ordering reverses at higher accumulations, where larger priority fractions can produce higher standard-class speeds. This reversal is associated with changes in traffic composition: as shown in \autoref{fig:multiclass_fraction_verification}, $\alpha_{\mathrm{pr}}^n$ decreases with increasing total accumulation, potentially causing $n_{\mathrm{pr}}$ to decline even as $n_{\mathrm{st}}$ increases. Consequently, the same $n_{\mathrm{st}}$ may correspond to different priority-class populations and aggregate congestion conditions. The case $\alpha_{\mathrm{pr}}^q=0.8$ exhibits qualitatively different behavior, with consistently lower standard-class speeds than the other three cases and a more restricted range of $n_{\mathrm{st}}$. Unlike the moderate-priority cases, $\alpha_{\mathrm{pr}}^n$ initially decreases but rises again beyond approximately $n_{\mathrm{tot}}=200$, suggesting that priority drones themselves become increasingly affected by congestion as same-class interactions intensify.

\begin{figure}[htbp]
    \centering
    \MCClassPanels%
        {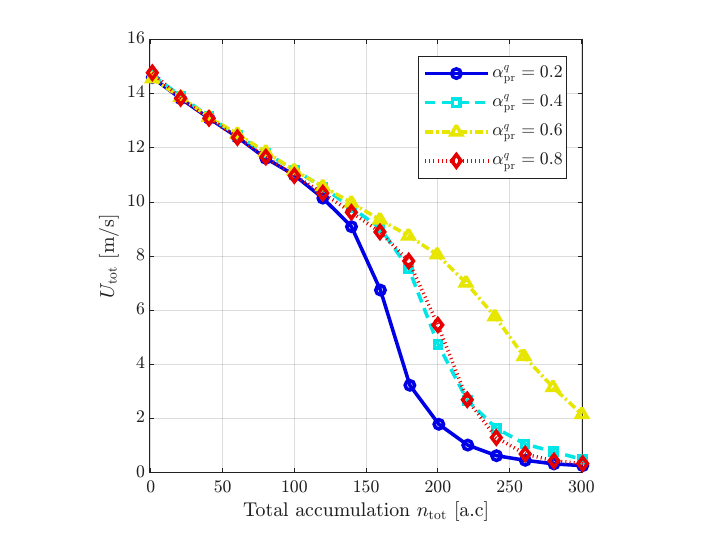}%
        {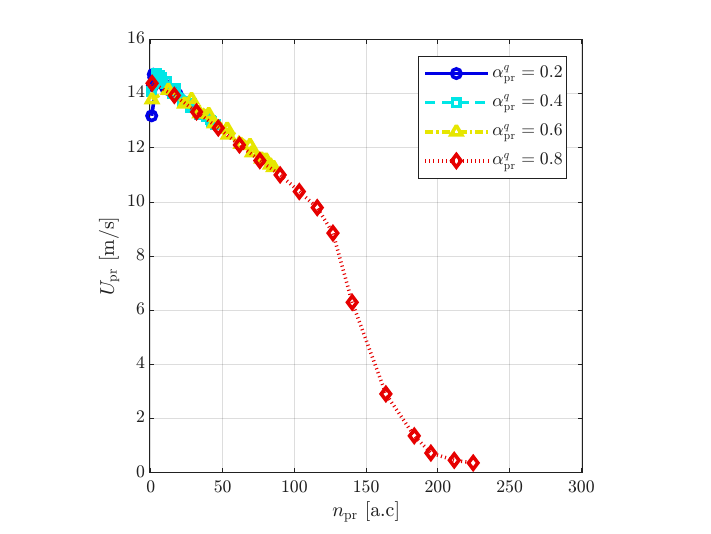}%
        {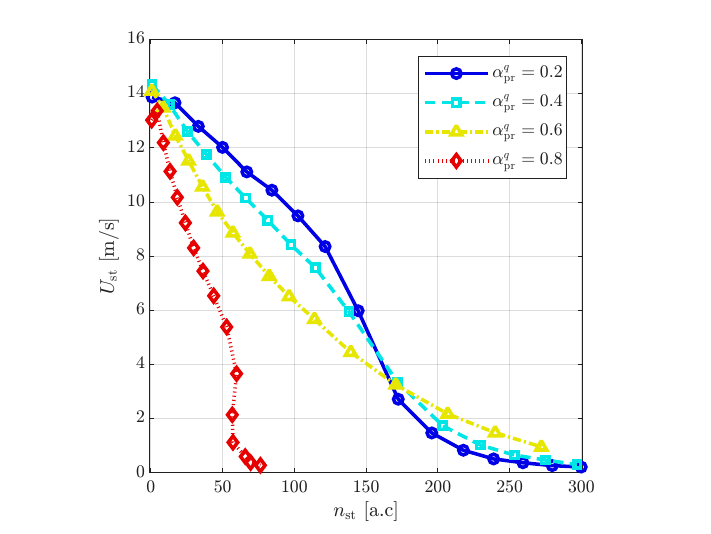}
    \caption{Space-mean projected-speed MFDs under generation priority fractions $\alpha_{\mathrm{pr}}^q\in\{0.2,0.4,0.6,0.8\}$. (a) Aggregate projected speed $U_{\mathrm{tot}}$ versus $n_{\mathrm{tot}}$. (b) Priority-class projected speed $U_{\mathrm{pr}}$ versus $n_{\mathrm{pr}}$. (c) Standard-class projected speed $U_{\mathrm{st}}$ versus $n_{\mathrm{st}}$.}
    \label{fig:multiclass_projected_speeds}
\end{figure}

\autoref{fig:multiclass_projected_speeds} presents the corresponding projected-speed MFDs, which exhibit patterns nearly identical to those of the true-speed MFDs in \autoref{fig:multiclass_true_speeds}. The main distinction emerges under severe congestion, where the projected speeds approach zero while the true speeds remain strictly positive, consistent with the findings in \autoref{subsubsec:speedMFD_singleclass}.

\begin{figure}[htbp]
    \centering
    \MCClassPanels%
        {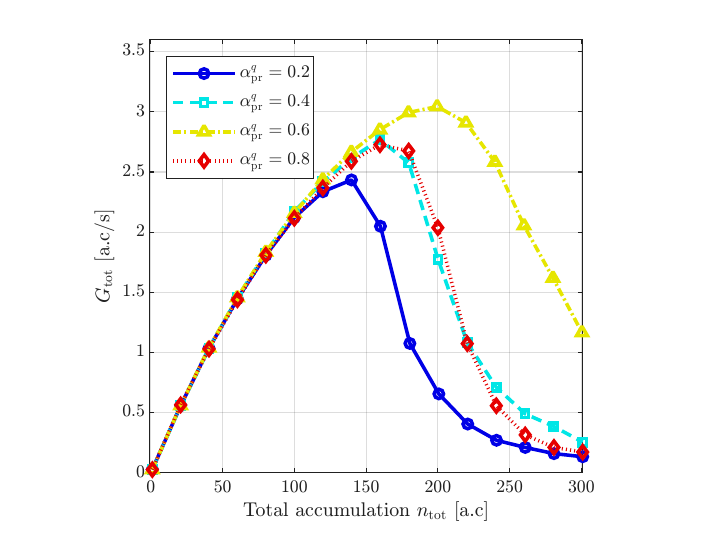}%
        {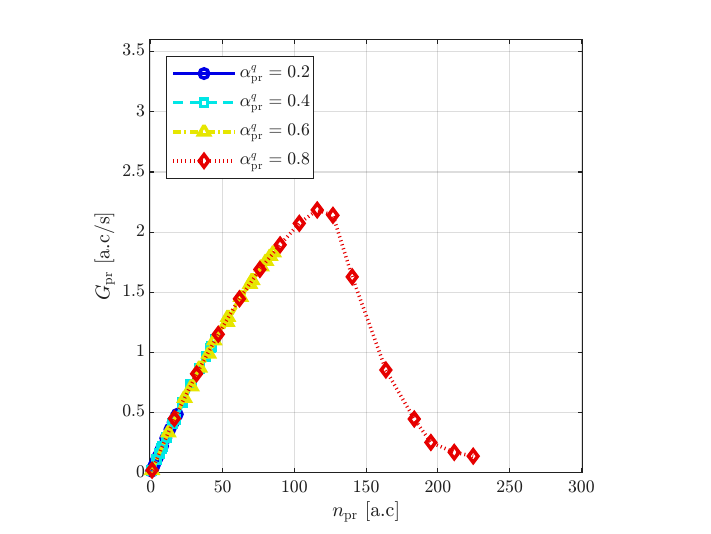}%
        {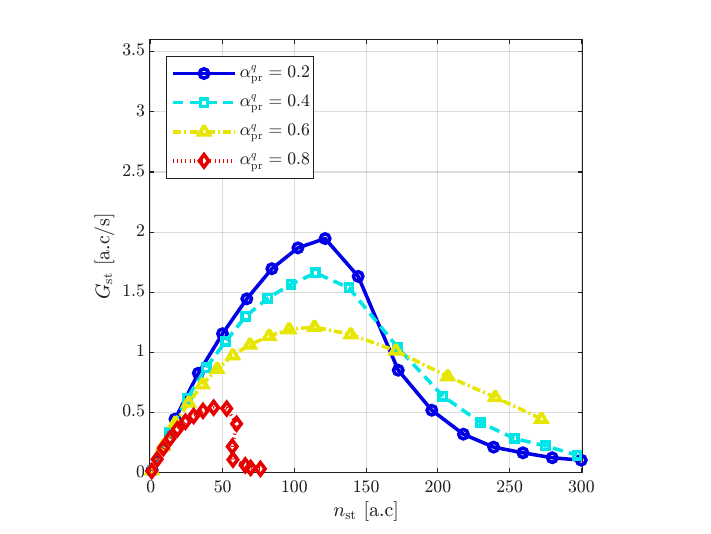}
    \caption{Aggregate and class-specific outflow MFDs under generation priority fractions $\alpha_{\mathrm{pr}}^q\in\{0.2,0.4,0.6,0.8\}$. (a) Aggregate outflow $G_{\mathrm{tot}}$ versus $n_{\mathrm{tot}}$. (b) Priority-class outflow $G_{\mathrm{pr}}$ versus $n_{\mathrm{pr}}$. (c) Standard-class outflow $G_{\mathrm{st}}$ versus $n_{\mathrm{st}}$.}
    \label{fig:multiclass_class_mfds}
\end{figure}

The corresponding outflow MFDs are shown in \autoref{fig:multiclass_class_mfds}. At the aggregate level, increasing the generation priority fraction initially improves both the maximum outflow and critical accumulation, whereas this improvement diminishes at $\alpha_{\mathrm{pr}}^q=0.8$. For the priority class, the outflow curves nearly coincide over their common accumulation ranges, exhibiting a pattern similar to the homogeneous single-class MFD. In particular, the wider accumulation range observed at $\alpha_{\mathrm{pr}}^q=0.8$ reveals a clear congested branch, which is less apparent at smaller priority fractions due to their limited priority-class accumulations. These observations suggest that the priority-class MFD is largely insensitive to the generation priority fraction under the adopted collision-avoidance mechanism, where priority drones primarily yield to drones of their own class.

The standard-class outflow MFDs exhibit a strong dependence on priority fraction, following trends similar to those observed in the true- and projected-speed MFDs. For $\alpha_{\mathrm{pr}}^q\in\{0.2,0.4,0.6\}$, larger priority fractions generally produce lower $G_{\mathrm{st}}$ at a given $n_{\mathrm{st}}$ over the low-to-moderate accumulation range, reflecting the greater avoidance burden imposed on standard drones. This ordering similarly reverses along the congested branches, where larger priority fractions can yield higher standard-class outflows. As discussed above, the reversal is associated with the declining priority-class accumulation under severe congestion, which alters the surrounding traffic conditions even at the same $n_{\mathrm{st}}$. At $\alpha_{\mathrm{pr}}^q=0.8$, standard-class outflows are substantially lower and the accumulation range is considerably restricted. Since $\alpha_{\mathrm{pr}}^n$ rises again under severe congestion rather than continuing to decrease, priority drones retain a substantial share of the active population, imposing persistent interference on standard drones and significantly reducing their MFD.

Overall, these results reveal a pronounced asymmetry in the effects of priority assignment. Under moderate priority fractions, priority drones maintain relatively consistent speed and outflow characteristics, whereas standard drones experience stronger congestion effects due to their greater avoidance responsibilities. At a sufficiently high priority fraction, however, priority drones themselves become congested, reducing the effectiveness of priority assignment. These class-specific interactions collectively influence the aggregate MFD, as illustrated by the differences in maximum outflow and critical accumulation across the four representative priority fractions. To further quantify these effects, the following subsection extends the aggregate outflow analysis to a wider range of priority compositions.


\subsubsection{Effects of priority composition on aggregate MFDs}
\label{subsubsec:mc_priority_performance}

As discussed previously, priority composition affects not only class-specific traffic characteristics but also the aggregate outflow MFD. To systematically quantify its influence on network performance, we extend the analysis from the four representative priority fractions to the full range $\alpha_{\mathrm{pr}}^q\in\{0,0.1,\ldots,1\}$. This finer parameter sweep allows us to characterize how the aggregate MFD evolves with the generation priority fraction and identify the priority composition that maximizes network throughput under the considered simulation setting.

\begin{figure}[htbp]
    \centering
    \begingroup
    \newlength{\mcgap}
    \newlength{\mctotalwidth}
    \newlength{\mcavailable}
    \newlength{\mcheight}
    \newlength{\mcwidthA}
    \newlength{\mcwidthB}

    \setlength{\mcgap}{0.015\linewidth}

    \settowidth{\mctotalwidth}{%
        \includegraphics[height=1cm]{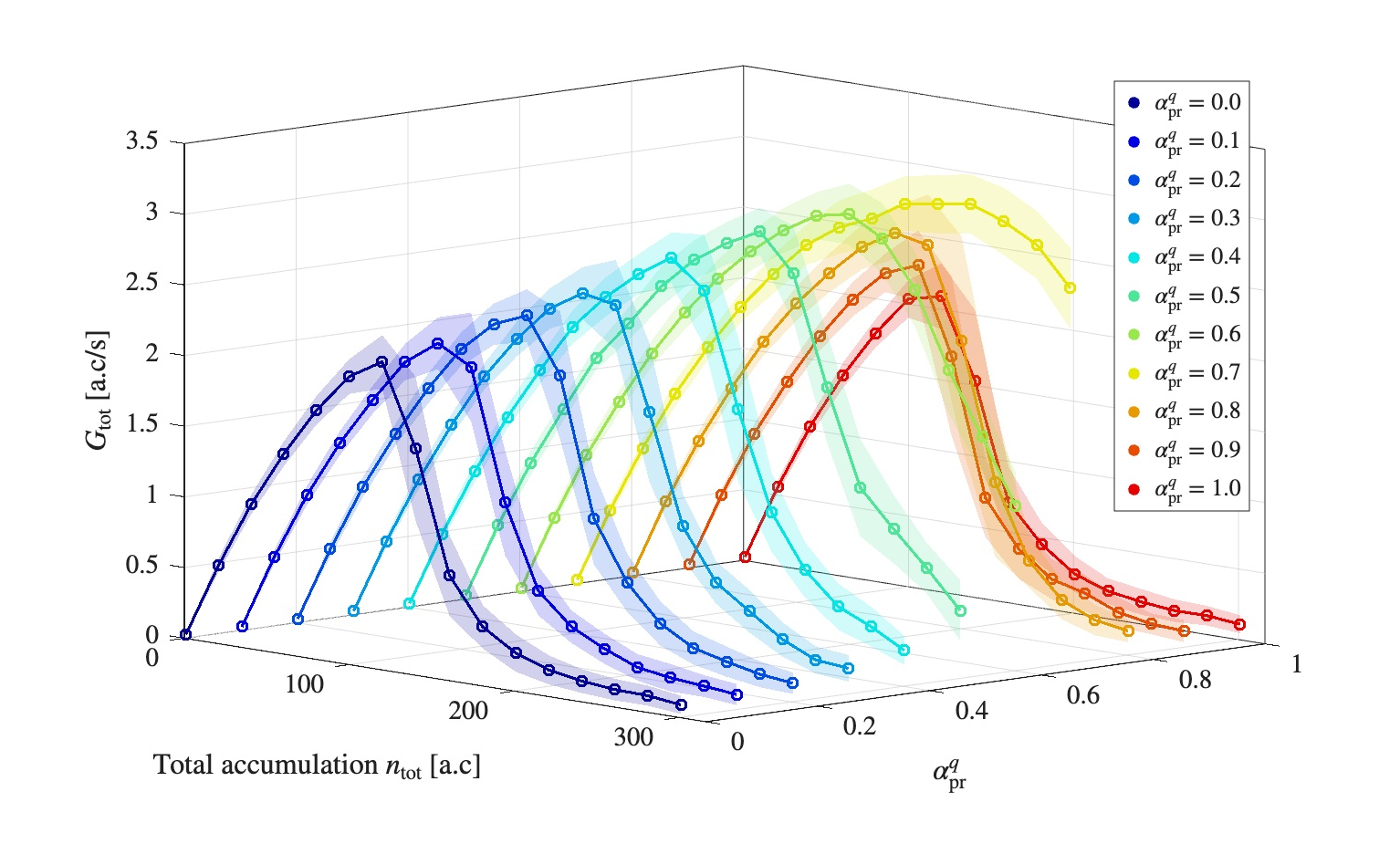}%
        \includegraphics[height=1cm]{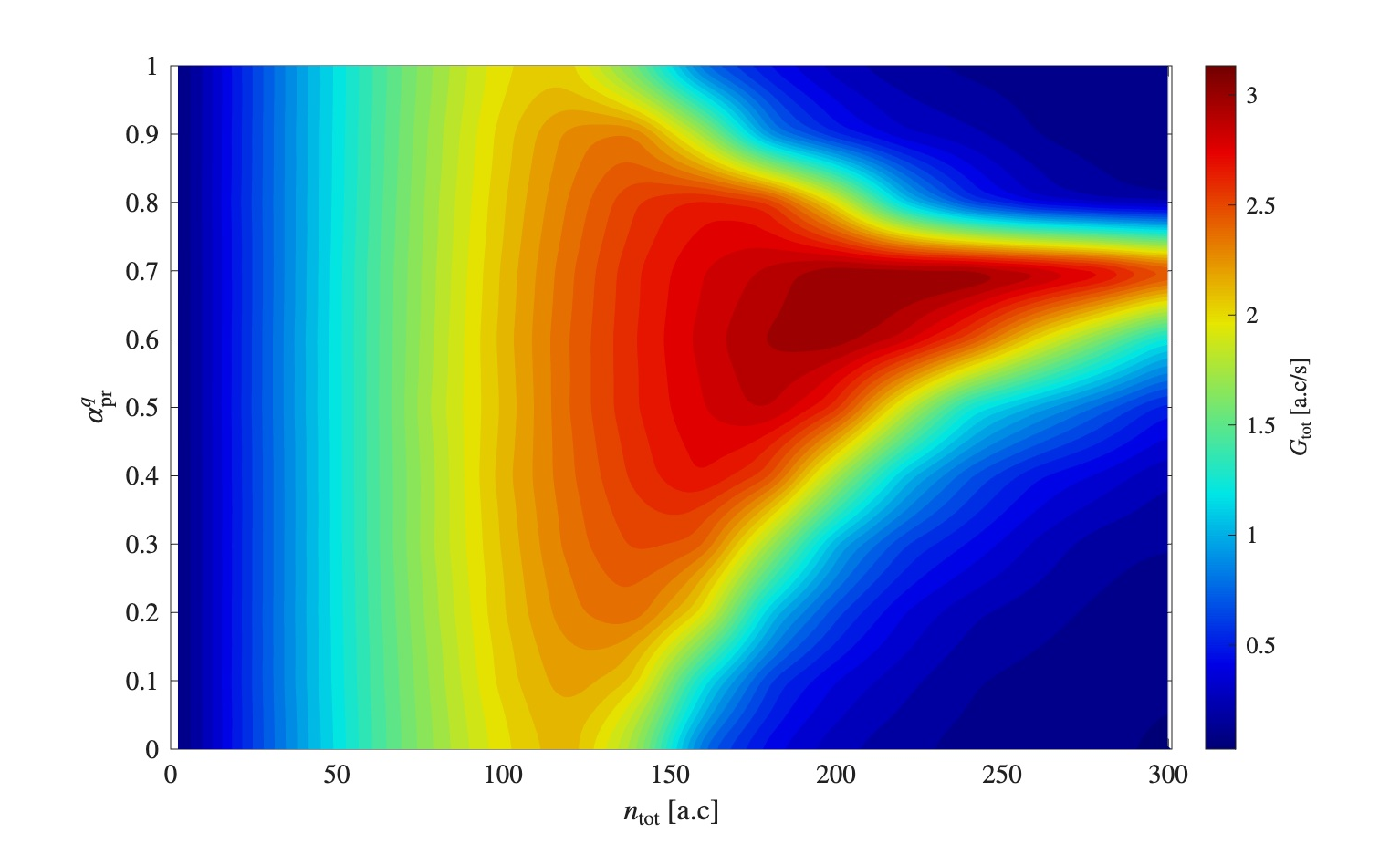}%
    }

    \setlength{\mcavailable}{0.98\linewidth-\mcgap}
    \setlength{\mcheight}{%
        1cm * \ratio{\mcavailable}{\mctotalwidth}%
    }

    \settowidth{\mcwidthA}{%
        \includegraphics[height=\mcheight]{images/multialphaMFD.pdf}%
    }
    \settowidth{\mcwidthB}{%
        \includegraphics[height=\mcheight]{images/multialphaMFD_contour.pdf}%
    }

    \makebox[\linewidth][c]{%
        \subcaptionbox{Three-dimensional aggregate MFDs.%
            \label{fig:multiclass_allalpha_3D}}[\mcwidthA]{%
            \includegraphics[height=\mcheight]{images/multialphaMFD.pdf}%
        }%
        \hspace{\mcgap}%
        \subcaptionbox{Aggregate outflow contour.%
            \label{fig:multiclass_allalpha_contour}}[\mcwidthB]{%
            \includegraphics[height=\mcheight]{images/multialphaMFD_contour.pdf}%
        }%
    }

    \caption{Aggregate multi-class MFDs under generation priority fractions $\alpha_{\mathrm{pr}}^q\in\{0,0.1,\ldots,1\}$, obtained from the mean observations of multiple independent simulation runs. (a) Three-dimensional representation of the aggregate outflow under different priority compositions. (b) Contour map of $G_{\mathrm{tot}}(n_{\mathrm{tot}},\alpha_{\mathrm{pr}}^q)$, illustrating the variation in network throughput with total accumulation and priority composition.}
    \label{fig:multiclass_allalpha_mfd}
    \endgroup
\end{figure}

\autoref{fig:multiclass_allalpha_mfd} presents the resulting aggregate MFDs as a function of total accumulation $n_{\mathrm{tot}}$ and generation priority fraction $\alpha_{\mathrm{pr}}^q$, using a three-dimensional representation and its corresponding contour plot. Consistent with the representative cases examined previously, increasing the priority fraction initially improves aggregate outflow and shifts the onset of congestion toward higher accumulations. However, this improvement is non-monotonic: beyond a certain priority fraction, both the maximum outflow and the critical accumulation decrease. The results therefore indicate that assigning priority to a larger proportion of drones does not necessarily improve overall network performance, as excessive priority assignment increases interactions among priority drones and reduces the benefits of differentiated collision avoidance.

To quantify these changes, we define the critical accumulation $n_{\mathrm{crit}}(\alpha_{\mathrm{pr}}^q)$ as the accumulation at which the aggregate outflow reaches its maximum, and the corresponding maximum outflow as $G_{\mathrm{crit}}(\alpha_{\mathrm{pr}}^q)$. Their estimated values are summarized in \autoref{tab:critical_points_priority} and illustrated in \autoref{fig:priority_critical_points}. Both quantities initially increase with the generation priority fraction, reaching their highest values among the tested compositions at $\alpha_{\mathrm{pr}}^q=0.7$, where $n_{\mathrm{crit}}=233$ and $G_{\mathrm{crit}}=3.12$ aircraft/s. Compared with the homogeneous standard-class case ($\alpha_{\mathrm{pr}}^q=0$), this represents an approximately $46\%$ increase in maximum outflow, while the critical accumulation more than doubles. Further increasing the priority fraction reduces both quantities, with the homogeneous priority-class case ($\alpha_{\mathrm{pr}}^q=1$) exhibiting performance comparable to the homogeneous standard-class case. These results suggest that the aggregate performance improvement arises primarily from the asymmetric interactions between the two classes, rather than from priority assignment alone.

The non-monotonic behavior can be interpreted in terms of the competing effects of priority assignment, as also illustrated by the projected-speed MFDs in \autoref{fig:multiclass_projected_speeds}. 
At moderate priority fractions, priority drones maintain relatively high projected speeds through asymmetric collision avoidance without having to yield to standard drones during mixed-class encounters, while standard drones accommodate the necessary avoidance maneuvers. This differentiation allows the system to sustain effective OD-directed progress at higher accumulations, thereby improving aggregate throughput. As the priority fraction increases further, however, standard drones experience increasingly restricted movement, while more frequent same-class encounters among priority drones also reduce their projected speeds. Meanwhile, the system gradually approaches the homogeneous priority-only case, in which the benefits of differentiated collision avoidance disappear. Consequently, the aggregate performance gains diminish, explaining why assigning priority to all drones does not maximize throughput.

\begin{table}[htbp]
    \centering
    \begingroup

    \newsavebox{\critTableBox}
    \begin{lrbox}{\critTableBox}
        \renewcommand{\arraystretch}{1.12}
        \setlength{\tabcolsep}{4.5pt}
        \footnotesize
        \begin{tabular}{ccc}
            \toprule
            $\alpha_{\mathrm{pr}}^q$ &
            $n_{\mathrm{crit}}$ &
            $G_{\mathrm{crit}}$ \\
            \midrule
            0.0 & 110 & 2.14 \\
            0.1 & 123 & 2.29 \\
            0.2 & 134 & 2.42 \\
            0.3 & 145 & 2.58 \\
            0.4 & 161 & 2.76 \\
            0.5 & 177 & 2.92 \\
            0.6 & 195 & 3.04 \\
            0.7 & 233 & 3.12 \\
            0.8 & 165 & 2.76 \\
            0.9 & 128 & 2.39 \\
            1.0 & 108 & 2.11 \\
            \bottomrule
        \end{tabular}
    \end{lrbox}

    \newlength{\critTableHeight}
    \setlength{\critTableHeight}{%
        \dimexpr\ht\critTableBox+\dp\critTableBox\relax
    }

    \noindent\makebox[\linewidth][c]{%

        \begin{minipage}[t]{0.40\linewidth}
            \vspace{0pt}
            \centering

            \usebox{\critTableBox}

            \par\vspace{2mm}

            \refstepcounter{table}%
            \label{tab:critical_points_priority}%
            \addcontentsline{lot}{table}{%
                \protect\numberline{\thetable}%
                Estimated critical accumulation and maximum outflow%
            }%

            \noindent
            \parbox[t]{\linewidth}{%
                \raggedright\footnotesize
                \textbf{\tablename~\thetable.}\enspace
                Estimated values of critical accumulation and maximum outflow for different generation priority fractions.
                \par
            }

        \end{minipage}%
        \hspace{0.02\linewidth}%
        \begin{minipage}[t]{0.56\linewidth}
            \vspace{0pt}
            \centering

            \includegraphics[
                height=\critTableHeight,
                width=\linewidth,
                keepaspectratio
            ]{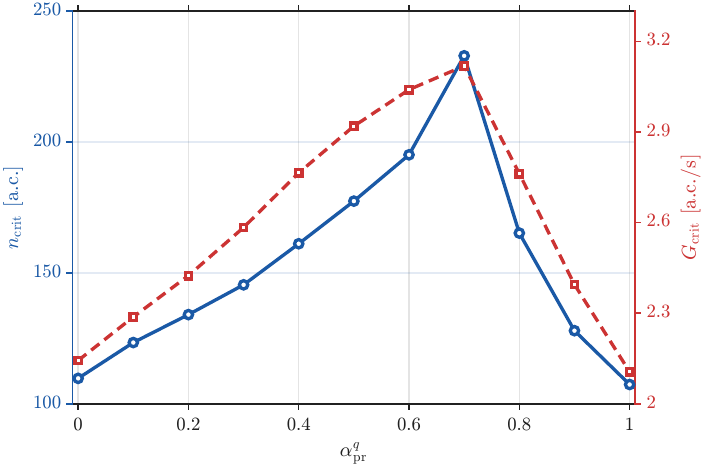}

            \par\vspace{2mm}

            \refstepcounter{figure}%
            \label{fig:priority_critical_points}%
            \addcontentsline{lof}{figure}{%
                \protect\numberline{\thefigure}%
                Variation of critical accumulation and maximum outflow%
            }%

            \noindent
            \parbox[t]{\linewidth}{%
                \raggedright\footnotesize
                \textbf{\figurename~\thefigure.}\enspace
                Variation of critical accumulation and outflow with the generation priority fraction.
                \par
            }
        \end{minipage}%
    }

    \endgroup
\end{table}

These findings suggest that a favorable generation priority fraction lies near $\alpha_{\mathrm{pr}}^q=0.7$ in the present scenario, although a finer parameter sweep would be necessary to identify the throughput-maximizing fraction more precisely. Moreover, such an optimum is scenario-dependent and may vary with network geometry, demand patterns, and collision-avoidance mechanisms. Nevertheless, the ability to modify both the maximum outflow and the critical accumulation offers an additional degree of freedom for macroscopic traffic control. Under a quasi-stationary aggregate approximation, the accumulation dynamics can be expressed as $\dot n_{\mathrm{tot}}=q_{\mathrm{tot}}-G_{\mathrm{tot}}(n_{\mathrm{tot}},\alpha_{\mathrm{pr}}^q)$. For fixed inflow and priority composition, equilibria on the increasing branch of the MFD are locally asymptotically stable, whereas those on the decreasing branch are unstable. By adjusting $\alpha_{\mathrm{pr}}^q$, it may therefore be possible to increase the outflow sufficiently to reduce an excessive accumulation and restore operation to a locally stable region, without necessarily restricting total trip generation. This potential motivates future investigation of priority allocation as an alternative or complementary control input to conventional departure regulation.

\section{Conclusion}
\label{sec:concl}

This paper develops a projected-space macroscopic modeling framework for Low-Altitude Air Transport (LAAT) systems to address the challenges posed by traffic-dependent trajectory deviations and heterogeneous drone interactions. The framework projects each drone's instantaneous velocity onto the fixed direction of its shortest origin--destination (OD) path, defining a projected speed that measures effective progression along this direction. In this representation, trip progression is characterized by the shortest OD distance, which is known at departure and remains independent of subsequent collision-avoidance maneuvers. The effects of these maneuvers are captured through projected speed, without requiring knowledge of the realized trip distance, which becomes available only upon trip completion. This provides an observable and tractable representation of trip progression for macroscopic modeling, including heterogeneous multi-class systems.

Building on this representation, we develop and analyze the corresponding macroscopic formulations under stationary, transient, and heterogeneous traffic conditions. First, we derive the projected accumulation-based (AB) outflow relation and demonstrate its equivalence to the conventional AB formulation under stationary conditions. Second, we formulate a projected trip-based (TB) model and derive finite-window outflow estimation error bounds for both conventional and projected TB formulations. Third, we extend the AB relation to multi-class systems, showing that aggregate outflow can be expressed using a common expected shortest OD distance, even when different classes exhibit distinct speeds and realized trip lengths. This avoids the need for class-specific realized travel distances in conventional multi-class formulations and provides a more tractable representation of heterogeneous LAAT traffic.

Numerical experiments using a microscopic LAAT simulation framework support the theoretical findings. Under approximately stationary conditions, the projected-space AB estimator closely reproduces the observed MFD across free-flow and congested regimes, outperforming the conventional estimator in finite-window outflow estimation, particularly under severe congestion. Under transient traffic conditions, the TB formulation provides a means of capturing time-dependent trip progression beyond the stationary assumptions of the AB models. The multi-class experiments further verify the projected-space aggregation relations and reveal how priority-dependent collision avoidance alters class-specific and aggregate macroscopic traffic characteristics.

Beyond model verification, the priority-based experiments demonstrate that traffic composition can substantially influence network performance. Appropriate mixtures of prioritized and standard drones can increase both the maximum network outflow and the critical accumulation compared with homogeneous traffic. In particular, the maximum outflow increases by approximately 46\% relative to the no-priority case, while the critical accumulation more than doubles under the investigated configurations. However, these benefits are not monotonic in the priority fraction, as excessive prioritization increases interactions among prioritized drones and reduces the advantages of differentiated traffic operations. These findings highlight the potential of priority allocation as a macroscopic traffic-management mechanism rather than solely an operational conflict-resolution rule.

Future research will examine the robustness of the proposed framework under different operational parameters and traffic conditions, as well as alternative collision-avoidance methods beyond the force-field controller considered in this study, such as velocity obstacles and integrated route guidance. Another direction is to explore its applicability to other multidirectional transportation systems, including pedestrian traffic, autonomous mobile robots, and lane-free road traffic, potentially through real-world experiments. More importantly, the identified priority-dependent MFD relationships provide a foundation for dynamic traffic management, including departure management in single-region airspace and perimeter control in multi-regional LAAT networks. Overall, the proposed projected-space formulations retain the essential structures of conventional AB and TB models while providing an observable and analytically tractable representation of trip progression, with particular advantages for multi-class modeling and future priority-based control.

\begin{appendices}

\section{Proofs of the accumulation-based relations}
\label{app:proof_AB}

\subsection{Proof of the conventional AB relation and distributional illustration}
\label{app:proof_AB_heterogeneous}

This appendix proves \autoref{prop:steady_AB_heterogeneous} by establishing the relationship between the generation and snapshot distributions under stationary operation. The proof accommodates heterogeneous trip lengths and time-varying within-trip speeds without requiring independence between trip characteristics. A numerical illustration is then provided to demonstrate how trip-duration weighting affects the representative quantities and how using expectations from inappropriate distributions can bias the estimated outflow.

\begin{proof}
Consider a stationary system with accumulation $n$ and constant trip generation rate $Q$. By flow conservation, the stationary generation and completion rates coincide:
\begin{equation}
\label{eq:Q=Gn}
    Q=G(n).
\end{equation}
Let $l$ and $T$ denote the realized length and travel time of a generated trip, respectively, with joint generation distribution $f_{l,T}(l,T\mid n)$. We assume that all generated trips eventually complete, with finite positive mean travel time and finite positive mean realized trip length. The expected travel time under the generation distribution is defined as
\begin{equation}
    \mathbb E_f[T\mid n]
    =
    \iint T\,f_{l,T}(l,T\mid n)\,dl\,dT.
\end{equation}
By Little's law, the stationary accumulation satisfies
\begin{equation}
\label{eq:n=QEfT}
    n
    =
    Q\,\mathbb E_f[T\mid n].
\end{equation}
Further, let $g_{l,T}(l,T\mid n)$ denote the joint snapshot distribution of realized trip length and travel time conditioned on the accumulation. Since a trip remains active for its duration $T$, its representation in the snapshot population is proportional to that duration. 
Following the stationary distributional argument of \citet{sirmatel2021modeling}, let $\mathcal B\subseteq\mathbb R_{++}^2$ be an arbitrary measurable set of trip types, each characterized by its realized trip length $l$ and travel time $T$. Under stationarity, we have
\begin{equation}
\label{eq:relation_f_g}
    n
    \iint_{(l,T)\in\mathcal B}
    g_{l,T}(l,T\mid n)\,dl\,dT
    =
    Q
    \iint_{(l,T)\in\mathcal B}
    T\,f_{l,T}(l,T\mid n)\,dl\,dT.
\end{equation}
For example, $\mathcal B$ may contain all trips with $l\in[l_1,l_2]$ and $T\in[T_1,T_2]$. Eq.~\eqref{eq:relation_f_g} states that, for any such subset, the expected number of active drones equals the generation rate of trips in that subset multiplied by their conditional mean travel time. Since this equality holds for every measurable set $\mathcal B$, the corresponding joint probability densities must satisfy, almost everywhere,
\begin{equation}
\label{eq:relation_f_g2}
    g_{l,T}(l,T\mid n)
    =
    \frac{QT}{n}
    f_{l,T}(l,T\mid n)
    =
    \frac{T}{\mathbb E_f[T\mid n]}
    f_{l,T}(l,T\mid n).
\end{equation}
Eq.~\eqref{eq:relation_f_g2} establishes that the snapshot distribution is obtained by weighting the generation distribution by trip duration. Consequently, trips with longer travel times are overrepresented in the active population, regardless of their realized lengths or speed profiles.

We next derive the representative snapshot speed. For any completed trip $i$, the distance identity in Eq.~\eqref{eq:real_length} implies that
\begin{equation}
    \frac{1}{T_i}
    \int_{t_{0,i}}^{t_{0,i}+T_i}
    v_i(\tau)\,d\tau
    =
    \frac{l_i}{T_i}.
\end{equation}
Accordingly, for trips characterized by realized length $l$ and duration $T$, the mean instantaneous speed under uniform sampling is $v=l/T$, regardless of within-trip speed variations. Taking the expectation over the snapshot distribution in Eq.~\eqref{eq:relation_f_g2} yields

\begin{equation}
\label{eq:relationEgvEfv}
\begin{aligned}
    \mathbb E_g[v\mid n]
    &=
    \iint
    \frac{l}{T}
    g_{l,T}(l,T\mid n)\,dl\,dT \\
    &=
    \iint
    \frac{l}{T}
    \frac{QT}{n}
    f_{l,T}(l,T\mid n)\,dl\,dT \\
    &=
    \frac{Q}{n}
    \mathbb E_f[l].
\end{aligned}
\end{equation}
Rearranging Eq.~\eqref{eq:relationEgvEfv} and recalling $Q=G(n)$ gives
\begin{equation}
    G(n)
    =
    n
    \frac{\mathbb E_g[v\mid n]}
    {\mathbb E_f[l]},
\end{equation}
which proves \autoref{prop:steady_AB_heterogeneous}.
\end{proof}

To illustrate the distinction between the generation and snapshot distributions, we consider a stationary system with accumulation $n=100$ and omit the conditioning on $n$ for notational simplicity. Suppose that the generated trip duration $T$ and an auxiliary distance variable $z$ are independent Gamma random variables:
\begin{equation}
    T\sim\mathrm{Gamma}(15,2),
    \qquad
    z\sim\mathrm{Gamma}(10,20),
\end{equation}
where $\mathrm{Gamma}(k,\vartheta)$ denotes the Gamma distribution with shape parameter $k$ and scale parameter $\vartheta$. The scale parameters are expressed in seconds and meters, respectively. We define the realized trip length as
\begin{equation}
    l=10T+z,
\end{equation}
where the coefficient $10$ has units of $\mathrm{m/s}$. Consequently, the generated travel time and trip length are positively correlated, with marginal distributions
\begin{equation}
    f_T(T)=\mathrm{Gamma}(15,2),\qquad f_l(l)=\mathrm{Gamma}(25,20).    
\end{equation}
Their means are then
\begin{equation}
            \mathbb E_f[T]=30~\mathrm{s}, \qquad \mathbb E_f[l]=500~\mathrm{m},
    \qquad
\end{equation}
Furthermore, according to Eq.~\eqref{eq:relation_f_g2}, the joint snapshot distribution of $(l,T)$ satisfies
\begin{equation}
    g_{l,T}(l,T)
    =
    \frac{T}{\mathbb E_f[T]}f_{l,T}(l,T).
\end{equation}
Since $T$ and $z$ are independent at generation, this weighting modifies the distribution of $T$ while leaving that of $z$ unchanged. In particular, the snapshot travel-time distribution becomes
\begin{equation}
    g_T(T)
    =
    \frac{T}{\mathbb E_f[T]}f_T(T)
    =
    \mathrm{Gamma}(16,2).
\end{equation}
Since $l=10T+z$, the corresponding snapshot distribution of realized trip length is
\begin{equation}
    g_l(l)
    =
    \mathrm{Gamma}(26,20).
\end{equation}
The representative snapshot quantities therefore satisfy
\begin{equation}
    \mathbb E_g[T]
    =
    16\times2
    =
    32~\mathrm{s},
    \qquad
    \mathbb E_g[l]
    =
    26\times20
    =
    520~\mathrm{m}.
\end{equation}

This illustration demonstrates how duration-weighted sampling changes the distributions of trip characteristics in the active population. Trips with longer travel times are overrepresented in the snapshot distribution, resulting in $\mathbb E_g[T]>\mathbb E_f[T]$. Since realized trip length and travel time are positively correlated in this example, the snapshot mean trip length also exceeds its generation mean, i.e., $\mathbb E_g[l]=520~\mathrm{m}>\mathbb E_f[l]=500~\mathrm{m}$. This highlights the importance of distinguishing the two distributions when evaluating the AB formulation.

\subsection{Proof of \autoref{prop:steady_projected_AB}}
\label{app:proof_projectedoutflow}

The proof follows similarly to that of \autoref{prop:steady_AB_heterogeneous}, with realized trip length and real speed replaced by shortest OD distance and projected speed, respectively.

\begin{proof}
Consider a stationary system with accumulation $n$ and trip generation rate $Q=G(n)$. By Little's law,
\begin{equation}
\label{eq:projected_little}
    n=Q\,\mathbb E_f[T\mid n].
\end{equation}
Let $f_{s,T}(s,T\mid n)$ and $g_{s,T}(s,T\mid n)$ denote the joint generation and snapshot distributions of shortest OD distance $s$ and travel time $T$, respectively. Following the duration-weighting relationship established in Eq.~\eqref{eq:relation_f_g2}, we have
\begin{equation}
\label{eq:projected_duration_bias}
    g_{s,T}(s,T\mid n)
    =
    \frac{T}{\mathbb E_f[T\mid n]}
    f_{s,T}(s,T\mid n).
\end{equation}
For each completed trip $i$, the projected-distance identity in Eq.~\eqref{eq:projected_preservation} gives
\begin{equation}
    \frac{1}{T_i}
    \int_{t_{0,i}}^{t_{0,i}+T_i}
    u_i(\tau)\,d\tau
    =
    \frac{s_i}{T_i}.
\end{equation}
Taking the expectation of the mean instantaneous projected speed $u=s/T$ over the snapshot distribution yields
\begin{equation}
\label{eq:Eg_projected_speed}
\begin{aligned}
    \mathbb E_g[u\mid n]
    &=
    \iint
    \frac{s}{T}
    g_{s,T}(s,T\mid n)\,ds\,dT \\
    &=
    \frac{1}{\mathbb E_f[T\mid n]}
    \iint
    s\,f_{s,T}(s,T\mid n)\,ds\,dT \\
    &=
    \frac{\mathbb E_f[s]}
    {\mathbb E_f[T\mid n]}.
\end{aligned}
\end{equation}
Combining Eq.~\eqref{eq:projected_little} and Eq.~\eqref{eq:Eg_projected_speed}, and recalling that $Q=G(n)$, gives
\begin{equation}
    G(n)
    =
    n
    \frac{\mathbb E_g[u\mid n]}
    {\mathbb E_f[s]}.
\end{equation}
Finally, under the state-invariant generated OD distribution, $\mathbb E_f[s]=S$. Therefore,
\begin{equation}
    G(n)
    =
    n
    \frac{U(n)}{S},
\end{equation}
which proves \autoref{prop:steady_projected_AB}.
\end{proof}

\section{Proofs of trip-based relations}
\label{app:proof_TB}

\subsection{Proof of \autoref{thm:TB_outflow_error_bound}}
\label{app:proof_TB_outflow_error_bound}

\begin{proof}
Let $c_i=t_{0,i}+T_i$ and $\hat c_i^\mathrm{conv}=t_{0,i}+\hat T_i^\mathrm{conv}$ denote the true and estimated completion times of trip $i$, respectively. By the true-motion distance identity and the estimated completion condition,
\begin{equation}
\label{eq:compcond}
    l_i
    =
    \int_{t_{0,i}}^{c_i}v_i(\tau)\,d\tau
    =
    \int_{t_{0,i}}^{\hat c_i^\mathrm{conv}}\hat V(\tau)\,d\tau.
\end{equation}
Define the representative-speed-induced distance discrepancy at the true completion time as
\begin{equation}
\label{eq:epsilon_l_i}
    \Delta l_i
    =
    \int_{t_{0,i}}^{c_i}
    \left[\hat V(\tau)-v_i(\tau)\right]d\tau.
\end{equation}
By the bounded speed discrepancy and travel time,
\begin{equation}
\label{eq:epsilon_l_bound}
    |\Delta l_i|
    \leq
    \varepsilon_V T_i
    \leq
    \varepsilon_V T_{\mathrm{ub}}.
\end{equation}
On the other hand, Eq.~\eqref{eq:compcond} implies
\begin{equation}
\label{eq:distance_time_error_identity}
    \Delta l_i
    =
    \int_{\hat c_i^\mathrm{conv}}^{c_i}
    \hat V(\tau)\,d\tau.
\end{equation}
Since $\hat V(\tau)\geq V_{\mathrm{lb}}>0$, we obtain
\begin{equation}
    |\Delta l_i|
    \geq
    V_{\mathrm{lb}}
    |c_i-\hat c_i^\mathrm{conv}|.
\end{equation}
Combining these inequalities yields the completion-time error bound
\begin{equation}
\label{eq:completion_time_error_bound}
    |c_i-\hat c_i^\mathrm{conv}|
    \leq
    \frac{\varepsilon_V T_{\mathrm{ub}}}
    {V_{\mathrm{lb}}}
    =
    \delta_r.
\end{equation}
Furthermore, define the indicator function
\begin{equation}
    \textbf{1}_{\mathcal S}(x)
    =
    \begin{cases}
        1, & x\in\mathcal S,\\
        0, & x\notin\mathcal S.
    \end{cases}
\end{equation}
Let
\begin{equation}
    N_C(t,W)
    =
    |\mathcal C(t,W)|,
    \qquad
    \hat N_C^{\mathrm{conv}}(t,W)
    =
    |\widehat{\mathcal C}_{\mathrm{TB}}^{\mathrm{conv}}(t,W)|
\end{equation}
denote the actual and estimated completion counts over $(t-W,t]$, respectively. Taking time zero as the beginning of the observation horizon, for any time boundary $\xi>0$, define the signed discrepancy between cumulative estimated and actual completion counts over $(0,\xi]$ as
\begin{equation}
\label{eq:Delta_NC}
\begin{aligned}
    \Delta N_C(\xi)
    &=
    \hat N_C^{\mathrm{conv}}(\xi,\xi)
    -
    N_C(\xi,\xi)\\
    &=
    \sum_i
    \left[
        \textbf{1}_{(0,\xi]}(\hat c_i^{\mathrm{conv}})
        -
        \textbf{1}_{(0,\xi]}(c_i)
    \right]\\
    &=
    \underbrace{
        \sum_i
        \textbf{1}_{(0,\xi]}(\hat c_i^{\mathrm{conv}})
        \textbf{1}_{(\xi,\infty)}(c_i)
    }_{\text{Trips with } \hat c_i^{\mathrm{conv}}\leq\xi<c_i}
    -
    \underbrace{
        \sum_i
        \textbf{1}_{(0,\xi]}(c_i)
        \textbf{1}_{(\xi,\infty)}(\hat c_i^{\mathrm{conv}})
    }_{\text{Trips with } c_i\leq\xi<\hat c_i^{\mathrm{conv}}}.
\end{aligned}
\end{equation}
By the completion-time error bound in Eq.~\eqref{eq:completion_time_error_bound}, for $\delta_r>0$, we have
\begin{equation}
\begin{aligned}
    0
    &\leq
    \textbf{1}_{(0,\xi]}(\hat c_i^{\mathrm{conv}})
    \textbf{1}_{(\xi,\infty)}(c_i)
    \leq
    \textbf{1}_{(\xi,\xi+\delta_r]}(c_i),\\
    0
    &\leq
    \textbf{1}_{(0,\xi]}(c_i)
    \textbf{1}_{(\xi,\infty)}(\hat c_i^{\mathrm{conv}})
    \leq
    \textbf{1}_{(\xi-\delta_r,\xi]}(c_i).
\end{aligned}
\end{equation}
Summing over all trips yields
\begin{equation}
\label{eq:boundary_count_bounds}
\begin{aligned}
    0
    \leq
    \sum_i
    \textbf{1}_{(0,\xi]}(\hat c_i^{\mathrm{conv}})
    \textbf{1}_{(\xi,\infty)}(c_i)\leq
    \sum_i
    \textbf{1}_{(\xi,\xi+\delta_r]}(c_i)
    =
    N_C(\xi+\delta_r,\delta_r),
\end{aligned}
\end{equation}
and similarly,
\begin{equation}
\begin{aligned}
    0
    \leq
    \sum_i
    \textbf{1}_{(0,\xi]}(c_i)
    \textbf{1}_{(\xi,\infty)}(\hat c_i^{\mathrm{conv}})\leq
    \sum_i
    \textbf{1}_{(\xi-\delta_r,\xi]}(c_i)
    =
    N_C(\xi,\delta_r).
\end{aligned}
\end{equation}
Since the two terms in Eq.~\eqref{eq:Delta_NC} are nonnegative and enter with opposite signs, their absolute difference is bounded by the larger of the two. Therefore,
\begin{equation}
\label{eq:boundary_completion_discrepancy}
\begin{aligned}
    |\Delta N_C(\xi)|
    &\leq
    \max\left\{
        N_C(\xi+\delta_r,\delta_r),
        N_C(\xi,\delta_r)
    \right\}\\
    &\leq
    G_{\max}\delta_r+B_{\max},
\end{aligned}
\end{equation}
where the last inequality follows from the completion-count envelope in Eq.~\eqref{eq:TB_completion_envelope}. In addition, with the convention $N_C(0,0)=\hat N_C^{\mathrm{conv}}(0,0)=0$, the completion count over $(t-W,t]$ can be expressed as the difference between the cumulative counts at the two window boundaries:
\begin{equation}
\begin{aligned}
    N_C(t,W)
    &=
    N_C(t,t)-N_C(t-W,t-W),\\
    \hat N_C^{\mathrm{conv}}(t,W)
    &=
    \hat N_C^{\mathrm{conv}}(t,t)
    -
    \hat N_C^{\mathrm{conv}}(t-W,t-W).
\end{aligned}
\end{equation}
Therefore, recalling the definition of $\Delta N_C(\xi)$ in Eq.~\eqref{eq:Delta_NC}, we obtain
\begin{equation}
\label{eq:window_completion_discrepancy}
\begin{aligned}
    &\left|
        \hat N_C^{\mathrm{conv}}(t,W)
        -
        N_C(t,W)
    \right|\\
    &=
    \left|
        \left[
            \hat N_C^{\mathrm{conv}}(t,t)-N_C(t,t)
        \right]
        -
        \left[
            \hat N_C^{\mathrm{conv}}(t-W,t-W)
            -
            N_C(t-W,t-W)
        \right]
    \right|\\
    &=
    |\Delta N_C(t)-\Delta N_C(t-W)|\\
    &\leq
    |\Delta N_C(t)|+|\Delta N_C(t-W)|\\
    &\leq
    2(G_{\max}\delta_r+B_{\max}),
\end{aligned}
\end{equation}
where the final inequality follows from Eq.~\eqref{eq:boundary_completion_discrepancy}. 
Dividing by the window length $W$ eventually proves
\begin{equation}
    \left|
        \hat G_{\mathrm{TB}}^{\mathrm{conv}}(t,W)
        -
        G(t,W)
    \right|
    \leq
    \frac{2(\delta_rG_{\max}+B_{\max})}{W}.
\end{equation}
In the special case $\delta_r=0$, the estimated completion times coincide with the true completion times, and the outflow-estimation error vanishes.
\end{proof}

\subsection{Proof of \autoref{thm:TB_outflow_error_bound_proj}}
\label{app:proof_TB_outflow_error_bound_proj}

The proof follows the same argument as the proof of \autoref{thm:TB_outflow_error_bound} in \autoref{app:proof_TB_outflow_error_bound}, with real distance and speed replaced by shortest OD distance and projected speed, respectively.

\begin{proof}
Let $c_i=t_{0,i}+T_i$ and $\hat c_i^{\mathrm{proj}}=t_{0,i}+\hat T_i^{\mathrm{proj}}$ denote the true and projected-space estimated completion times of trip $i$, respectively. By the projected-distance identity and the estimated first-passage condition,
\begin{equation}
\label{eq:compcond_p}
    s_i
    =
    \int_{t_{0,i}}^{c_i}
    u_i(\tau)\,d\tau
    =
    \int_{t_{0,i}}^{\hat c_i^{\mathrm{proj}}}
    \hat U(\tau)\,d\tau.
\end{equation}
Define the representative projected-speed-induced progress discrepancy at the true completion time as
\begin{equation}
\label{eq:delta_s_i}
    \Delta s_i
    =
    \int_{t_{0,i}}^{c_i}
    \left[
        \hat U(\tau)-u_i(\tau)
    \right]d\tau.
\end{equation}
Since $u_i(\tau)=v_i(\tau)\cos\theta_i(\tau)$ and $|\cos\theta_i(\tau)|\leq1$, the speed bounds in Eq.~\eqref{eq:projected_speed_bounds_p} imply
\begin{equation}
    -V_{\max}
    \leq
    u_i(\tau)
    \leq
    V_{\max},
    \qquad
    0<
    U_{\mathrm{lb}}
    \leq
    \hat U(\tau)
    \leq
    V_{\max}.
\end{equation}
Consequently, the projected-speed discrepancy satisfies
\begin{equation}
    \varepsilon_U
    =
    \sup_{\tau:\mathcal A(\tau)\neq\varnothing}
    \max_{i\in\mathcal A(\tau)}
    |\hat U(\tau)-u_i(\tau)|
    \leq
    2V_{\max}.
\end{equation}
It follows that
\begin{equation}
\label{eq:delta_s_i_upper_bound}
    |\Delta s_i|
    \leq
    \varepsilon_U T_i
    \leq
    \varepsilon_U T_{\mathrm{ub}}.
\end{equation}
On the other hand, Eq.~\eqref{eq:compcond_p} implies
\begin{equation}
\label{eq:projected_distance_time_error_identity}
    \Delta s_i
    =
    \int_{\hat c_i^{\mathrm{proj}}}^{c_i}
    \hat U(\tau)\,d\tau.
\end{equation}
Since $\hat U(\tau)\geq U_{\mathrm{lb}}>0$, we obtain
\begin{equation}
\label{eq:delta_s_i_lower_bound}
    |\Delta s_i|
    \geq
    U_{\mathrm{lb}}
    |c_i-\hat c_i^{\mathrm{proj}}|.
\end{equation}
Combining Eqs.~\eqref{eq:delta_s_i_upper_bound} and~\eqref{eq:delta_s_i_lower_bound} yields the uniform completion-time error bound
\begin{equation}
\label{eq:completion_time_error_bound_p}
\begin{aligned}
    |c_i-\hat c_i^{\mathrm{proj}}|
    &\leq
    \delta_p
    =
    \frac{\varepsilon_U T_{\mathrm{ub}}}
    {U_{\mathrm{lb}}}.
\end{aligned}
\end{equation}
Next, let
\begin{equation}
    \hat N_C^{\mathrm{proj}}(t,W)
    =
    \left|
        \widehat{\mathcal C}_{\mathrm{TB}}^{\mathrm{proj}}(t,W)
    \right|
\end{equation}
denote the estimated completion count over $(t-W,t]$. Taking time zero as the beginning of the observation horizon, define the signed cumulative completion-count discrepancy at any boundary $\xi\geq0$ as
\begin{equation}
\label{eq:Delta_NC_p}
    \Delta N_C^{\mathrm{proj}}(\xi)
    =
    \hat N_C^{\mathrm{proj}}(\xi,\xi)
    -
    N_C(\xi,\xi),
\end{equation}
with $\Delta N_C^{\mathrm{proj}}(0)=0$. Following the indicator-function argument in \autoref{app:proof_TB_outflow_error_bound}, the completion-time bound in Eq.~\eqref{eq:completion_time_error_bound_p} implies, for $\delta_p>0$,
\begin{equation}
\label{eq:boundary_completion_discrepancy_p}
\begin{aligned}
    |\Delta N_C^{\mathrm{proj}}(\xi)|
    &\leq
    \max\left\{
        N_C(\xi+\delta_p,\delta_p),
        N_C(\xi,\delta_p)
    \right\}\\
    &\leq
    G_{\max}\delta_p+B_{\max},
\end{aligned}
\end{equation}
where the last inequality follows from the completion-count envelope in Eq.~\eqref{eq:TB_completion_envelope}. Since the completion count over $(t-W,t]$ equals the difference between the cumulative counts at its two boundaries, for $t\geq W>0$ we have
\begin{equation}
\label{eq:projected_completion_count_error_bound}
\begin{aligned}
    \left|
        \hat N_C^{\mathrm{proj}}(t,W)
        -
        N_C(t,W)
    \right|&=
    \left|
        \Delta N_C^{\mathrm{proj}}(t)
        -
        \Delta N_C^{\mathrm{proj}}(t-W)
    \right|\\
    &\leq
    2(G_{\max}\delta_p+B_{\max}).
\end{aligned}
\end{equation}
Dividing by $W$ proves
\begin{equation}
    \left|
        \hat G_{\mathrm{TB}}^{\mathrm{proj}}(t,W)
        -
        G(t,W)
    \right|
    \leq
    \frac{2(\delta_pG_{\max}+B_{\max})}{W},
\end{equation}
Again, if $\delta_p=0$, the outflow-estimation error vanishes.
\end{proof}

\section{Proofs of the multi-class projected-space relations}
\label{app:proof_multi_class}

\subsection{Proof of \autoref{prop:multi_class_projected_AB}}
\label{app:proof_multi_class_AB}

\begin{proof}
Under the common-distance assumption $S_m=S$, the class-specific projected-space AB relation in Eq.~\eqref{eq:Gkbfn} gives
\begin{equation}
    G_m(\mathbf n)
    =
    \frac{n_mU_m(\mathbf n)}{S},
    \qquad m\in\mathcal M.
\end{equation}
Summing over all classes, the aggregate outflow is
\begin{equation}
\label{eq:G_multiclass}
    G(\mathbf n)
    =
    \sum_{m\in\mathcal M}G_m(\mathbf n)
    =
    \frac{1}{S}
    \sum_{m\in\mathcal M}n_mU_m(\mathbf n).
\end{equation}
Since the active population is partitioned into mutually exclusive classes, the aggregate snapshot distribution is a mixture of the class-specific snapshot distributions, weighted by their accumulation fractions. Hence, the representative projected speed over the entire active population satisfies
\begin{equation}
\label{eq:aggregate_projected_speed_multiclass}
    U(\mathbf n)
    =
    \sum_{m\in\mathcal M}
    \frac{n_m}{n_{\mathrm{tot}}}
    U_m(\mathbf n),
\end{equation}
where $n_{\mathrm{tot}}=\sum_{m\in\mathcal M}n_m>0$. Substituting this identity into Eq.~\eqref{eq:G_multiclass} yields
\begin{equation}
    G(\mathbf n)
    =
    \frac{n_{\mathrm{tot}}U(\mathbf n)}{S},
\end{equation}
which establishes the aggregate projected-space AB relation.
\end{proof}

\subsection{Proof of \autoref{prop:multi_class_fraction_relation}}
\label{app:proof_multi_class_fractions}

\begin{proof}
Under stationary conditions, Little's law applied separately to each class gives
\begin{equation}
\label{eq:multiclass_little_law}
    n_m
    =
    q_m\mathbb E_f[T\mid\mathbf n,m],
    \qquad m\in\mathcal M.
\end{equation}
For each class with a positive stationary generation rate, flow conservation and the class-specific projected-space AB relation in Eq.~\eqref{eq:Gkbfn} imply
\begin{equation}
    q_m
    =
    G_m(\mathbf n)
    =
    n_m\frac{U_m(\mathbf n)}{S_m}.
\end{equation}
Combining these identities yields
\begin{equation}
\label{eq:class_mean_travel_time_projected}
    \mathbb E_f[T\mid\mathbf n,m]
    =
    \frac{S_m}{U_m(\mathbf n)}.
\end{equation}
Substituting this expression into Eq.~\eqref{eq:multiclass_little_law} and using $q_m=q_{\mathrm{tot}}\alpha_m^q$ gives
\begin{equation}
\label{eq:nk}
    n_m
    =
    q_m\frac{S_m}{U_m(\mathbf n)}
    =
    q_{\mathrm{tot}}\alpha_m^q
    \frac{S_m}{U_m(\mathbf n)}.
\end{equation}
Summing over all classes yields
\begin{equation}
\label{eq:ntot}
    n_{\mathrm{tot}}
    =
    q_{\mathrm{tot}}
    \sum_{l\in\mathcal M}
    \alpha_l^q
    \frac{S_l}{U_l(\mathbf n)}.
\end{equation}
Therefore, dividing Eq.~\eqref{eq:nk} by Eq.~\eqref{eq:ntot}, we obtain
\begin{equation}
    \alpha_m^n
    =
    \frac{n_m}{n_{\mathrm{tot}}}
    =
    \frac{
        \displaystyle
        \alpha_m^q\frac{S_m}{U_m(\mathbf n)}
    }{
        \displaystyle
        \sum_{l\in\mathcal M}
        \alpha_l^q\frac{S_l}{U_l(\mathbf n)}
    },
\end{equation}
which proves Eq.~\eqref{eq:multi_class_fraction_forward}. When all classes share the same expected shortest OD distance $S_m=S$, the common factor cancels, giving
\begin{equation}
    \alpha_m^n
    =
    \frac{
        \alpha_m^q/U_m(\mathbf n)
    }{
        \displaystyle
        \sum_{l\in\mathcal M}
        \alpha_l^q/U_l(\mathbf n)
    },
\end{equation}
as stated in Eq.~\eqref{eq:multi_class_fraction_common_S}.
\end{proof}

\end{appendices}

\bibliographystyle{abbrvnat}
\bibliography{cas-refs}

\begin{thebibliography}{70}
\providecommand{\natexlab}[1]{#1}
\providecommand{\url}[1]{\texttt{#1}}
\expandafter\ifx\csname urlstyle\endcsname\relax
  \providecommand{\doi}[1]{doi: #1}\else
  \providecommand{\doi}{doi: \begingroup \urlstyle{rm}\Url}\fi

\bibitem[Alahi et~al.(2014)Alahi, Bierlaire, and Vandergheynst]{alahi2014robust}
Alahi, A., Bierlaire, M., and Vandergheynst, P., 2014.
\newblock Robust real-time pedestrians detection in urban environments with low-resolution cameras.
\newblock \emph{Transportation Research Part C: Emerging Technologies}, 39:\penalty0 113--128.
\newblock \doi{10.1016/j.trc.2013.11.019}.

\bibitem[Archetti et~al.(2022)Archetti, Peirano, and Speranza]{archetti2022optimization}
Archetti, C., Peirano, L., and Speranza, M.~G., 2022.
\newblock Optimization in multimodal freight transportation problems: A survey.
\newblock \emph{European Journal of Operational Research}, 299\penalty0 (1):\penalty0 1--20.

\bibitem[Arnott(2013)]{arnott2013bathtub}
Arnott, R., 2013.
\newblock A bathtub model of downtown traffic congestion.
\newblock \emph{Journal of Urban Economics}, 76:\penalty0 110--121.

\bibitem[Barmpounakis and Geroliminis(2020)]{barmpounakis2020new}
Barmpounakis, E. and Geroliminis, N., 2020.
\newblock On the new era of urban traffic monitoring with massive drone data: The pneuma large-scale field experiment.
\newblock \emph{Transportation research part C: emerging technologies}, 111:\penalty0 50--71.

\bibitem[Batista et~al.(2019)Batista, Leclercq, and Geroliminis]{batista2019estimation}
Batista, S., Leclercq, L., and Geroliminis, N., 2019.
\newblock Estimation of regional trip length distributions for the calibration of the aggregated network traffic models.
\newblock \emph{Transportation Research Part B: Methodological}, 122:\penalty0 192--217.

\bibitem[Batista et~al.(2022)Batista, Tilg, and Men{\'e}ndez]{batista2022exploring}
Batista, S., Tilg, G., and Men{\'e}ndez, M., 2022.
\newblock Exploring the potential of aggregated traffic models for estimating network-wide emissions.
\newblock \emph{Transportation Research Part D: Transport and Environment}, 109:\penalty0 103354.

\bibitem[Beojone and Geroliminis(2023)]{beojone2023dynamic}
Beojone, C.~V. and Geroliminis, N., 2023.
\newblock A dynamic multi-region {MFD} model for ride-sourcing with ridesplitting.
\newblock \emph{Transportation Research Part B: Methodological}, 177:\penalty0 102821.

\bibitem[Beza et~al.(2025)Beza, Xie, Ramezani, and Levinson]{beza2025lanefree}
Beza, A.~D., Xie, Z., Ramezani, M., and Levinson, D., 2025.
\newblock From lane-less to lane-free: Implications in the era of automated vehicles.
\newblock \emph{Transportation Research Part C: Emerging Technologies}, 170:\penalty0 104898.
\newblock \doi{10.1016/j.trc.2024.104898}.

\bibitem[Brunner et~al.(2024)Brunner, Ni, Kouvelas, and Makridis]{brunner2024microscopic}
Brunner, J.~S., Ni, Y.-C., Kouvelas, A., and Makridis, M.~A., 2024.
\newblock Microscopic simulation of bicycle traffic flow incorporating cyclists’ heterogeneous dynamics and non-lane-based movement strategies.
\newblock \emph{Simulation Modelling Practice and Theory}, 135:\penalty0 102986.

\bibitem[Carrami{\~n}ana et~al.(2025)Carrami{\~n}ana, Besada, and Bernardos]{carraminana2025authorization}
Carrami{\~n}ana, D., Besada, J.~A., and Bernardos, A.~M., 2025.
\newblock A fair and congestion-aware flight authorization framework for unmanned traffic management.
\newblock \emph{Aerospace}, 12\penalty0 (10):\penalty0 881.
\newblock \doi{10.3390/aerospace12100881}.

\bibitem[Cassidy and Rudjanakanoknad(2002)]{cassidy2002empirical}
Cassidy, M.~J. and Rudjanakanoknad, J., 2002.
\newblock Empirical study of ramp metering and capacity.

\bibitem[Chen and Geroliminis(2026)]{chen2026perimeter}
Chen, C. and Geroliminis, N., 2026.
\newblock A perimeter control-route guidance framework of urban networks with priorities.
\newblock \emph{Transportation Research Part B: Methodological}, 214:\penalty0 103589.

\bibitem[Chen et~al.(2022)Chen, Huang, Lam, Pan, Hsu, Sumalee, and Zhong]{chen2022data}
Chen, C., Huang, Y., Lam, W., Pan, T., Hsu, S., Sumalee, A., and Zhong, R., 2022.
\newblock Data efficient reinforcement learning and adaptive optimal perimeter control of network traffic dynamics.
\newblock \emph{Transportation Research Part C: Emerging Technologies}, 142:\penalty0 103759.

\bibitem[Chen et~al.(2024)Chen, Geroliminis, and Zhong]{chen2024iterative}
Chen, C., Geroliminis, N., and Zhong, R., 2024.
\newblock An iterative adaptive dynamic programming approach for macroscopic fundamental diagram-based perimeter control and route guidance.
\newblock \emph{Transportation Science}, 58\penalty0 (4):\penalty0 896--918.

\bibitem[Chen et~al.(2026)Chen, Zhu, Fu, and Lam]{chen2026medical}
Chen, J., Zhu, N., Fu, H., and Lam, W. H.~K., 2026.
\newblock Drone pre-positioning for cardiac arrest response: A feature-driven distributionally robust chance constrained programming.
\newblock \emph{Transportation Research Part B: Methodological}, 211:\penalty0 103525.
\newblock \doi{10.1016/j.trb.2026.103525}.

\bibitem[Cummings and Mahmassani(2024)]{cummings2024airspace}
Cummings, C. and Mahmassani, H., 2024.
\newblock Airspace congestion, flow relations, and 4-d fundamental diagrams for advanced urban air mobility.
\newblock \emph{Transportation Research Part C: Emerging Technologies}, 159:\penalty0 104467.
\newblock \doi{10.1016/j.trc.2023.104467}.

\bibitem[Daganzo(2007)]{daganzo2007urban}
Daganzo, C.~F., 2007.
\newblock Urban gridlock: Macroscopic modeling and mitigation approaches.
\newblock \emph{Transportation Research Part B: Methodological}, 41\penalty0 (1):\penalty0 49--62.

\bibitem[Daganzo and Lehe(2015)]{daganzo2015distance}
Daganzo, C.~F. and Lehe, L.~J., 2015.
\newblock Distance-dependent congestion pricing for downtown zones.
\newblock \emph{Transportation Research Part B: Methodological}, 75:\penalty0 89--99.

\bibitem[Ding et~al.(2026)Ding, Zeng, Zheng, Zheng, Bai, and Hou]{ding2026state}
Ding, H., Zeng, R., Zheng, N., Zheng, X., Bai, H., and Hou, C., 2026.
\newblock A state-degradation-risk-embedded sliding mode perimeter control considering state transition for regional road networks under cyberattacks.
\newblock \emph{Transportation Research Part C: Emerging Technologies}, 193:\penalty0 105944.
\newblock \doi{10.1016/j.trc.2026.105944}.

\bibitem[Du et~al.(2025)Du, Zhong, Wang, Wu, Zhang, and Xue]{du2025mission}
Du, S., Zhong, G., Wang, F., Wu, L., Zhang, H., and Xue, D., 2025.
\newblock A framework for collaborative {UAM} traffic flow optimization with mission preferences: Incorporating customized strategy synergy into strategic conflict management.
\newblock \emph{Transportation Research Part E: Logistics and Transportation Review}, 202:\penalty0 104326.
\newblock \doi{10.1016/j.tre.2025.104326}.

\bibitem[Edie et~al.(1963)]{edie1963discussion}
Edie, L.~C. et~al., 1963.
\newblock Discussion of traffic stream measurements and definitions.

\bibitem[Fosgerau(2015)]{fosgerau2015congestion}
Fosgerau, M., 2015.
\newblock Congestion in the bathtub.
\newblock \emph{Economics of Transportation}, 4\penalty0 (4):\penalty0 241--255.

\bibitem[Fragapane et~al.(2021)Fragapane, De~Koster, Sgarbossa, and Strandhagen]{fragapane2021planning}
Fragapane, G., De~Koster, R., Sgarbossa, F., and Strandhagen, J.~O., 2021.
\newblock Planning and control of autonomous mobile robots for intralogistics: Literature review and research agenda.
\newblock \emph{European journal of operational research}, 294\penalty0 (2):\penalty0 405--426.

\bibitem[Fu et~al.(2026)Fu, Safadi, Quan, and Haddad]{fu2026sky}
Fu, R., Safadi, Y., Quan, Q., and Haddad, J., 2026.
\newblock Sky highway: An air traffic structure for low-altitude heterogeneous vtol aircraft.
\newblock \emph{Transportation Research Part C: Emerging Technologies}, 185:\penalty0 105531.
\newblock \doi{10.1016/j.trc.2026.105531}.

\bibitem[Geroliminis and Daganzo(2008)]{geroliminis2008existence}
Geroliminis, N. and Daganzo, C.~F., 2008.
\newblock Existence of urban-scale macroscopic fundamental diagrams: Some experimental findings.
\newblock \emph{Transportation Research Part B: Methodological}, 42\penalty0 (9):\penalty0 759--770.

\bibitem[Geroliminis et~al.(2014)Geroliminis, Zheng, and Ampountolas]{geroliminis2014bimodal}
Geroliminis, N., Zheng, N., and Ampountolas, K., 2014.
\newblock A three-dimensional macroscopic fundamental diagram for mixed bi-modal urban networks.
\newblock \emph{Transportation Research Part C: Emerging Technologies}, 42:\penalty0 168--181.
\newblock \doi{10.1016/j.trc.2014.03.004}.

\bibitem[Haddad(2015)]{haddad2015robust}
Haddad, J., 2015.
\newblock Robust constrained control of uncertain macroscopic fundamental diagram networks.
\newblock \emph{Transportation Research Procedia}, 7:\penalty0 669--688.

\bibitem[Haddad and Zheng(2020)]{haddad2020adaptive}
Haddad, J. and Zheng, Z., 2020.
\newblock Adaptive perimeter control for multi-region accumulation-based models with state delays.
\newblock \emph{Transportation Research Part B: Methodological}, 137:\penalty0 133--153.

\bibitem[Haddad et~al.(2021)Haddad, Mirkin, and Assor]{haddad2021traffic}
Haddad, J., Mirkin, B., and Assor, K., 2021.
\newblock Traffic flow modeling and feedback control for future low-altitude air city transport: An mfd-based approach.
\newblock \emph{Transportation Research Part C: Emerging Technologies}, 133:\penalty0 103380.

\bibitem[Hoogendoorn et~al.(2018)Hoogendoorn, Daamen, Knoop, Steenbakkers, and Sarvi]{hoogendoorn2018macroscopic}
Hoogendoorn, S.~P., Daamen, W., Knoop, V.~L., Steenbakkers, J., and Sarvi, M., 2018.
\newblock Macroscopic fundamental diagram for pedestrian networks: theory and applications.
\newblock \emph{Transportation research part C: emerging technologies}, 94:\penalty0 172--184.

\bibitem[Huang et~al.(2022)Huang, Sun, and Zhang]{huang2022bicycle}
Huang, Y., Sun, D.~J., and Zhang, S., 2022.
\newblock Three-dimensional macroscopic fundamental diagram for car and bicycle heterogeneous traffic.
\newblock \emph{Transportmetrica B: Transport Dynamics}, 10\penalty0 (1):\penalty0 312--339.
\newblock \doi{10.1080/21680566.2021.1994050}.

\bibitem[Jover et~al.(2022)Jover, Berm{\'u}dez, and Casado]{jover2022priority}
Jover, J., Berm{\'u}dez, A., and Casado, R., 2022.
\newblock Priority-aware conflict resolution for {U-Space}.
\newblock \emph{Electronics}, 11\penalty0 (8):\penalty0 1225.
\newblock \doi{10.3390/electronics11081225}.

\bibitem[Kim et~al.(2024)Kim, Tak, Barmpounakis, and Geroliminis]{kim2024monitoring}
Kim, S., Tak, Y., Barmpounakis, E., and Geroliminis, N., 2024.
\newblock Monitoring outdoor parking in urban areas with unmanned aerial vehicles.
\newblock \emph{IEEE Transactions on Intelligent Transportation Systems}, 25\penalty0 (10):\penalty0 13393--13406.

\bibitem[Knoop et~al.(2025)Knoop, Ellerbroek, ter Heide, and Hoogendoorn]{knoop2025airplane}
Knoop, V.~L., Ellerbroek, J., ter Heide, M., and Hoogendoorn, S., 2025.
\newblock Macroscopic fundamental diagram for airplane traffic: Empirical findings.
\newblock \emph{Transportation Research Record}, 2679\penalty0 (2):\penalty0 960--970.
\newblock \doi{10.1177/03611981241265683}.

\bibitem[Lamotte and Geroliminis(2018)]{lamotte2018morning}
Lamotte, R. and Geroliminis, N., 2018.
\newblock The morning commute in urban areas with heterogeneous trip lengths.
\newblock \emph{Transportation Research Part B: Methodological}, 117:\penalty0 794--810.

\bibitem[Lamotte et~al.(2018)Lamotte, Murashkin, Kouvelas, and Geroliminis]{lamotte2018dynamic}
Lamotte, R., Murashkin, M., Kouvelas, A., and Geroliminis, N., 2018.
\newblock Dynamic modeling of trip completion rate in urban areas with mfd representations.
\newblock In \emph{2018 TRB Annual Meeting Online}, pages 18--06192. Transportation Research Board.

\bibitem[Leclercq and Paipuri(2019)]{leclercq2019macroscopic}
Leclercq, L. and Paipuri, M., 2019.
\newblock Macroscopic traffic dynamics under fast-varying demand.
\newblock \emph{Transportation Science}, 53\penalty0 (6):\penalty0 1526--1545.
\newblock \doi{10.1287/trsc.2019.0908}.

\bibitem[Leclercq et~al.(2014)Leclercq, Chiabaut, and Trinquier]{leclercq2014macroscopic}
Leclercq, L., Chiabaut, N., and Trinquier, B., 2014.
\newblock Macroscopic fundamental diagrams: A cross-comparison of estimation methods.
\newblock \emph{Transportation Research Part B: Methodological}, 62:\penalty0 1--12.

\bibitem[Leclercq et~al.(2017)Leclercq, S{\'e}n{\'e}cat, and Mariotte]{leclercq2017dynamic}
Leclercq, L., S{\'e}n{\'e}cat, A., and Mariotte, G., 2017.
\newblock Dynamic macroscopic simulation of on-street parking search: A trip-based approach.
\newblock \emph{Transportation Research Part B: Methodological}, 101:\penalty0 268--282.

\bibitem[Li et~al.(2020)Li, Qu, Li, Huang, and Wei]{li2020overview}
Li, C.-l., Qu, W.-q., Li, Y.-d., Huang, L.-y., and Wei, P., 2020.
\newblock Overview of traffic management of urban air mobility (uam) with evtol aircraft.
\newblock \emph{J. Traffic Transp. Eng}, 20\penalty0 (4):\penalty0 35--54.

\bibitem[Li et~al.(2021)Li, Mohajerpoor, and Ramezani]{li2021perimeter}
Li, Y., Mohajerpoor, R., and Ramezani, M., 2021.
\newblock Perimeter control with real-time location-varying cordon.
\newblock \emph{Transportation Research Part B: Methodological}, 150:\penalty0 101--120.

\bibitem[Li et~al.(2026)Li, Gu, Zheng, Ma, and Liu]{li2026two}
Li, Y., Gu, Z., Zheng, N., Ma, W., and Liu, Z., 2026.
\newblock Two-region mixed autonomy network with park-and-ride: Macroscopic modeling, microscopic validation, and multi-scenario analysis.
\newblock \emph{Transportation Research Part A: Policy and Practice}, 211:\penalty0 105084.

\bibitem[Liu et~al.(2026)Liu, Cai, Zhu, and Liu]{liu2026priority}
Liu, Z., Cai, K., Zhu, Y., and Liu, K., 2026.
\newblock Priority-based conflict resolution for manned and unmanned aircraft integrated operations.
\newblock \emph{IEEE Transactions on Intelligent Transportation Systems}, 27\penalty0 (10):\penalty0 12043--12062.
\newblock \doi{10.1109/TITS.2026.3719353}.

\bibitem[Loder et~al.(2022)Loder, Bliemer, and Axhausen]{loder2022optimal}
Loder, A., Bliemer, M.~C., and Axhausen, K.~W., 2022.
\newblock Optimal pricing and investment in a multi-modal city-{Introducing} a macroscopic network design problem based on the {MFD}.
\newblock \emph{Transportation Research Part A: Policy and Practice}, 156:\penalty0 113--132.

\bibitem[Ma et~al.(2026)Ma, Huang, Zheng, Pan, and Zhong]{ma2026calibration}
Ma, W., Huang, Y., Zheng, N., Pan, T., and Zhong, R., 2026.
\newblock Calibration and uncertainty quantification of macroscopic fundamental diagrams.
\newblock \emph{Transportation Research Part A: Policy and Practice}, 211:\penalty0 105127.

\bibitem[Maiti et~al.(2025)Maiti, Seppecher, and Leclercq]{maiti2025scaling}
Maiti, N., Seppecher, M., and Leclercq, L., 2025.
\newblock Scaling methods for estimating macroscopic fundamental diagrams in urban networks with sparse stationary sensor coverage.
\newblock \emph{Transportation Research Part C: Emerging Technologies}, 178:\penalty0 105213.
\newblock \doi{10.1016/j.trc.2025.105213}.

\bibitem[Mariotte et~al.(2017)Mariotte, Leclercq, and Laval]{mariotte2017macroscopic}
Mariotte, G., Leclercq, L., and Laval, J.~A., 2017.
\newblock Macroscopic urban dynamics: Analytical and numerical comparisons of existing models.
\newblock \emph{Transportation Research Part B: Methodological}, 101:\penalty0 245--267.

\bibitem[Mart{\'i}nez and Jin(2024)]{martinez2024relative}
Mart{\'i}nez, I. and Jin, W.-L., 2024.
\newblock Priority queue formulation of agent-based bathtub model for network trip flows in the relative space.
\newblock \emph{Transportation Research Part C: Emerging Technologies}, 168:\penalty0 104765.
\newblock \doi{10.1016/j.trc.2024.104765}.

\bibitem[{National Aeronautics and Space Administration}(2026)]{nasa2026priority}
{National Aeronautics and Space Administration}, Apr. 2026.
\newblock {NASA} explores prioritizing first response drones in crowded skies.
\newblock URL \url{https://www.nasa.gov/centers-and-facilities/ames/nasa-explores-prioritizing-first-response-drones-in-crowded-skies/}.
\newblock NASA Ames Research Center.

\bibitem[Ni et~al.(2024)Ni, Makridis, and Kouvelas]{ni2024bicycle}
Ni, Y.-C., Makridis, M.~A., and Kouvelas, A., 2024.
\newblock Bicycle as a traffic mode: From microscopic cycling behavior to macroscopic bicycle flow.
\newblock \emph{Journal of Cycling and Micromobility Research}, 2:\penalty0 100022.

\bibitem[Paipuri et~al.(2021)Paipuri, Barmpounakis, Geroliminis, and Leclercq]{paipuri2021empirical}
Paipuri, M., Barmpounakis, E., Geroliminis, N., and Leclercq, L., 2021.
\newblock Empirical observations of multi-modal network-level models: Insights from the pneuma experiment.
\newblock \emph{Transportation Research Part C: Emerging Technologies}, 131:\penalty0 103300.

\bibitem[Papageorgiou et~al.(2021)Papageorgiou, Mountakis, Karafyllis, Papamichail, and Wang]{papageorgiou2021lane}
Papageorgiou, M., Mountakis, K.-S., Karafyllis, I., Papamichail, I., and Wang, Y., 2021.
\newblock Lane-free artificial-fluid concept for vehicular traffic.
\newblock \emph{Proceedings of the IEEE}, 109\penalty0 (2):\penalty0 114--121.

\bibitem[Patire et~al.(2016)Patire, Butler, Xuan, Schreiters, Gomes, Dion, Gabet, and Winik]{I210}
Patire, A.~D., Butler, J., Xuan, E., Schreiters, T., Gomes, G., Dion, F., Gabet, P., and Winik, F., 2016.
\newblock I-210 pilot: Analysis, modeling, and simulation phase 1 report.
\newblock Technical report, University of California, Berkeley.

\bibitem[Quan et~al.(2021)Quan, Fu, and Cai]{quan2021practical}
Quan, Q., Fu, R., and Cai, K.-Y., 2021.
\newblock Practical control for multicopters to avoid non-cooperative moving obstacles.
\newblock \emph{IEEE Transactions on Intelligent Transportation Systems}, 23\penalty0 (8):\penalty0 10839--10857.

\bibitem[Ramezani and Nourinejad(2018)]{ramezani2018dynamic}
Ramezani, M. and Nourinejad, M., 2018.
\newblock Dynamic modeling and control of taxi services in large-scale urban networks: A macroscopic approach.
\newblock \emph{Transportation Research Part C: Emerging Technologies}, 94:\penalty0 203--219.

\bibitem[Safadi et~al.(2023)Safadi, Fu, Quan, and Haddad]{safadi2023macroscopic}
Safadi, Y., Fu, R., Quan, Q., and Haddad, J., 2023.
\newblock Macroscopic fundamental diagrams for low-altitude air city transport.
\newblock \emph{Transportation Research Part C: Emerging Technologies}, 152:\penalty0 104141.

\bibitem[Sengupta et~al.(2025)Sengupta, Bulusu, Mballo, Onat, and Cao]{sengupta2025uam}
Sengupta, R., Bulusu, V., Mballo, C.~E., Onat, E.~B., and Cao, S., 2025.
\newblock Urban air mobility research challenges and opportunities.
\newblock \emph{Annual Review of Control, Robotics, and Autonomous Systems}, 8:\penalty0 407--431.
\newblock \doi{10.1146/annurev-control-022823-031353}.

\bibitem[Sirmatel and Geroliminis(2018)]{sirmatel2018economic}
Sirmatel, I.~I. and Geroliminis, N., 2018.
\newblock Economic model predictive control of large-scale urban road networks via perimeter control and regional route guidance.
\newblock \emph{IEEE Transactions on Intelligent Transportation Systems}, 19\penalty0 (4):\penalty0 1112--1121.

\bibitem[Sirmatel and Geroliminis(2019)]{sirmatel2019nonlinear}
Sirmatel, I.~I. and Geroliminis, N., 2019.
\newblock Nonlinear moving horizon estimation for large-scale urban road networks.
\newblock \emph{IEEE Transactions on Intelligent Transportation Systems}, 21\penalty0 (12):\penalty0 4983--4994.

\bibitem[Sirmatel et~al.(2021)Sirmatel, Tsitsokas, Kouvelas, and Geroliminis]{sirmatel2021modeling}
Sirmatel, I.~I., Tsitsokas, D., Kouvelas, A., and Geroliminis, N., 2021.
\newblock Modeling, estimation, and control in large-scale urban road networks with remaining travel distance dynamics.
\newblock \emph{Transportation Research Part C: Emerging Technologies}, 128:\penalty0 103157.

\bibitem[Straubinger et~al.(2020)Straubinger, Rothfeld, Shamiyeh, B{\"u}chter, Kaiser, and Pl{\"o}tner]{straubinger2020overview}
Straubinger, A., Rothfeld, R., Shamiyeh, M., B{\"u}chter, K.-D., Kaiser, J., and Pl{\"o}tner, K.~O., 2020.
\newblock An overview of current research and developments in urban air mobility--setting the scene for uam introduction.
\newblock \emph{Journal of Air Transport Management}, 87:\penalty0 101852.

\bibitem[Sun et~al.(2027)Sun, Chen, Li, Li, Qu, and Nie]{sun2027sharp}
Sun, H., Chen, S., Li, Y., Li, Q., Qu, X., and Nie, B., 2027.
\newblock A sharp capacity limit for low-altitude urban airspace and its management implications.
\newblock \emph{Transportation Research Part C: Emerging Technologies}, 194:\penalty0 106029.

\bibitem[Trinh et~al.(2022)Trinh, Ekstr{\"o}m, and C{\"u}r{\"u}kl{\"u}]{trinh2022navigation}
Trinh, L.~A., Ekstr{\"o}m, M., and C{\"u}r{\"u}kl{\"u}, B., 2022.
\newblock Dependable navigation for multiple autonomous robots with petri nets based congestion control and dynamic obstacle avoidance.
\newblock \emph{Journal of Intelligent \& Robotic Systems}, 104:\penalty0 69.
\newblock \doi{10.1007/s10846-022-01589-1}.

\bibitem[Wang et~al.(2026)Wang, Ding, Zheng, Ma, and Zheng]{wang2026regional}
Wang, L., Ding, H., Zheng, N., Ma, W., and Zheng, X., 2026.
\newblock Regional route guidance with multi-party perspective under disinformation attacks: An {MFD}-based approach.
\newblock \emph{Transportation Research Part C: Emerging Technologies}, 186:\penalty0 105626.
\newblock \doi{10.1016/j.trc.2026.105626}.

\bibitem[Wang et~al.(2025)Wang, Ding, Chen, Zhang, Fang, and Ma]{wang2025detour}
Wang, Q., Ding, H., Chen, J., Zhang, J., Fang, H., and Ma, J., 2025.
\newblock Detour characteristics of bicycles for collision avoidance in the bicycle-pedestrian mixed traffic.
\newblock \emph{Transportmetrica A: Transport Science}.
\newblock \doi{10.1080/23249935.2025.2487133}.

\bibitem[Weng et~al.(2025)Weng, Pan, Chen, and Zhong]{weng2025urban}
Weng, C., Pan, T., Chen, C., and Zhong, R., 2025.
\newblock Urban low-altitude air transport management: Bridging dynamic traffic control and static network equilibrium.
\newblock \emph{Transportation Research Part C: Emerging Technologies}, 178:\penalty0 105237.

\bibitem[Weng et~al.(2026)Weng, Chen, Tan, Pan, and Zhong]{weng2026real}
Weng, C., Chen, C., Tan, J., Pan, T., and Zhong, R., 2026.
\newblock Real-time traffic simulation and management for large-scale urban air mobility: Integrating route guidance and collision avoidance.
\newblock \emph{Transportation Research Part C: Emerging Technologies}, 183:\penalty0 105477.

\bibitem[Xiong et~al.(2025)Xiong, Fonod, Alahi, and Geroliminis]{xiong2025multi}
Xiong, W., Fonod, R., Alahi, A., and Geroliminis, N., 2025.
\newblock Multi-source urban traffic flow forecasting with drone and loop detector data.
\newblock \emph{IEEE Transactions on Intelligent Transportation Systems}.

\bibitem[Yang et~al.(2019)Yang, Menendez, and Zheng]{yang2019heterogeneity}
Yang, K., Menendez, M., and Zheng, N., 2019.
\newblock Heterogeneity aware urban traffic control in a connected vehicle environment: A joint framework for congestion pricing and perimeter control.
\newblock \emph{Transportation Research Part C: Emerging Technologies}, 105:\penalty0 439--455.

\bibitem[Zhao et~al.(2026)Zhao, Wang, Zhang, Chen, and Qu]{zhao2026joint}
Zhao, Z., Wang, K., Zhang, W., Chen, S., and Qu, X., 2026.
\newblock Joint planning of vertiports and backup landing sites for urban air mobility networks.
\newblock \emph{Transportation Research Part E: Logistics and Transportation Review}, 214:\penalty0 105075.

\end{thebibliography}
\end{sloppypar}
\end{document}